\documentclass[aps,prd,twocolumn,nofootinbib,superscriptaddress]{revtex4-2}

\usepackage{amsmath,amssymb,amsfonts}
\usepackage{graphicx}
\usepackage{dcolumn}
\usepackage{bm}
\usepackage{hyperref}
\usepackage{xcolor}
\usepackage{booktabs}
\usepackage{threeparttable}

\newcommand{\Mpl}{M_{\rm Pl}}
\newcommand{\fPBH}{f_{\rm PBH}}

\newcommand{\Trh}{T_{\rm rh}}

\newcommand{\bgdelta}{\Delta}
\newcommand{\CITE}[1]{\cite{#1}} 

\begin{document}

\title{Are Primordial Black Holes a Natural Dark Matter Candidate?}

\author{Stefano Profumo}
\affiliation{Department of Physics and Santa Cruz Institute for Particle Physics,\\
             University of California, Santa Cruz, California 95064, USA}
\email{profumo@ucsc.edu}

\date{\today}

\begin{abstract}
Primordial black holes (PBHs) in the asteroid-mass window
($10^{17}$--$10^{22}$~g) can account for all of the dark matter
without violating any observational constraint, yet are routinely
dismissed as fine-tuned. I put that dismissal to the test by
applying three complementary fine-tuning measures uniformly across
a broad landscape: three non-inflationary PBH production mechanisms,
six classes of inflationary PBH models, and seven particle dark
matter benchmarks, all evaluated against the same observable target.
Three distinct naturalness universality classes emerge, determined
entirely by the analytic structure of the abundance map rather than
by the nature of the dark matter candidate. Biased-domain-wall PBHs
are as natural as off-resonance weakly interacting massive particles
and freeze-in particles; early-matter-domination and first-order
phase transition PBH mechanisms occupy an intermediate tier alongside
coannihilating WIMPs, unified by a structural identity in which the
fine-tuning measure equals the logarithm of the ratio of the
formation scale to the matter--radiation equality scale; and
single-field ultra-slow-roll inflationary collapse is severely tuned
for a distinct reason: a double exponential in which the power
spectrum amplitude is itself exponentially sensitive to the inflaton
potential coefficients, on top of the exponential collapse
sensitivity of the abundance map. My main conclusion is that {\em the claim that PBH dark matter is
generically fine-tuned conflates the worst case with a landscape
spanning every naturalness tier}. The three-measure protocol also
resolves a tension in the recent literature: the Barbieri--Giudice
and Iovino--Riotto fine-tuning measures answer complementary
questions and are reconciled within the two-layer decomposition
developed here.
\end{abstract}


\maketitle


\section{\label{sec:intro}Introduction}

The identity of dark matter remains one of the central open
questions of physics.
Among the candidates that have attracted sustained interest
are weakly interacting massive particles (WIMPs)
\CITE{Jungman:1996, Bertone:2005},
axions \CITE{Peccei:1977, Weinberg:1978, Wilczek:1978},
feebly interacting (``frozen-in'') particles
\CITE{Hall:2010, McDonald:2002},
and asymmetric dark matter \CITE{Nussinov:1985, Kaplan:2009}.
More recently, primordial black holes (PBHs) have re-emerged
as a viable dark matter candidate \CITE{Carr:1974, Green:1997},
particularly in the asteroid-mass window
($10^{17} \lesssim M_{\rm PBH}/{\rm g} \lesssim 10^{22}$)
where Hawking evaporation is negligible, microlensing and
disruption constraints are weak, and the PBH fraction
$\fPBH \equiv \Omega_{\rm PBH}/\Omega_{\rm DM}$ can equal
unity \CITE{Carr:2021, Green:2021, Escriva:2022}.

A common objection to PBH dark matter is that it requires
fine-tuning: in inflationary PBH production the probability
of collapse is exponentially sensitive to the primordial
curvature power spectrum, which must be arranged to produce
the right number density in the right mass window.
This argument is frequently made qualitatively, but rarely
subjected to the same quantitative scrutiny applied to WIMP
dark matter, where a similar exponential sensitivity is implicitly 
encoded, for instance, in the fabled thermal freeze-out mechanism.
The WIMP miracle, the coincidence that a particle with
electroweak-scale mass and coupling naturally freezes out
with $\Omega h^2 \approx 0.12$, is itself a statement
about naturalness, with a well-defined quantitative
expression that deserves comparison.

The fine-tuning of dark matter relic abundances has attracted
quantitative attention most prominently in the supersymmetric WIMP
context.
Grothaus, Lindner, and Takanishi~\CITE{Grothaus:2012} applied
the Barbieri--Giudice (BG) measure~\cite{Barbieri:1988},
which quantifies how sensitively a physical observable responds
to fractional variations of the underlying input parameters,
$\Delta \equiv \max_i |\partial \ln \Omega / \partial \ln x_i|$,
to neutralino dark matter in the MSSM, showing that
direct-detection exclusions preferentially eliminate untuned
parameter regions.
The most comprehensive MSSM treatment was given by Cabrera
\textit{et al.}~\CITE{Cabrera:2016}, who studied all major
neutralino annihilation mechanisms under a unified BG
framework; their finding that funnel and coannihilation
scenarios carry the highest combined tuning is fully
consistent with the WIMP benchmarks of the present paper.
King and Roberts~\CITE{King:2006} coined the term
``supernatural dark matter'' for MSSM regions where the
relic abundance is achieved with no fine-tuning at all.

For PBH dark matter, the fine-tuning discussion has been
framed almost entirely around the inflationary-collapse
mechanism.
Several studies have quantified the sensitivity of the PBH
abundance to the inflaton potential shape in the
ultra-slow-roll regime~\CITE{Hertzberg:2017, Stamou:2024}.
Kalaja \textit{et al.}~\CITE{Kalaja:2019} showed that the
curvature power-spectrum variance $\sigma$ must lie within
a $\sim 7\%$ fractional window to produce all of the dark
matter as PBHs under Gaussian statistics, a result broadly
consistent with the quantitative fine-tuning estimates
of Refs.~\CITE{Hertzberg:2017, Stamou:2024}.
Iovino and Riotto~\CITE{Iovino:2024}, by contrast, apply
Wilson's naturalness criterion to single-field inflationary
PBH models and conclude that they are not technically
unnatural when the measure is formulated symmetrically
around the target abundance --- a conclusion that sits
in apparent tension with the sensitivity estimates above.
The present paper resolves this tension: the two approaches
answer complementary questions and are reconciled within
the two-layer decomposition developed here.

Three important gaps remain in the existing literature that
this paper fills.
First, no prior work has compared PBH and particle dark
matter fine-tuning under a \emph{common} measure applied
to a \emph{common} observable target ($\Omega_{\rm DM} h^2 = 0.120$);
the two communities have developed parallel but disconnected
naturalness vocabularies.
Second, the non-inflationary PBH formation mechanisms, including
biased domain walls, first-order phase transitions, and early
matter domination, have never been analyzed from a
quantitative fine-tuning perspective.
Third, no systematic comparison of fine-tuning \emph{across}
inflationary PBH model classes, from curvaton and
spectator-field models through hybrid inflation to single-field
ultra-slow-roll, has been presented within a unified, internally consistent fine-tuning measure  
framework.
The present paper addresses all three gaps.

In what follows I apply the BG measure to the parameter spaces
of three non-inflationary PBH formation mechanisms, six classes
of inflationary PBH model, and seven particle dark matter
benchmarks, using an identical protocol for each: $\Delta = 1$
means the abundance scales linearly with the parameter
(perfectly natural), $\Delta \gg 1$ means it is exponentially
sensitive (fine-tuned).

The principal results of this analysis are:
\begin{enumerate}
\item The fine-tuning of any dark matter construction
  is set by the analytic structure of its abundance map,
  \emph{not} by whether the dark matter is a particle
  or a gravitational relic.
  Three universality classes emerge, and PBH constructions
  appear in all three.

\item Biased-domain-wall PBHs are as natural as
  off-resonance WIMPs, freeze-in particles, and
  post-inflationary axions, and substantially more
  natural than the coannihilating or resonant-funnel
  WIMPs that survive current direct-detection limits.
  Within the PBH paradigm alone the naturalness hierarchy
  spans more than seven orders of magnitude, from the
  gravity-fixed domain-wall construction at one extreme
  to single-field ultra-slow-roll inflationary collapse
  at the other.

\item Constructions with a single exponential factor in their abundance
universally satisfy $\Delta \approx 14$--$50$ on the observed
dark matter contour, independently of the underlying microscopic
physics; the precise value is set by the logarithm of the ratio
of the formation temperature to the matter--radiation equality
temperature.
Coannihilating WIMPs and early-matter-domination PBHs occupy
this universality class robustly.
First-order phase transition PBHs also fall here within a
standard single-exponential approximation for the collapse
probability, but a more accurate treatment raises the
fine-tuning estimate significantly, and tracing the abundance
back to the underlying scalar potential reveals a further
double-exponential structure with $\Delta \sim 10^3$--$10^4$.
Curvaton inflationary PBH models occupy this intermediate tier
only for low reheating temperatures; for standard high-reheating
scenarios they are elevated to the most-tuned class, with the
reheating temperature setting the boundary between the two.
The off-resonance WIMP ``miracle'', by contrast, falls in the
natural class: when the freeze-out exponential is absorbed into
the measured annihilation cross section, the abundance map
reduces to a pure power law with $\Delta = 2$.

\item Resonance- or cancellation-dependent constructions
  (e.g. Higgs-funnel WIMP) and single-field inflationary
  collapse models are both highly tuned, but for
  distinct reasons: the former through a narrow
  annihilation pole, the latter through a double
  exponential in which the power spectrum amplitude
  is itself exponentially sensitive to inflaton
  potential coefficients, on top of an additional exponential sensitivity in the condition for 
  collapse.
  Multi-field and spectator-field inflationary models
  avoid the second exponential and fall into the
  intermediate universality class.
\end{enumerate}

This paper is organized as follows.
In Sec.~\ref{sec:measure} I define the fine-tuning measures
and establish the protocol applied uniformly to all scenarios.
In Sec.~\ref{sec:pbh} I derive the BG measure for three
non-inflationary PBH formation mechanisms
(Secs.~\ref{subsec:dw}--\ref{subsec:fopt}), presenting
analytic results and parameter-space heatmaps, and then map
the framework onto six classes of inflationary PBH model
(Sec.~\ref{subsec:inflation_ft}), including a systematic
comparison of their two-layer tuning structure.
In Sec.~\ref{sec:particle} I apply the same analysis to
seven particle dark matter benchmarks.
In Sec.~\ref{sec:comparison} I compare the full set of
results, establish the three universality classes, and
prove the structural identity $\Delta \approx \ln(\text{scale
ratio})$ that unifies the Class~II scenarios; the tier
classification is shown to be robust to the choice of
fine-tuning convention by presenting $\bgdelta_{\rm BG}$
alongside alternate fine-tuning measures, namely the Strumia--Rattazzi measure $\bgdelta_{\rm SR}$
and the island half-width $\epsilon$.
Section~\ref{sec:discussion} places the results in broader
context, discusses limitations, and identifies directions
for follow-up.
Conclusions are in Sec.~\ref{sec:conclusions}.

\section{\label{sec:measure}The Barbieri--Giudice Measure
         Applied to Relic Abundances}

\subsection{Definition and conventions}
\label{subsec:bg_def}

Fine-tuning in the context of new physics models was first
quantified systematically by Barbieri and Giudice~\CITE{Barbieri:1988},
who sought to measure how sensitively the electroweak scale depends
on parameters in the supersymmetric spectrum.
Given a physical observable $\mathcal{O}$ and a set of
independent input parameters $\{x_i\}$, the Barbieri--Giudice (BG)
measure is defined as
\begin{equation}
  \bgdelta \;\equiv\; \max_i \,\bgdelta_i, \qquad
  \bgdelta_i \equiv \left| \frac{\partial \ln \mathcal{O}}
                                 {\partial \ln x_i} \right|.
  \label{eq:bg}
\end{equation}
The interpretation is direct: $\bgdelta_i = n$ means that a
fractional change $\epsilon$ in the parameter $x_i$ produces
a fractional change $n\epsilon$ in the observable $\mathcal{O}$.
A construction is considered to be ``natural'' if $\bgdelta \sim \mathcal{O}(1)$,
``moderately tuned'' if $\bgdelta \sim \mathcal{O}(10)$--$\mathcal{O}(100)$,
and ``highly tuned'' if $\bgdelta \gg 100$; the requirement
$\bgdelta < 10$ is commonly adopted as a naturalness
criterion in the SUSY literature \CITE{Barbieri:1988, Ellis:1986},
though no sharp threshold carries fundamental significance.

In this paper the observable is the dark matter relic abundance,
$\mathcal{O} \equiv \Omega h^2$, evaluated at the benchmark values
of the model parameters and compared against the Planck~2018
measurement $\Omega_{\rm DM} h^2 = 0.1200 \pm 0.0012$
\CITE{Planck:2018}.
The sensitivity $\Delta_i$ measures how precisely one must
specify $x_i$ in order for the predicted abundance to land
within, say, a factor of two of $\Omega_{\rm DM} h^2$:
a value $\Delta_i = 35$ means that a $\sim 2\%$ variation
in $x_i$ takes the predicted abundance outside this
window.\footnote{A fractional shift $\epsilon$ in $x_i$
produces a fractional shift $\Delta_i \cdot \epsilon$ in
$\Omega h^2$; requiring the abundance to remain within a
factor of two of the target imposes
$\Delta_i \cdot \epsilon \lesssim \ln 2 \approx 0.69$,
hence $\epsilon \lesssim 0.69/\Delta_i$.
For $\Delta_i = 35$ this gives $\epsilon \lesssim 2\%$;
more generally, the fractional precision requirement is
$\epsilon \approx 69\%/\Delta_i$.}

The BG measure is a \emph{local} quantity: it characterizes
the sensitivity of $\Omega_{\rm DM} h^2$ to small variations around a
chosen benchmark point in parameter space.
It should be distinguished from related but inequivalent notions
of fine-tuning, including the prior probability of landing on the
$\Omega = \Omega_{\rm DM}$ surface from a random draw of
parameters, and the fractional volume of the natural region
relative to the prior volume.
I return to the distinction between these concepts, and to
the specific case of misalignment-axion dark matter where it is
particularly consequential, in Sec.~\ref{subsec:prior}.

The BG measure is symmetric in the sense that $\bgdelta_i$
takes the same value whether one uses $\mathcal{O} = \Omega h^2$
or $\mathcal{O} = 1/(\Omega h^2)$ as the observable,
since $|\partial \ln \mathcal{O} / \partial \ln x_i|$ is
unchanged by this inversion.
For PBH scenarios, replacing $\Omega h^2$ with $\fPBH$
also leaves $\bgdelta$ unchanged, since $\fPBH = \Omega_{\rm PBH} h^2 /
\Omega_{\rm DM} h^2$ differs only by the constant $\Omega_{\rm DM} h^2$.
It is \emph{not} symmetric under reparametrizations of the
parameter space: $\bgdelta$ evaluated with respect to parameters
$\{x_i\}$ and $\{y_j = f_j(x)\}$ will generally differ if
$f_j$ is nonlinear.
The physical content of $\bgdelta$ therefore depends
on the identification of fundamental parameters, which
I address in Sec.~\ref{subsec:parameters}.

For reference, a pure power-law abundance map of the form
\begin{equation}
  \Omega h^2 \propto \prod_i x_i^{\,\alpha_i}
  \label{eq:powerlaw}
\end{equation}
yields $\bgdelta_i = |\alpha_i|$ identically
everywhere in parameter space.
For an abundance set by a single exponential,
\begin{equation}
  \Omega h^2 \propto e^{-A\,x} \cdot x^\beta,
  \label{eq:explaw}
\end{equation}
the dominant contribution on the
$\Omega h^2 = \Omega_{\rm DM} h^2$ surface is
\begin{equation}
  \bgdelta_x \;\approx\; A\,x \;=\; \ln\!\left[
    \frac{\Omega_{\rm natural}}{\Omega_{\rm DM} h^2}\right],
  \label{eq:expft}
\end{equation}
where $\Omega_{\rm natural}$ is the abundance that would obtain
if the exponential were replaced by unity.
The full derivative is $\bgdelta_x = Ax + \beta$, so the
approximation $\bgdelta_x \approx Ax$ overestimates the
true value by $-\beta/(Ax)$; for the early-matter-domination
benchmark where $\beta = 5$ and $Ax \approx 9$--$19$, the
correction is $20$--$35\%$ and is the reason the benchmark
$\bgdelta_\sigma = 14.4$ (which includes the prefactor term
$+5$) is larger than the pure-exponential approximation
$Ax \approx 9$--$10$ at $\sigma = 0.055$.
This result, which is exact when $\beta = 0$ and approximate
otherwise, is central to understanding why the three
``single-exponential'' dark matter constructions discussed
in Sec.~\ref{sec:comparison} yield a universal value
$\bgdelta \approx \ln(T_{\rm form}/T_{\rm eq}) \approx 14$--$50$.
Constructions with a Breit--Wigner resonance yield
$\bgdelta \sim M^2/(\Gamma M) \gg \ln(\text{scale ratio})$,
a qualitatively distinct and much more severe tuning.

Two complementary measures are computed alongside BG
throughout this paper.
The \emph{Strumia--Rattazzi} (SR) measure
\CITE{Strumia:2000} replaces the maximum with a
quadrature sum over all parameter directions:
\begin{equation}
  \bgdelta_{\rm SR}
  \;\equiv\;
  \sqrt{\sum_i \bgdelta_i^2}
  \;=\;
  \left|\nabla_{\ln x}\, \ln \Omega h^2\right|.
  \label{eq:sr}
\end{equation}
This equals the magnitude of the log-gradient vector and
is sensitive to fine-tuning distributed across multiple
parameters simultaneously.
For single-parameter scenarios, obviously, $\bgdelta_{\rm SR} = \bgdelta_{\rm BG}$;
for scenarios where all parameters are equally sensitive
(e.g.\ ADM with $\bgdelta_{M} = \bgdelta_R = 1$),
$\bgdelta_{\rm SR} = \sqrt{n}\,\bgdelta_{\rm BG}$ where
$n$ is the number of parameters.

The \emph{island half-width} $\epsilon$ in the most sensitive
parameter direction provides the most operationally transparent
measure of tuning.
Rather than asking how steeply the abundance responds to a
parameter shift (BG) or summing sensitivities in quadrature
(SR), $\epsilon$ asks directly: what is the fractional range
of the most sensitive parameter that is compatible with the
observed dark matter density?
Formally,
\begin{equation}
  \epsilon
  \;\equiv\;
  \frac{\Delta x_{\rm sens}}{x_{\rm sens}}\bigg|_{\rm island}
  \;\approx\;
  \frac{\ln 2}{\bgdelta_{\rm BG}}
  \;\approx\;
  \frac{0.693}{\bgdelta_{\rm BG}},
  \label{eq:eps}
\end{equation}
where $\Delta x_{\rm sens}$ is the full fractional width
of the natural island in the direction of the most sensitive
parameter $x_{\rm sens}$ (the parameter with the largest
$|\bgdelta_i|$), and the approximation holds for a
factor-of-2 abundance band.
The equivalence $\epsilon \approx \ln 2 / \bgdelta_{\rm BG}$
follows directly from the definition of $\bgdelta_{\rm BG}$:
a fractional shift $\delta x / x$ changes $\ln \Omega$ by
$\bgdelta_{\rm BG} \cdot \delta x / x$, so the island
boundary is reached when $\bgdelta_{\rm BG} \cdot \epsilon
\approx \ln 2$ (a factor of 2 in $\Omega$).
The result is a single percentage that answers the most
experimentally concrete version of the naturalness question.
To illustrate with two benchmarks defined below: benchmark
B3 (coannihilation, Sec.~\ref{subsubsec:b3}, $\bgdelta = 48$)
requires $M_{\rm NLSP}$ to be specified to within
$\epsilon = 1.4\%$ to remain in the natural island;
benchmark B5 (asymmetric dark matter, Sec.~\ref{subsec:adm},
$\bgdelta = 1$) tolerates a factor of two in either direction
($\epsilon = 75\%$).

The three measures, BG, SR, and $\epsilon$, are not
independent; they are related by construction and probe the
same underlying sensitivity from different angles.
BG isolates the worst-case direction; SR weights all
directions equally; $\epsilon$ converts the worst-case
sensitivity into an experimentally interpretable precision
requirement.
In single-parameter scenarios all three agree exactly.
In multi-parameter scenarios BG and SR diverge by at most
$\sqrt{n}$ (where $n$ is the number of equally sensitive
parameters), and $\epsilon$ always tracks BG through
Eq.~(\ref{eq:eps}).
The three measures have been applied and compared in the
context of supersymmetric dark matter and electroweak
naturalness by several authors.
Anderson and Castano~\CITE{Anderson:1995} introduced a
normalised variant of BG that accounts for the expected
sensitivity of the prior; Strumia~\CITE{Strumia:2000}
proposed the quadrature measure adopted here as $\bgdelta_{\rm SR}$;
Cabrera, Casas, and Ruiz de Austri~\CITE{Cabrera:2008}
applied both BG and Bayesian posterior-volume measures
to MSSM dark matter and found them to agree on tier
assignments while disagreeing on absolute values by
factors of a few.

A separate line of work has addressed fine-tuning measures
specifically in the context of inflationary PBH production.
Iovino and Riotto~\CITE{Iovino:2024} identified a
pathology of the BG measure when applied to the inflaton
potential coefficients $\{c_k\}$: as the target PBH
abundance is reduced below the observed dark matter density
(i.e.\ as $\fPBH \to 0$), the BG measure grows without
bound even though a model predicting fewer PBHs is
physically less constrained, not more.
This occurs because BG measures the local log-slope
$|\partial\ln\Omega/\partial\ln c_k|$ without reference
to whether $\Omega$ is above or below the target; as
the exponential suppression is tuned further, the
log-slope at any fixed point in parameter space continues
to grow.
Iovino and Riotto proposed Wilson's naturalness criterion
as a remedy: a parameter is natural if its value is stable
under small variations of the UV completion, independent
of the abundance it produces.
Their measure assigns a value close to unity to essentially
all inflationary PBH models, in sharp contrast to the BG
values of $10^2$--$10^8$ discussed in Sec.~\ref{subsec:inflation_ft}.

The present paper avoids this pathology by construction,
because all three measures (BG, SR, $\epsilon$) are
evaluated \emph{on} the $\fPBH = 1$ contour rather
than at an arbitrary point in parameter space.
The relevant question is not ``how sensitive is the model
at some generic parameter point?'' but ``how precisely
must each parameter be specified to maintain the observed
dark matter density?''; and this question is only
meaningful on the $\fPBH = 1$ contour.
Evaluated there, the BG measure is finite and well-defined
for all scenarios in the comparison, including inflationary
collapse (Eq.~\ref{eq:delta_inf_total}), and the
Iovino--Riotto pathology does not arise.
The Wilson question, ``does a technically natural UV
completion exist for the required parameter values?'',
is complementary and physically meaningful, but it
answers a different question than the one asked here;
Sec.~\ref{subsec:prior} returns to this distinction in
the context of the prior-probability vs.\ local-sensitivity
debate.
To the best of my knowledge, this is the first study to apply
all three measures simultaneously to a cross-paradigm comparison
that includes PBH formation mechanisms alongside
particle dark matter candidates.

\subsection{Choice of fundamental parameters}
\label{subsec:parameters}

Because the BG measure is not reparametrization-invariant,
the choice of what to call a ``fundamental'' parameter
must be specified and physically motivated.
I adopt the following convention throughout this paper:
\emph{the fundamental parameters are the independently
  specified inputs to the UV Lagrangian or to the cosmological
  initial conditions of the model under consideration},
evaluated at the relevant renormalization scale.
For a particle physics model these are the pole masses and
Lagrangian coupling constants; in supersymmetric models
they are the soft supersymmetry-breaking masses at the mediation scale; for
cosmological constructions they are the scales and couplings
of the relevant phase transition or symmetry-breaking potential,
together with any field-space initial conditions.

This convention has concrete implications for several of the
scenarios analyzed below.

\textit{Coannihilation.}---
In the Minimal Supersymmetric Standard Model
(MSSM)~\CITE{Martin:1997}, supersymmetry is broken in a
hidden sector and the breaking is transmitted to the visible
sector through a mediation mechanism (gravity, gauge, or
anomaly mediation) at a high scale $M_{\rm med}$.
The low-energy spectrum is then determined by a set of
soft-breaking mass terms in the Lagrangian; crucially,
each soft mass is an \emph{independent} parameter at
$M_{\rm med}$, and there is no symmetry or dynamics
that correlates them in the absence of additional model
structure.
For the coannihilation scenario, the two relevant soft
masses are $M_1$ (the bino mass parameter, which sets
$M_\chi \equiv |M_1|$ at low energies) and the soft mass
$m_{\tilde{f}}$ of the lightest sfermion (stop, stau,
or sneutrino), which sets $M_{\rm NLSP}$
(I take the bino as the lightest neutralino and the
lightest sfermion as the NLSP; the analysis is qualitatively
identical for any spectrum where the two lightest states
are nearly degenerate).
Coannihilation is relevant when $M_{\rm NLSP}$ and
$M_\chi$ are nearly degenerate: the NLSP is present in
the thermal bath at freeze-out and its annihilation
cross section $\langle\sigma v\rangle_{\rm NLSP}$
contributes to the effective rate that sets the relic
density.
The contribution is Boltzmann-suppressed by a factor
$e^{-x_F \delta}$ (Eq.~\ref{eq:sveff_coann}), where
$x_F \equiv m_\chi/T_F \approx 25$ is the ratio of the
dark matter mass to the freeze-out temperature
$T_F$~\CITE{Griest:1991, Edsjo:1997} and
$\delta \equiv (M_{\rm NLSP} - M_\chi)/M_\chi$ is the
fractional mass splitting.
Since $T_F \simeq m_\chi/x_F$ is set by the freeze-out
condition $\langle\sigma v\rangle n_\chi = H$, and
$x_F \approx 20$--$30$ for weak-scale dark matter
regardless of the exact cross section~\CITE{Griest:1991},
the exponential suppression can be substantial even
for splitting $\delta \sim 0.05$--$0.10$.

The fractional mass splitting $\delta$ is a
\emph{derived} quantity: it is the difference of two
Lagrangian parameters, not itself a fundamental input.
Taking BG derivatives with respect to $\delta$ at fixed
$M_\chi$ gives
\begin{equation}
  \bgdelta_\delta \;=\; x_F\,\delta \;+\; \mathcal{O}(\delta^2)
  \;\approx\; 1\text{--}2
  \quad (x_F\,\delta \ll 1),
  \label{eq:coann_wrong}
\end{equation}
which substantially underestimates the tuning because
the derivative $\partial\ln\Omega/\partial\ln\delta$
vanishes as $\delta \to 0$ by construction: the coordinate
$\delta$ crosses zero at the threshold where the two
masses are degenerate, and multiplicative log-derivatives
are pathologically small near any additive zero.
Taking derivatives with respect to the independent
Lagrangian parameters $(M_\chi, M_{\rm NLSP})$ at fixed
mediation-scale boundary conditions instead yields
\begin{equation}
  \bgdelta_{M_{\rm NLSP}} \;=\; x_F\,\frac{M_{\rm NLSP}}{M_\chi}
  \;\approx\; 26\text{--}50,
  \label{eq:coann_correct}
\end{equation}
which correctly captures the sensitivity of the
Boltzmann-suppressed coannihilation rate to the NLSP
pole mass: a 2\% shift in $M_{\rm NLSP}$ at fixed
$M_\chi$ changes $\delta$ by 2\%, which changes
$e^{-x_F\delta}$ by a factor $e^{\pm 0.5} \approx 1.6$,
shifting $\Omega h^2$ by $\sim 50\%$ and taking the
model outside the natural island.
The MSSM coannihilation benchmark B3 uses the
parametrization $(M_\chi, M_{\rm NLSP})$ and yields
$\bgdelta \approx 48$ as a consequence
(Eq.~\ref{eq:coann_correct}; see also
Sec.~\ref{subsubsec:b3}).
This distinction between derived and fundamental
parameters has been noted in the electroweak fine-tuning
literature~\CITE{Baer:2012} but is rarely applied
consistently to dark matter relic calculations.

\textit{Axion misalignment.}---
The initial misalignment angle $\theta_i \in [-\pi, \pi]$
is a cosmological initial condition, not a Lagrangian
parameter, and is independently specified from $f_a$, the axion decay constant.
It is therefore treated as a separate fundamental input.
The important distinction between the BG measure evaluated
with respect to $\theta_i$ (which gives $\bgdelta \approx 2$
for $\theta_i \sim 1$) and the prior probability of $\theta_i$
being small (which gives a prior tuning $\sim 1/\theta_i$ for
a flat distribution) is discussed in Sec.~\ref{subsec:prior}.

\textit{PBH formation rates.}---
For the PBH mechanisms analyzed in Sec.~\ref{sec:pbh},
the fundamental parameters are the scales and couplings
appearing in the potential or power spectrum of the
underlying model, e.g. the symmetry-breaking scale $\eta$ and
bias energy density $V_b$ for domain walls, the modulus mass
$m_\phi$ and curvature perturbation amplitude $\sigma$ for
early matter domination, and the transition temperature $T_*$,
nucleation rate parameter $\beta/H$, and latent-heat parameter
$\alpha$ for the FOPT.
All three sets of parameters are taken as independent inputs,
and BG derivatives are evaluated holding all other
parameters fixed.

\subsection{Numerical protocol}
\label{subsec:protocol}

All BG derivatives in this paper are evaluated numerically
via central finite differences in log-space,
\begin{equation}
  \bgdelta_i \;\approx\;
    \frac{\left|\ln \Omega(x_i e^h) - \ln \Omega(x_i e^{-h})\right|}{2h},
    \qquad h = 10^{-4}.
  \label{eq:numderiv}
\end{equation}
This step size is small enough to suppress second-order
corrections to better than 1 part in $10^6$ for all smooth
abundance functions considered, yet large enough to avoid
floating-point cancellation errors.
Analytic closed-form expressions for $\bgdelta_i$ are
derived independently in Appendix~\ref{app:analytic} for all
cases where the abundance map is sufficiently simple;
numerical and analytic results agree to better than
0.1\% in every case.

Parameter-space heatmaps are computed on grids of
$400 \times 400$ points, logarithmically spaced over the
ranges specified in the captions of Figs.~\ref{fig:pbh}--\ref{fig:axions}.
The color scale is identical across all panels:
$\log_{10} \bgdelta \in [0, 3.5]$ on the \texttt{viridis\_r}
colormap, saturating to dark purple for $\bgdelta > 3000$.
This shared scale is here the principal visual device for
comparing fine-tuning across production paradigms.

The \emph{natural island} in each panel is defined as the
region of parameter space satisfying
\begin{equation}
  0.06 \;\leq\; \Omega h^2 \;\leq\; 0.24
  \label{eq:island}
\end{equation}
(i.e.\ within a factor of two of $\Omega_{\rm DM} h^2 = 0.120$)
together with all constraints listed in
Table~\ref{tab:constraints}.
The factor-of-two band is chosen to include representative
theoretical uncertainties in the relic-abundance calculation
(QCD uncertainties in $g_*$, perturbative corrections to
annihilation cross sections, etc.)\ without enlarging the
island to the point of obscuring the tuning structure.
Results are qualitatively insensitive to widening this band to
a factor of 3 or narrowing it to 30\%.
For PBH mechanisms, $\Omega h^2$ is replaced by $\fPBH$
(both measure the fraction of dark matter), and the
gray dashed contours in the PBH heatmaps show the
wider $\pm 1$~dex sidebands $\fPBH \in [0.1, 10]$
as visual guides; all BG derivatives and island
half-widths are computed using the factor-of-two
band of Eq.~(\ref{eq:island}) applied to $\fPBH$
(i.e.\ $\fPBH \in [0.06/0.12, 0.24/0.12] \equiv
[0.5, 2]$), consistent with the particle DM protocol.

\begin{table}[htb]
\centering
\caption{Observational constraints applied in this paper and
         the scenarios to which they are relevant.
         The stochastic gravitational-wave background from the
         FOPT itself is not listed here because it constrains the
         GW signal rather than $\fPBH$ directly; this constraint is
         discussed qualitatively in Sec.~\ref{subsec:fopt} and
         Sec.~\ref{subsec:implications_pbh}.
         All PBH abundance constraints are already encoded in the
         $\fPBH = 1$ requirement and the asteroid-mass window.
         \label{tab:constraints}}
\begin{ruledtabular}
\begin{tabular}{lll}
Constraint & Value & Applies to \\
\hline
LZ-2024 SI cross section & see \CITE{LZ:2024} & WIMPs \\
Planck 2018 $\Omega_{\rm DM} h^2$ & $0.1200 \pm 0.0012$ & all \\
PBH asteroid window & $10^{17}$--$10^{22}$~g & PBHs \\
Ly$\alpha$ forest & $M_{\rm DM} \geq 3$~keV & FIMPs \\
\end{tabular}
\end{ruledtabular}
\end{table}

The direct-detection constraint for WIMPs is implemented
via the spin-independent cross section per nucleon,
\begin{equation}
  \sigma_{\rm SI}(M_S, \lambda) \;=\;
  \frac{\lambda^2 f_N^2 m_N^2 \mu_{SN}^2}
       {\pi\, M_S^2\, M_h^4},
  \label{eq:sigSI}
\end{equation}
where $f_N = 0.30$ is the nucleon Higgs form factor
\CITE{Alarcon:2011},
$m_N = 0.939$~GeV, and
$\mu_{SN} = m_N M_S/(m_N + M_S)$ is the singlet--nucleon
reduced mass.
The LZ~2024 sensitivity curve is approximated by a
piecewise power law calibrated to the published limit
\CITE{LZ:2024};
this prescription overestimates exclusion at $M_S \lesssim 10$~GeV
and underestimates it above $\sim$10~TeV, but neither
regime contains any of the benchmarks analyzed in this paper.
The Lyman-$\alpha$ forest constraint on the freeze-in
DM mass ($M_{\rm DM} \gtrsim 3$~keV for production from a
100-GeV mediator \CITE{Ballesteros:2020, Decant:2022}) is implemented as a hard
boundary in the FIMP parameter plane.

\section{\label{sec:pbh}Primordial Black Hole Dark Matter}

The formation of PBHs in the asteroid-mass window has been
studied in a wide variety of cosmological scenarios.
Three non-inflationary mechanisms have received particular
attention for their ability to produce $\fPBH \approx 1$
without additional cosmological ingredients: a network of
biased domain walls~\CITE{Ferrer:2019, Gouttenoire:2024, Ferreira:2024},
an early
matter-dominated era driven by a decaying modulus
field~\CITE{Georg:2016, Georg:2017, Dalianis:2019, Carr:2021b},
and a strongly first-order phase transition in a hidden
sector~\CITE{Liu:2022, Lewicki:2023, Gehrman:2023}.
The idea that biased domain walls can collapse to form a
PBH dark matter population was developed in
Ref.~\CITE{Ferrer:2019} in the context of QCD axion domain
wall networks, building on the long-known instability of
biased networks~\CITE{Vilenkin:2000}; it was subsequently
revisited and extended by several authors, including
the network-collapse treatments of
Refs.~\CITE{Gouttenoire:2024, Ferreira:2024}.
The most extensively discussed mechanism in the literature
is single-field inflation with an enhanced small-scale
power spectrum, generated by an inflection point or
ultra-slow-roll phase in the inflaton
potential~\CITE{Hertzberg:2017, Franciolini:2022, Ozsoy:2021};
multi-field and spectator-field variants offer partial relief
from the severe inflationary fine-tuning~\CITE{Stamou:2024}.
Concretely, Ref.~\CITE{Geller:2022} and
\cite{Qin:2023} demonstrated that
the USR potential shape arises naturally in multifield models with
nonminimal couplings to gravity, and performed the first full MCMC
analysis of CMB constraints and PBH formation in this setting,
identifying a degeneracy direction along which parameter shifts of
$\sim 10\%$ preserve the fit to Planck data.
More recently, Ref.~\cite{Lorenzoni:2025, McDonough:2025} showed that adding a non-interacting
spectator field to a single-field USR potential changes the formation
mechanism entirely: for example, the system no longer enters USR, and the
resulting model predictions become exponentially more robust to small
parameter changes, substantially reducing the Layer~2
fine-tuning.
Within the framework of this paper, these constructions populate the
lower end of the multi-field stochastic inflation range in
Fig.~\ref{fig:inf_comparison}, and move toward the boundary between
Class~II and Class~III for low-reheating scenarios. 
A distinct inflationary mechanism is the collapse of Higgs
field fluctuations near the electroweak vacuum instability
scale~\CITE{Espinosa:2018}.

For the three non-inflationary mechanisms, I derive the
abundance map $\fPBH(\{x_i\})$ from the underlying model
parameters, compute $\bgdelta$ analytically, and display
the result as a parameter-space heatmap
(Secs.~\ref{subsec:dw}--\ref{subsec:fopt}).
Inflationary mechanisms are treated separately in
Sec.~\ref{subsec:inflation_ft}: rather than a single
heatmap, I map the BG framework onto six representative
inflationary model classes and compare their two-layer
tuning structure systematically
(Fig.~\ref{fig:inf_comparison}).
This distinction in treatment reflects a genuine
physical difference.
For the non-inflationary mechanisms, the fundamental
Lagrangian parameters map onto $\fPBH$ through a single,
well-calibrated chain with no hidden exponentials;
the BG measure can therefore be computed precisely
and displayed as a smooth landscape.
For inflationary models, the mapping from inflaton
potential coefficients $\{c_k\}$ to $\sigma$ to $\fPBH$
involves two successive exponentials whose ratio depends
sensitively on the potential shape, leading to estimates
that span several orders of magnitude depending on the
model class.
Furthermore, as noted by Iovino and
Riotto~\CITE{Iovino:2024}, applying BG to the potential
coefficients off the $\fPBH = 1$ contour raises a
subtlety: $\bgdelta$ grows without bound as $\fPBH \to 0$
even for variations that are physically harmless.
The approach adopted throughout this paper, i.e. evaluating
all measures \emph{on} the $\fPBH = 1$ contour,
avoids this pathology by construction
(see Sec.~\ref{subsec:param_choice}).

All four mechanisms produce PBHs whose mass is set by
the Hubble horizon at formation,
\begin{equation}
  M_{\rm PBH}
  \;\simeq\;
  \frac{3.8\,\Mpl^3}{T_{\rm form}^2},
  \label{eq:Mpbh_general}
\end{equation}
where $T_{\rm form}$ is the temperature at which PBH
formation occurs (in natural units $c = k_B = 1$).
The asteroid-mass window $10^{17}$--$10^{22}$~g requires
formation temperatures
$10^4 \lesssim T_{\rm form}/{\rm GeV} \lesssim 10^{6.5}$,
shown as gold vertical lines in Fig.~\ref{fig:pbh}.

\subsection{Biased domain walls}
\label{subsec:dw}

\subsubsection{Formation mechanism and abundance map}
\label{subsubsec:dw_mechanism}

Consider a scalar field $\phi$ with a discrete $\mathbb{Z}_2$
symmetry $\phi \to -\phi$ that is spontaneously broken when
$\phi$ acquires a vacuum expectation value $\pm\eta$ at a
phase transition in the early Universe.
The resulting potential has two degenerate minima separated by
a potential barrier, and the field takes different values in
causally disconnected regions; the boundaries between these
regions form a network of two-dimensional topological defects
called \emph{domain walls}~\CITE{Kibble:1976, Vilenkin:2000}.
Each wall carries a surface energy density (tension)
$\sigma_w \sim \eta^3$, where $\eta$ is the symmetry-breaking
scale; the exact coefficient is model-dependent and of order
unity, and is absorbed into $\eta$ throughout.

In the exact $\mathbb{Z}_2$ limit the two vacua are
energetically degenerate and the wall network is
cosmologically stable~\CITE{Kibble:1976}, which would lead to
domain-wall domination and conflict with
cosmology~\CITE{Vilenkin:2000}.
A small explicit breaking of the $\mathbb{Z}_2$ symmetry
lifts this degeneracy by an energy density $V_{\rm bias}$
between the two vacua.
Such a bias arises naturally from Planck-suppressed operators
in the scalar potential~\CITE{Barr:1992, Holman:1992},
or can be introduced by hand; in either case it can be
parametrized as $V_{\rm bias} = V_b^4$, where $V_b$ is an
effective bias energy scale.
The pressure difference $V_{\rm bias}$ across each wall
provides a volume force that drives the lower-energy vacuum
to expand, causing the network to collapse and annihilate
at a temperature $T_{\rm ann}$ set by the competition
between this volume pressure and the Hubble friction
that resists wall motion~\CITE{Vilenkin:2000, Ferrer:2019}:
\begin{equation}
  T_{\rm ann}
  \;\simeq\;
  \left(\frac{90}{\pi^2 g_*}\right)^{1/4}
  \sqrt{\frac{c_t\,\Mpl\,V_b^4}{\eta^3}},
  \label{eq:Tann}
\end{equation}
where $g_* \approx 100$ counts relativistic degrees of
freedom and $c_t \approx 0.30$ is a numerical coefficient
characterizing the wall-network annihilation dynamics
(see Ref.~\CITE{Ferrer:2019}, which estimated the resulting
PBH abundance using the earlier field-theoretic
wall-network simulations of Ref.~\CITE{Kawasaki:2015}; a
dedicated set of network-collapse simulations was performed
subsequently in Ref.~\CITE{Ferreira:2024}).
At $T_{\rm ann}$, a fraction $p \approx 10^{-3}$ of
horizon-sized wall segments collapse gravitationally into
PBHs~\CITE{Ferrer:2019, Gouttenoire:2024}.
This collapse probability reflects
the small fraction of Hubble volumes in which the
converging wall geometry generates a local overdensity
sufficient to exceed the gravitational collapse threshold;
as discussed in Ref.~\CITE{Ferrer:2019} and quantified by
the simulations of Ref.~\CITE{Ferreira:2024}, the PBH mass
is robust but the abundance itself carries large
uncertainties (see Sec.~\ref{subsec:caveats}).
The resulting PBH mass and formation fraction are
\begin{align}
  M_{\rm PBH}
  &\;\simeq\;
  \frac{3.8\,\Mpl^3}{T_{\rm ann}^2}
  \;\propto\;
  \frac{\eta^3}{V_b^4},
  \label{eq:Mpbh_dw}
  \\[4pt]
  \beta_{\rm PBH}
  &\;\equiv\;
  \frac{\rho_{\rm PBH}}{\rho_{\rm tot}}\bigg|_{T_{\rm ann}}
  \;\simeq\;
  \frac{p\,V_b^4}{\tfrac{\pi^2}{30}g_*\,T_{\rm ann}^4}.
  \label{eq:beta_dw}
\end{align}
The present-day PBH dark matter fraction is obtained by
redshifting $\beta_{\rm PBH}$ from $T_{\rm ann}$ to
matter--radiation equality at $T_{\rm eq} = 0.75$~eV:
\begin{equation}
  \fPBH
  \;\simeq\;
  \beta_{\rm PBH}\,\frac{T_{\rm ann}}{0.84\,T_{\rm eq}},
  \label{eq:fpbh_dw_full}
\end{equation}
where the factor $0.84$ accounts for the change in
$g_*$ between $T_{\rm ann}$ and $T_{\rm eq}$~\CITE{Carr:2021}.
Substituting Eqs.~(\ref{eq:Tann})--(\ref{eq:beta_dw})
into Eq.~(\ref{eq:fpbh_dw_full}) and collecting powers,
one finds
\begin{equation}
  \boxed{
  \fPBH
  \;\propto\;
  \frac{\eta^{9/2}}{V_b^2},
  }
  \label{eq:fpbh_dw}
\end{equation}
a \emph{pure power-law} in the two fundamental
parameters $(\eta, V_b)$ with no exponential factor.
This is the central structural result for this construction:
the abundance map is a monomial, and the BG fine-tuning
measure is therefore constant across the entire parameter space.

The asteroid-mass window $10^{17}$--$10^{22}$~g requires,
via Eq.~(\ref{eq:Mpbh_dw}),
\begin{equation}
  T_{\rm ann} \;\in\; [10^5,\;3\times10^7]~{\rm GeV}
  \;\approx\; [10^{4.9},\;10^{7.5}]~{\rm GeV}.
  \label{eq:Tann_range}
\end{equation}
These boundaries follow directly from inverting
$M_{\rm PBH} = 3.8\,\Mpl^3/T_{\rm ann}^2$: at
$T_{\rm ann} = 10^5$~GeV one obtains
$M_{\rm PBH} \approx 10^{22}$~g (upper edge), and at
$T_{\rm ann} = 3\times10^7$~GeV one obtains
$M_{\rm PBH} \approx 10^{17}$~g (lower edge).

The benchmark $\eta = 2.4\times10^6$~GeV, $V_b = 2$~TeV
is chosen as the simultaneous intersection of three
conditions.
First, $\fPBH \approx 1$, so PBHs account for all of the
dark matter.
Second, $T_{\rm ann} \approx 5\times10^5$~GeV, placing
$M_{\rm PBH} \approx 4\times10^{20}$~g well within the
asteroid window.
Third, $V_b = 2$~TeV is within 20\% of the gravity-induced
bias value $V_b^{\rm grav} = (\eta^5/\Mpl)^{1/4} \approx
2.4$~TeV at this $\eta$, so the bias is naturally
generated by Planck-suppressed operators of the form
$\phi^5/\Mpl$ evaluated at $\phi = \eta$, i.e.\
$V_{\rm bias} \sim \eta^5/\Mpl$~\CITE{Barr:1992, Holman:1992},
without introducing a new free scale.
\begin{equation}
  V_b^{\rm grav}(\eta) \;\equiv\; (\eta^5/\Mpl)^{1/4}.
  \label{eq:Vb_grav}
\end{equation}
These three conditions together identify $\eta \approx
2.4\times10^6$~GeV as a theoretically motivated ``sweet
spot'': a single symmetry-breaking scale that simultaneously
accounts for all of the dark matter, targets the asteroid-mass
window, and requires no bias scale beyond what
quantum gravity generates automatically.
A particularly clean sub-class is the ``gravity-induced
bias'' scenario in which one sets $V_b = V_b^{\rm grav}$
exactly, reducing the construction to a single-parameter
family parametrized by $\eta$ alone.

\subsubsection{Analytic fine-tuning structure}
\label{subsubsec:dw_ft}

From Eq.~(\ref{eq:fpbh_dw}) the BG sensitivities follow
immediately by log-differentiation:
\begin{equation}
  \bgdelta_\eta
  \;=\;
  \left|\frac{\partial \ln \fPBH}{\partial \ln \eta}\right|
  \;=\; \frac{9}{2},
  \qquad
  \bgdelta_{V_b}
  \;=\;
  \left|\frac{\partial \ln \fPBH}{\partial \ln V_b}\right|
  \;=\; 2.
  \label{eq:delta_dw}
\end{equation}
Both are \emph{exact constants}, independent of the values
of $\eta$ and $V_b$ and of the location in parameter space.
The overall measure is
\begin{equation}
  \bgdelta_{\rm DW}
  \;=\; \max\!\left(\bgdelta_\eta,\, \bgdelta_{V_b}\right)
  \;=\; \frac{9}{2}
  \quad \forall\;(\eta, V_b).
  \label{eq:delta_dw_total}
\end{equation}

For the gravity-induced-bias sub-case, substituting
$V_b \propto \eta^{5/4}\Mpl^{-1/4}$ into
Eq.~(\ref{eq:fpbh_dw}) gives
$\fPBH \propto \Mpl^{1/2}\,\eta^2$,
so the single-parameter measure reduces to
$\bgdelta_\eta^{(\rm grav)} = 2$,
the minimum value compatible with a non-trivial
dependence on any dimensionful scale.

As shown below, this result places biased-domain-wall PBHs in the
same naturalness tier as off-resonance heavy WIMPs and
freeze-in particles (all with $\bgdelta \approx 2$),
and substantially below the coannihilation and FOPT
constructions discussed below.
The quantitative answer to the title question is clear for
this mechanism: biased-domain-wall PBHs are \emph{not}
fine-tuned by any reasonable standard.

\subsubsection{Parameter-space heatmap}
\label{subsubsec:dw_map}

\begin{figure*}[t]
  \includegraphics[width=\textwidth]{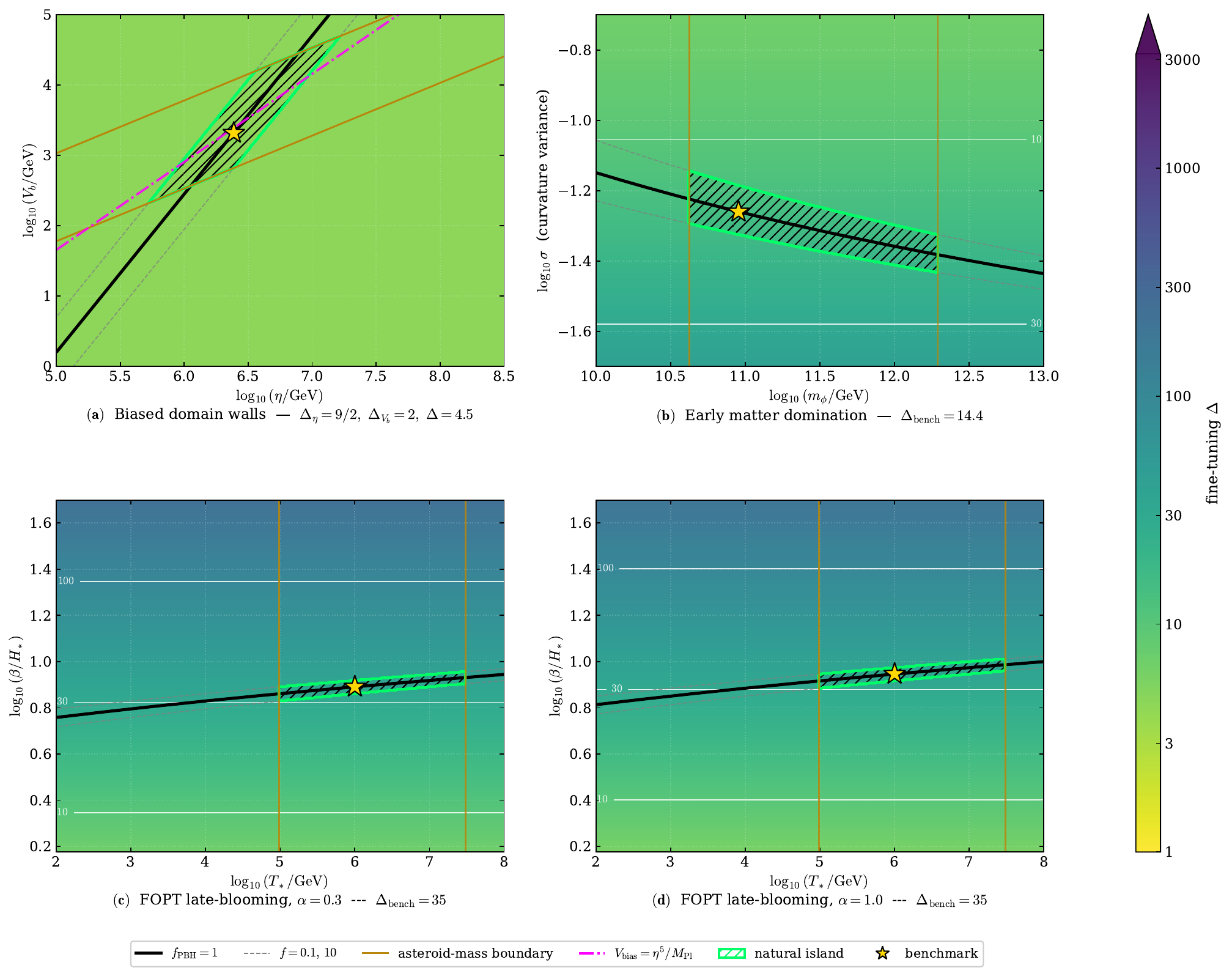}
  \caption{%
    Fine-tuning of the three PBH dark-matter formation
    mechanisms analyzed in this paper.
    \textit{Background color}: $\log_{10}\bgdelta$ on a shared
    scale $\bgdelta \in [1, 3000]$ (pale yellow = natural,
    dark purple = highly tuned; see colorbar at right).
    \textit{Thick black contour}: $\fPBH = 1$.
    \textit{Gray dashed contours}: $\fPBH = 0.1$ and $10$
    ($\pm 1$~dex sidebands).
    \textit{Gold/orange vertical lines}: asteroid-mass
    boundaries $M_{\rm PBH} = 10^{17}$~g and $10^{22}$~g.
    \textit{Green hatching}: natural island, defined as
    $0.1 \leq \fPBH \leq 10$ within the asteroid-mass window
    and satisfying all observational constraints.
    \textit{Gold star}: benchmark point.
    \textbf{(a)}~Biased domain walls in the $(\eta, V_b)$
    plane ($\log_{10}$ axes; $\eta$ = symmetry-breaking scale,
    $V_b$ = bias energy scale; Eq.~\protect\ref{eq:fpbh_dw}).
    The magenta dash-dotted line is the gravity-induced-bias
    locus $V_b = (\eta^5/\Mpl)^{1/4}$.
    The panel is uniformly pale yellow, reflecting
    $\bgdelta = 9/2$ everywhere (Eq.~\protect\ref{eq:delta_dw_total}).
    Benchmark: $\eta = 2.4\times10^6$~GeV, $V_b = 2$~TeV.
    \textbf{(b)}~Early matter domination in the $(m_\phi, \sigma)$
    plane ($m_\phi$ = modulus mass, $\sigma$ = curvature
    perturbation variance;
    Eqs.~\protect\ref{eq:beta_AM}--\protect\ref{eq:fpbh_md}).
    The color gradient from yellow-green to teal reflects
    the rising $\bgdelta_\sigma = 5 + 0.196/\sigma^{4/3}$
    (Eq.~\protect\ref{eq:delta_md_sigma}) as $\sigma$ decreases.
    The asteroid-mass boundaries correspond to
    $m_\phi \in [4\times10^{10},\, 2\times10^{12}]$~GeV
    (Eq.~\protect\ref{eq:Mpbh_md}).
    Benchmark: $m_\phi = 9\times10^{10}$~GeV, $\sigma = 0.055$,
    giving $M_{\rm PBH} \approx 10^{21}$~g; $\bgdelta = 14.4$.
    \textbf{(c, d)}~FOPT late-blooming in the $(T_*, \beta/H_*)$
    plane for $\alpha = 0.3$ (panel c) and $\alpha = 1.0$ (panel d).
    The two panels are nearly identical, confirming the
    $\alpha$-independence of the key identity
    Eq.~(\protect\ref{eq:delta_fopt_universal}).
    The $\fPBH = 1$ contour coincides with the $\bgdelta = 30$
    contour throughout, making
    $\bgdelta_{\beta/H} = \ln(T_*/0.84\,T_{\rm eq}) \approx 35$
    directly visible.
    Benchmark: $T_* = 10^6$~GeV, $\beta/H_* = 7.78$ ($\alpha = 0.3$)
    and $8.84$ ($\alpha = 1.0$); both at $\bgdelta = 35$.
    \label{fig:pbh}
  }
\end{figure*}

Figure~\ref{fig:pbh}(a) shows $\log_{10}\bgdelta$ in
the $(\log_{10}\eta,\, \log_{10}V_b)$ plane.
The entire panel is uniformly pale yellow, the lowest
color on the shared scale, reflecting the analytic result
$\bgdelta = 9/2$ everywhere, a direct consequence of
the pure power-law structure
$\fPBH \propto \eta^{9/2}/V_b^2$
(Eq.~\ref{eq:fpbh_dw}).
The thick black contour marks $\fPBH = 1$; gray dashed
contours at $\fPBH = 0.1$ and $10$ define the $\pm1$~dex
sidebands that bound the natural island (green hatching).
The two gold-orange lines show the asteroid-mass boundaries
$M_{\rm PBH} = 10^{17}$~g and $10^{22}$~g; these are
nearly parallel to the $\fPBH = 1$ contour because both
$\fPBH$ and $M_{\rm PBH}$ are controlled by the same
parameter combination $\eta^3/V_b^4$
(Eqs.~\ref{eq:Mpbh_dw} and~\ref{eq:fpbh_dw}).
The magenta dash-dotted line shows the gravity-induced-bias
locus $V_b = (\eta^5/\Mpl)^{1/4}$ (Eq.~\ref{eq:Vb_grav}),
along which the bias is generated by
Planck-suppressed operators rather than freely tuned.
This locus passes through the natural island and through
the benchmark point ($\eta = 2.4\times10^6$~GeV,
$V_b = 2$~TeV; gold star), confirming that the
Planck-suppressed scenario naturally populates the
asteroid-mass window without any additional parameter adjustment.

\subsection{Early matter domination}
\label{subsec:md}

\subsubsection{Formation mechanism and abundance map}
\label{subsubsec:md_mechanism}
A natural setting for an early matter-dominated era is
provided by string-theoretic moduli: scalar fields $\phi$
with only gravitational-strength couplings to ordinary matter
that are ubiquitous in compactifications of extra
dimensions~\CITE{Acharya:2008, Allahverdi:2020}.
More broadly, any scalar field that begins to oscillate
coherently in the early Universe when the Hubble rate drops
below its mass, $H \sim m_\phi$, behaves as a pressureless
fluid and temporarily dominates the energy density before
decaying~\CITE{Allahverdi:2020}.
Such a field decays with a rate set by dimensional analysis
for a gravitationally coupled particle,
$\Gamma_\phi \sim c_\phi\,m_\phi^3/\Mpl^2$,
where $c_\phi \sim 10^{-3}$ encodes loop and phase-space
suppression factors~\CITE{Acharya:2008}.
When $H$ drops to $\Gamma_\phi$, the field decays and
reheats the Universe to a temperature~\CITE{Georg:2016,
Georg:2017, Allahverdi:2020}
\begin{equation}
  \Trh
  \;\simeq\;
  0.55\,\left(\frac{c_\phi\,m_\phi^3}{\Mpl}\right)^{1/2}.
  \label{eq:Trh}
\end{equation}

The period of matter domination preceding reheating alters
cosmological perturbation growth in a way that is crucial
for PBH production.
In a radiation-dominated background, sub-Hubble density
perturbations oscillate and decay; in a matter-dominated
background, the same perturbations grow as $\delta\rho/\rho
\propto a$ (where $a$ is the scale factor), and sufficiently
large fluctuations can overcome pressure support and collapse
gravitationally before reheating occurs~\CITE{Khlopov:1980,
Georg:2016}.
The relevant perturbations are curvature perturbations
inherited from inflation, characterized by their
root-mean-square amplitude $\sigma$ on the scale that
re-enters the Hubble horizon during the matter-dominated era.
The fraction of the energy density that collapses into PBHs
depends sensitively on how many Hubble patches contain a
perturbation large enough to collapse; this collapse
criterion in a pressureless medium was derived analytically
and verified numerically by Harada, Yoo, Kohri, and
Nakama (HYKN)~\CITE{Harada:2016}, who found
\begin{equation}
  \beta_{\rm AM}(\sigma)
  \;\simeq\;
  4.6\times10^{-6}\,\sigma^5\,
  \exp\!\left(-\frac{0.147}{\sigma^{4/3}}\right).
  \label{eq:beta_AM}
\end{equation}
Here $\beta_{\rm AM}$ is the mass fraction of the Universe
that collapses into PBHs at the time of formation.
The $\sigma^5$ prefactor reflects the five-dimensional
phase-space volume for a fluctuation to exceed the collapse
threshold in a pressureless fluid (compared to the
exponential-only Press--Schechter form in radiation
domination~\CITE{Press:1974}), while the exponential
encodes the probability tail of a Gaussian distribution
above the threshold density contrast $\delta_c \simeq 0.05$
appropriate to matter domination~\CITE{Harada:2016}.
The two contributions are equal at $\sigma \approx 0.050$,
close to the natural-island benchmark, so neither can be
neglected.

The PBH mass equals the Hubble-horizon mass at the time
of collapse, which for this mechanism occurs just before
reheating:
\begin{equation}
  M_{\rm PBH}
  \;\simeq\;
  \frac{3.8\,\Mpl^3}{\Trh^2}
  \;\propto\;
  m_\phi^{-3},
  \label{eq:Mpbh_md}
\end{equation}
where the proportionality uses Eq.~(\ref{eq:Trh}).
The asteroid-mass window $M_{\rm PBH} \in [10^{17}, 10^{22}]$~g
therefore maps to a modulus-mass range
$m_\phi \in [4\times10^{10},\, 2\times10^{12}]$~GeV,
as shown by the gold vertical lines in Fig.~\ref{fig:pbh}(b).
The present-day dark matter fraction is obtained by
redshifting $\beta_{\rm AM}$ from the formation epoch to
matter--radiation equality at $T_{\rm eq} = 0.75$~eV:
\begin{equation}
  \fPBH(m_\phi, \sigma)
  \;\simeq\;
  \beta_{\rm AM}(\sigma)\,
  \frac{\Trh(m_\phi)}{0.84\,T_{\rm eq}}.
  \label{eq:fpbh_md}
\end{equation}
The factor $0.84$ accounts for the change in relativistic
degrees of freedom $g_*$ between $\Trh$ and $T_{\rm eq}$~\CITE{Carr:2021}.
The dependence on $m_\phi$ enters solely through $\Trh$
and is a pure power law,
$\fPBH \propto m_\phi^{3/2}\,\beta_{\rm AM}(\sigma)$,
while the $\sigma$ dependence through Eq.~(\ref{eq:beta_AM})
is non-trivial, containing both the polynomial prefactor
and the exponential that will give rise to the characteristic
Class~II fine-tuning structure analyzed in
Sec.~\ref{subsubsec:md_ft}.

\subsubsection{Analytic fine-tuning structure}
\label{subsubsec:md_ft}

Differentiating Eq.~(\ref{eq:fpbh_md}) with respect to
$\ln m_\phi$ and $\ln\sigma$:
\begin{align}
  \bgdelta_{m_\phi}
  &\;=\;
  \left|\frac{\partial \ln \fPBH}{\partial \ln m_\phi}\right|
  \;=\; \frac{3}{2},
  \label{eq:delta_md_m}
  \\[4pt]
  \bgdelta_\sigma
  &\;=\;
  \left|\frac{\partial \ln \fPBH}{\partial \ln\sigma}\right|
  \;=\; 5 \;+\; \frac{0.196}{\sigma^{4/3}}.
  \label{eq:delta_md_sigma}
\end{align}
The modulus-mass sensitivity $\bgdelta_{m_\phi} = 3/2$
is a constant, reflecting the simple power-law connection
$\Trh \propto m_\phi^{3/2}$.
The curvature-variance sensitivity $\bgdelta_\sigma$ is
dominated by the exponential factor in $\beta_{\rm AM}$
at small $\sigma$ and by the prefactor at large $\sigma$;
the two contributions are equal at $\sigma^{4/3} = 0.196/5
\approx 0.039$, i.e.\ $\sigma \approx 0.050$.
Since the natural island (defined by $\fPBH \approx 1$
in the asteroid window) requires $\sigma \in [0.037,
0.071]$ at the benchmark $m_\phi = 9\times10^{10}$~GeV,
the relevant range is
\begin{equation}
  \bgdelta_\sigma^{(\rm island)}
  \;\in\; [12,\, 21],
  \label{eq:delta_md_range}
\end{equation}
with a benchmark value $\bgdelta_\sigma = 14.4$ at
$\sigma = 0.055$.

The benchmark value $\sigma = 0.055$ is chosen because it
sits near $\sigma \approx 0.050$ where the exponential
and polynomial terms in Eq.~(\ref{eq:beta_AM}) contribute
equally ($0.147/\sigma^{4/3} \approx 5$), making it a
physically representative point at which neither term
dominates and the value $\bgdelta_\sigma = 14.4$ is
typical of the natural island.
It corresponds to a curvature power spectrum amplitude
$\mathcal{P}_\mathcal{R}(k_{\rm form}) \approx \sigma^2
\approx 3\times10^{-3}$ on the asteroid-mass scale,
a factor of $\sim 10^6$ above the CMB amplitude.
Values $\sigma \lesssim 0.04$ exponentially suppress
$\beta_{\rm AM}$ and require an unrealistically high
$m_\phi$ to compensate; values $\sigma \gtrsim 0.10$
overproduce PBHs and push $m_\phi$ below the asteroid-mass
lower boundary.
The natural island in $\sigma$ is therefore bounded from
below by the exponential tail of the HYKN formula and from
above by the asteroid-mass constraint, precisely the
two-sided confinement that makes $\bgdelta_\sigma \approx 14$
a robust result within this window.

The overall measure $\bgdelta_{\rm MD} = \max(3/2,\,
\bgdelta_\sigma) = \bgdelta_\sigma$ across the entire
natural island, placing this construction firmly in
the intermediate ``single-exponential'' class despite
the $\sigma^5$ prefactor moderating the growth rate.

The value $\sigma \approx 0.055$ required for $f_{\rm PBH} = 1$
in the asteroid-mass window corresponds to a primordial curvature
power spectrum amplitude $\mathcal{P}_\mathcal{R} \approx
\sigma^2 \approx 3\times10^{-3}$ at the formation scale
$k_{\rm form}$, some seven orders of magnitude above the
CMB amplitude $\mathcal{P}_\mathcal{R}^{\rm CMB} \approx
2\times10^{-9}$.
Generating such an enhancement without disturbing CMB scales
requires additional model ingredients --- a spectator field,
a feature in the inflationary potential, or resonant particle
production~\CITE{Kohri:2018, Ozsoy:2021, Franciolini:2022} ---
whose model-building cost I fold into the tuning through the
single effective parameter $\sigma$, rather than through the
inflaton potential shape. This treatment is appropriate for the early-matter-domination
mechanism, where $\sigma$ is a free parameter of the
cosmological initial conditions; for inflationary models,
the additional cost of generating the required $\sigma$
from the inflaton potential is treated separately in
Sec.~\ref{subsec:inflation_ft} as a second layer of tuning.

\subsubsection{Parameter-space heatmap}
\label{subsubsec:md_map}

Figure~\ref{fig:pbh}(b) shows $\log_{10}\bgdelta$ in
the $(m_\phi, \sigma)$ plane, with the x-axis spanning
the modulus masses relevant for asteroid-mass PBH
production, $m_\phi \in [10^{10}, 10^{13}]$~GeV.
In contrast to the domain-wall panel, a clear color
gradient is visible: the panel darkens from yellow-green
at large $\sigma$ (low tuning, $\bgdelta \approx 7$)
toward teal at small $\sigma$ (higher tuning,
$\bgdelta \approx 20$), reflecting the rising exponential
sensitivity $\bgdelta_\sigma = 5 + 0.196/\sigma^{4/3}$
(Eq.~\ref{eq:delta_md_sigma}) as the Gaussian tail
that controls collapse is pushed further into the rare-event
regime.
The $\bgdelta = 10$ and $\bgdelta = 30$ white contours
bracket the natural island, confirming the analytic range
$\bgdelta_\sigma \in [12, 21]$ of
Eq.~(\ref{eq:delta_md_range}).
The vertical gold lines mark the asteroid-mass boundaries
$M_{\rm PBH} = 10^{17}$~g and $10^{22}$~g, which
correspond via Eq.~(\ref{eq:Mpbh_md}) to
$m_\phi \in [4\times10^{10},\, 2\times10^{12}]$~GeV;
the constraint is purely horizontal because $M_{\rm PBH}$
depends on $m_\phi$ alone and not on $\sigma$.
The natural island (green hatching) is a diagonal strip
running along the $\fPBH = 1$ contour.
Its tilt reflects the interplay between the two parameters:
increasing $\sigma$ raises $\beta_{\rm AM}$ through the
HYKN formula, which would overproduce PBHs unless
compensated by a larger $m_\phi$ (hence higher $\Trh$ and
stronger dilution from formation to equality).
The benchmark $m_\phi = 9\times10^{10}$~GeV,
$\sigma = 0.055$ (gold star) lies within the
asteroid-mass natural island at
$M_{\rm PBH} \approx 10^{21}$~g and $\bgdelta = 14.4$.

A guaranteed observational consequence of the enhanced
curvature power spectrum $\mathcal{P}_\mathcal{R}(k_{\rm form})
\approx \sigma^2 \approx 3\times10^{-3}$ required for
asteroid-mass PBH production is a stochastic gravitational-wave
background (SGWB) sourced at second order in perturbation
theory by the same scalar perturbations~\CITE{Ananda:2007,
Baumann:2007}.
The peak frequency of this scalar-induced GW (SIGW) background
is set by the formation temperature,
$f_{\rm peak} \approx 3\times10^{-3}\,{\rm Hz}\,
(T_{\rm form}/10^6\,{\rm GeV})$,
placing it in the mHz--Hz band for the asteroid-mass
window, precisely the target range of LISA and
DECIGO~\CITE{Caprini:2020}.
The SIGW amplitude scales as $\Omega_{\rm GW}h^2 \sim
\mathcal{P}_\mathcal{R}^2 \sim 10^{-6}$--$10^{-4}$
at the benchmark $\sigma = 0.055$, which is within the
projected LISA sensitivity for transition temperatures
$T_{\rm form} \gtrsim 10^5$~GeV~\CITE{Caprini:2020}.
Current bounds from Planck on the effective number of
relativistic species, $\Delta N_{\rm eff} \lesssim 0.3$,
translate into an integrated upper bound
$\int \Omega_{\rm GW}\,d\ln f \lesssim 5\times10^{-6}$
that does not constrain the benchmark point but begins
to disfavor $\sigma \gtrsim 0.15$ across the full
asteroid-mass range~\CITE{Caprini:2020}.
The SIGW signal is therefore a robust and near-future
testable prediction of the early-matter-domination PBH
mechanism, and constitutes an independent observational
channel complementary to microlensing searches.

\subsection{First-order phase transition}
\label{subsec:fopt}

\subsubsection{Formation mechanism and abundance map}
\label{subsubsec:fopt_mechanism}

A first-order phase transition (FOPT) occurs when a field
$\Phi$ in a hidden sector tunnels from a metastable
false vacuum to the true vacuum of its potential, rather
than rolling continuously as in a second-order transition.
The transition proceeds by the stochastic nucleation of
bubbles of the true vacuum within the false-vacuum background;
these bubbles expand at close to the speed of light,
collide, and eventually convert the entire Universe to the
new phase~\CITE{Linde:1981, Witten:1984}.
I consider a FOPT occurring at a temperature $T_*$ in a
hidden sector that is thermally coupled to the Standard Model
plasma, so that the transition temperature is a physical
parameter that sets the PBH mass through the Hubble horizon
at that time.

Two dimensionless parameters characterize the transition
completely for the purposes of PBH production.
The \emph{transition strength} $\alpha$ is the ratio
of the latent heat released during nucleation to the
ambient radiation energy density,
\begin{equation}
  \alpha \;\equiv\; \frac{\Delta V - T_*\,\partial\Delta V/\partial T_*}
                         {\rho_{\rm rad}(T_*)},
  \label{eq:alpha_def}
\end{equation}
where $\Delta V$ is the difference in free-energy density
between the two phases~\CITE{Espinosa:2010}.
A small $\alpha$ corresponds to a weak transition in which
latent heat is a minor perturbation to the radiation bath;
a large $\alpha$ corresponds to a strong, potentially
runaway transition that can dramatically reheat the Universe.
The \emph{transition rate parameter} $\beta/H_*$ measures
how rapidly the nucleation rate grows relative to the Hubble
rate at $T_*$; equivalently, $H_*/\beta$ is the duration
of the transition in Hubble units, with small $\beta/H_*$
corresponding to a slow, prolonged transition and large
$\beta/H_*$ to an almost instantaneous one.

Regions of the false vacuum that are not reached by any
bubble wall before the transition ends are compressed
by the surrounding expanding walls.
If the resulting overdensity exceeds the gravitational
collapse threshold, these regions form PBHs~\CITE{Hawking:1982,
Moss:1994, Kodama:1982, Liu:2022, Lewicki:2023, Gehrman:2023}.
The probability that a given Hubble-volume region remains
in the false vacuum until the end of the transition is
determined by the nucleation statistics; for an exponentially
growing nucleation rate $\Gamma \propto e^{\beta t}$
(the standard approximation~\CITE{Caprini:2020}), the
fraction of the Universe that collapses into PBHs is
\begin{equation}
  \beta_{\rm PBH}
  \;=\;
  \exp\!\left[-S(\alpha)\,\frac{\beta}{H_*}\right],
  \label{eq:beta_fopt}
\end{equation}%
where $S(\alpha)$ encodes how efficiently the transition
converts false-vacuum regions into PBHs as a function
of its strength\footnote{Equation~(\ref{eq:beta_fopt}) is the standard
phenomenological approximation in which the false-vacuum survival
fraction is written as a single exponential in the
nucleation-rate parameter $\beta/H_*$, treating $\beta/H_*$ as a
fundamental input.
A more complete analysis of the collapse probability in the
supercooled limit, tracking the full past-light-cone percolation
integral, shows that $\mathcal{P}_{\rm coll}$ is super-exponential
in $\beta/H$~\cite{Gouttenoire:2024}: the BG derivative
evaluated from that formula at the benchmark $\beta/H_*\approx 8$
gives $\bgdelta_{\beta/H} \approx 265$ rather than $\approx 35$,
placing FOPT-PBH in Class~III even at the phenomenological level
(Sec.~\ref{subsec:caveats}, Eq.~\ref{eq:delta_fopt_GV_bench}).
At a deeper level, tracing $f_{\rm PBH}$ back to the microphysical scalar
potential through the Euclidean action $S_3(T_p)/T_p$ yields
a double-exponential (``superexponential'') structure,
$f_{\rm PBH} \simeq M\exp(-Q\exp(-S_3(T_p)/T_p))$, with
$Q \sim 10^{77}$~\cite{Wu:2025}, giving $\bgdelta \sim 10^3$--$10^4$.
The fine-tuning analysis in
Eqs.~(\ref{eq:delta_fopt_T})--(\ref{eq:delta_fopt_universal})
uses the single-exponential approximation and therefore
gives a \emph{lower bound} on the true BG measure; see
Sec.~\ref{subsec:caveats} for the full assessment.}.
$S(\alpha)$ is model-dependent, but can be computed
analytically for a given scalar potential; for the
hidden-sector benchmark model of Ref.~\CITE{Gehrman:2023}
the result is well approximated by the interpolating
formula
\begin{equation}
  S(\alpha)
  \;\simeq\;
  5.5 \;-\; \frac{2\alpha}{\alpha + 0.3},
  \label{eq:Salpha}
\end{equation}
which decreases monotonically from $S(0) = 5.5$ in the
weak-transition limit (where the radiation bath is barely
disturbed and the false-vacuum survival probability is
relatively insensitive to $\alpha$) to $S(\infty) = 3.5$
in the ultra-strong limit (where the latent heat dominates
and bubble walls run away~\CITE{Espinosa:2010, Caprini:2020}).
The coefficients in Eq.~(\ref{eq:Salpha}) are specific to
that model; different bubble-wall velocities, plasma
couplings, or reheating dynamics would yield different
values.
As I show below in Eq.~(\ref{eq:delta_fopt_universal}),
however, the key fine-tuning identity
$S(\alpha)\,\beta/H_*\big|_{\fPBH = 1} =
\ln[T_*/(0.84\,T_{\rm eq})]$
follows from setting $\fPBH = 1$ in Eq.~(\ref{eq:fpbh_fopt})
and holds for \emph{any} monotone $S(\alpha)$; the
fine-tuning result is therefore model-independent even
though $S(\alpha)$ is not.
The PBH mass is set by the Hubble horizon at $T_*$,
$M_{\rm PBH} \simeq 3.8\Mpl^3/T_*^2$, and the
present-day dark matter fraction is
\begin{equation}
  \fPBH(T_*, \beta/H_*, \alpha)
  \;=\;
  \exp\!\left[-S(\alpha)\,\frac{\beta}{H_*}\right]
  \cdot
  \frac{T_*}{0.84\,T_{\rm eq}}.
  \label{eq:fpbh_fopt}
\end{equation}
The parameter space of interest for this mechanism
is defined entirely by the requirement $\fPBH \approx 1$
within the asteroid-mass window, which fixes
$\beta/H_* \sim 5$--$15$ at the benchmark temperatures
$T_* \sim 10^5$--$10^7$~GeV.
It is a non-trivial consequence that this same range
of $\beta/H_*$ lies within the sensitivity reach of LISA
and next-generation ground-based gravitational-wave
detectors for the stochastic background produced by the
transition itself~\CITE{Caprini:2020, Hindmarsh:2017}:
FOPT-PBH dark matter is therefore simultaneously a target
of gravitational-wave observatories and microlensing
surveys, not because the two observations are by
construction linked, but because both probe the same
underlying parameter $\beta/H_*$.

\subsubsection{Analytic fine-tuning structure}
\label{subsubsec:fopt_ft}

Differentiating Eq.~(\ref{eq:fpbh_fopt}):
\begin{align}
  \bgdelta_{T_*}
  &\;=\;
  \left|\frac{\partial \ln \fPBH}{\partial \ln T_*}\right|
  \;=\; 1,
  \label{eq:delta_fopt_T}
  \\[4pt]
  \bgdelta_{\beta/H}
  &\;=\;
  \left|\frac{\partial \ln \fPBH}{\partial \ln(\beta/H_*)}\right|
  \;=\; S(\alpha)\,\frac{\beta}{H_*},
  \label{eq:delta_fopt_bH}
  \\[4pt]
  \bgdelta_\alpha
  &\;=\;
  \left|\frac{\partial \ln \fPBH}{\partial \ln\alpha}\right|
  \;=\;
  \frac{\beta}{H_*}\cdot\frac{0.6\,\alpha}{(\alpha+0.3)^2}.
  \label{eq:delta_fopt_alpha}
\end{align}
The temperature sensitivity $\bgdelta_{T_*} = 1$ is
a constant, since $\fPBH \propto T_*$ through the
dilution factor alone.
The dominant contribution is $\bgdelta_{\beta/H}$,
which grows linearly with $\beta/H_*$.

The key result emerges upon imposing the condition
$\fPBH = 1$, that is, solving
Eq.~(\ref{eq:fpbh_fopt}) for $\beta/H_*$ at fixed
$T_*$ and $\alpha$:
\begin{equation}
  S(\alpha)\,\frac{\beta}{H_*}
  \;\bigg|_{\fPBH = 1}
  \;=\;
  \ln\!\left[\frac{T_*}{0.84\,T_{\rm eq}}\right].
  \label{eq:fopt_contour}
\end{equation}
Substituting Eq.~(\ref{eq:fopt_contour}) into Eq.~(\ref{eq:delta_fopt_bH}),
\begin{equation}
  \bgdelta_{\beta/H}\bigg|_{\fPBH = 1}
  \;=\;
  \ln\!\left[\frac{T_*}{0.84\,T_{\rm eq}}\right]
  \;\approx\; 30\text{--}37
  \label{eq:delta_fopt_universal}
\end{equation}
for $T_* \in [10^4,\, 10^{6.5}]$~GeV
(the asteroid-mass formation window).
This result is exact for any $\alpha$ and any
$T_*$ in the relevant range, and is independent of
the specific value of $S(\alpha)$:
\emph{the fine-tuning of FOPT-PBH dark matter is
  determined entirely by the logarithm of the
  dilution ratio from formation to equality, and
  is insensitive to the strength or dynamics of the
  phase transition.}
The nucleation-rate parameter $\beta/H_*$ must be tuned
to $\sim 3\%$ precision to land on the $\fPBH = 1$ contour.

Equation~(\ref{eq:delta_fopt_universal}) is the
gravitational-sector analog of the thermal-freeze-out
tuning $\bgdelta \approx x_F \simeq 25$ for
standard WIMPs (see Sec.~\ref{subsubsec:b3});
both express the cost of exponentially suppressing a
natural formation rate to match the observed
dark matter density.
This structural identity is the central observation of
Sec.~\ref{sec:comparison}.

\subsubsection{Parameter-space heatmap}
\label{subsubsec:fopt_map}

Figures~\ref{fig:pbh}(c) and~\ref{fig:pbh}(d) show
$\log_{10}\bgdelta$ in the $(T_*, \beta/H_*)$ plane
for $\alpha = 0.3$ (moderately strong transition,
$S(0.3) \approx 4.1$) and $\alpha = 1.0$ (strong
transition, $S(1.0) \approx 4.0$), respectively.
Both panels share the same qualitative color structure:
a uniform teal background corresponding to
$\bgdelta \approx 35$ throughout most of the plane,
with darker (more tuned) shading at large $\beta/H_*$
where the exponential suppression of $\beta_{\rm PBH}$
becomes more severe, and lighter shading at small
$\beta/H_*$ where PBH overproduction would occur.
The $\bgdelta = 30$ white contour runs nearly horizontally
across each panel and coincides almost exactly with
the $\fPBH = 1$ contour (thick black), making the
key identity $\bgdelta_{\beta/H} = \ln[T_*/(0.84\,
T_{\rm eq})] \approx 35$ (Eq.~\ref{eq:delta_fopt_universal})
directly visible as a geometric feature of both plots.

The two panels are strikingly similar despite the
factor-of-three difference in $\alpha$: the $\fPBH = 1$
contour shifts upward by only $\Delta\log_{10}(\beta/H_*)
\approx 0.07$ between panels (c) and (d),
corresponding to $\beta/H_* = 7.78$ vs.\ $8.84$ at the
benchmark temperature $T_* = 10^6$~GeV.
This near-degeneracy is the visual confirmation of the
$\alpha$-independence established analytically in
Eq.~(\ref{eq:delta_fopt_universal}): since
$S(0.3) \approx 4.1 \approx S(1.0) \approx 4.0$,
the required value of $\beta/H_*$ to achieve $\fPBH = 1$
at fixed $T_*$ changes by only $\sim 14\%$, and the
fine-tuning $\bgdelta = 35$ is identical in both cases.

The natural island (green hatching, $0.1 \leq \fPBH \leq 10$
within the asteroid-mass window) appears in both panels
as a narrow, nearly horizontal strip at
$\log_{10}(\beta/H_*) \approx 0.9$.
Its horizontal extent spans the full asteroid-mass
temperature range $\log_{10}(T_*/{\rm GeV}) \in [4.7,\, 7.6]$
(bounded by the vertical gold lines), a factor of
$\sim 800$ in $T_*$.
This wide coverage means that a single value of
$\beta/H_*$, independently of the transition temperature,
places the mechanism in the asteroid-mass window;
the PBH mass is selected by $T_*$ (with $\bgdelta_{T_*} = 1$,
a one-parameter freedom) while the abundance is controlled
by $\beta/H_*$ (with $\bgdelta_{\beta/H} \approx 35$
within the single-exponential approximation used here,
and $\approx 265$ from the more accurate collapse formula
of Ref.~\cite{Gouttenoire:2024}; see Sec.~\ref{subsec:caveats}).
The color scale in both panels therefore represents a
\emph{lower bound} on the true BG measure.
The benchmark $T_* = 10^6$~GeV is marked by gold stars
in both panels; the required nucleation rate is
$\beta/H_* = 7.78$ in panel~(c) and $8.84$ in panel~(d),
and both stars lie precisely on the $\fPBH = 1$ contour
at $\bgdelta = 35$.

\subsection{Inflationary collapse: a two-layer structure}
\label{subsec:inflation_ft}

The three mechanisms analyzed above (biased domain walls,
early matter domination, and first-order phase transitions)
do not exhaust the list of PBH formation channels.
The most extensively studied mechanism in the literature is
the collapse of large-amplitude curvature perturbations
generated during single-field inflation.
In the standard picture, inflation produces a nearly
scale-invariant power spectrum of curvature perturbations
$\mathcal{P}_\mathcal{R}(k)$ whose amplitude on CMB scales
is $\mathcal{P}_\mathcal{R} \approx 2\times10^{-9}$.
PBH formation requires $\mathcal{P}_\mathcal{R} \sim
10^{-2}$--$10^{-3}$ on sub-CMB scales corresponding to
the asteroid-mass Hubble horizon, an enhancement of
$\mathcal{O}(10^6$--$10^7)$ relative to the CMB
amplitude~\CITE{Franciolini:2022, Ozsoy:2021}.
Such an enhancement is typically achieved through a
feature in the single-field inflaton potential, most
commonly an inflection point or a phase of ultra-slow-roll
(USR), during which the inflaton velocity is briefly
reduced to near zero, allowing quantum fluctuations to grow
to large amplitude on the scales that cross the Hubble
horizon at that moment~\CITE{Hertzberg:2017, Ozsoy:2021}.
The curvature perturbation variance $\sigma$ that enters
the collapse formula (Eq.~\ref{eq:beta_AM}) is related to
the power spectrum by
$\sigma^2 \approx \int_{k_{\rm form}} \mathcal{P}_\mathcal{R}(k)
\,W^2(k)\,{\rm d}\ln k$,
where $W(k)$ is a window function on the formation scale
$k_{\rm form}$; for a narrow peak at $k_{\rm form}$,
$\sigma \approx \mathcal{P}_\mathcal{R}^{1/2}(k_{\rm form})$,
though more generally $\sigma$ depends on the full
spectral shape through the window-function integral.
Since the present analysis treats $\sigma$ as a free
effective parameter, the precise form of this relation
does not affect the fine-tuning conclusions; the
model-building cost of generating the required large
$\sigma$ from the inflaton potential is instead
captured by the Layer~2 analysis below.

I do not analyze inflationary PBH production in the same
detail as the previous three mechanisms, partly because
its fine-tuning has already received dedicated attention
in the literature~\CITE{Kalaja:2019, Hertzberg:2017,
Stamou:2024, Iovino:2024}, and partly for the methodological
reason explained in the opening of this section.
Instead, I develop a two-layer decomposition of the
inflationary fine-tuning: a first layer capturing the
sensitivity of the PBH abundance to the curvature power
spectrum amplitude $\sigma$, whose value depends critically
on the reheating history, and a second layer capturing the
additional sensitivity of $\sigma$ to the inflaton
potential coefficients.
This decomposition places the six inflationary model
classes into a systematic hierarchy and clarifies why
their fine-tuning spans more than seven orders of magnitude.

The inflationary mechanism involves two successive
parameter-to-observable maps, each contributing
independently to the total fine-tuning.

\textit{Layer~1: $\sigma \to f_{\rm PBH}$.}
The form of the collapse fraction $\beta(\sigma)$ depends
on the equation of state at the epoch when the enhanced
perturbations re-enter the Hubble horizon, and this
distinction is physically important for the fine-tuning
estimate.

For the early-matter-domination mechanism of
Sec.~\ref{subsec:md}, collapse occurs during a
pressureless era and is described by the HYKN
formula~(\ref{eq:beta_AM}) with threshold
$\delta_c \approx 0.05$, giving
$\Delta_\sigma^{\rm (MD)} \approx 14$
(Eq.~\ref{eq:delta_md_sigma}).
I refer to this as the Layer~1 floor for the MD case:
it sets a minimum fine-tuning cost independent of
whatever mechanism generates the required $\sigma$.

For inflationary PBH models the background at re-entry
depends critically on the reheating temperature $T_{\rm reh}$.
Two limiting cases arise.

\textit{Low reheating ($T_{\rm reh} \lesssim T_{\rm form}$).}
If the inflaton decay rate is small, the inflaton oscillates
coherently after inflation ends and the universe is
matter-dominated at the epoch when the formation-scale
perturbations re-enter the
horizon~\CITE{Georg:2017, Allahverdi:2020}.
In this case the collapse occurs in an MD background and
the HYKN formula with $\delta_c \approx 0.05$ applies,
giving a Layer~1 floor identical to the non-inflationary
early-MD mechanism:
\begin{equation}
  \Delta_\sigma^{\rm (MD)}
  \;\approx\; 14,
  \label{eq:delta_sigma_layer1_MD_inf}
\end{equation}
identical to Eq.~(\ref{eq:delta_md_sigma}).
For asteroid-mass PBHs with
$T_{\rm form} \sim 10^4$--$10^7$~GeV,
this requires $T_{\rm reh} \lesssim 10^7$~GeV, achievable
in models with a weakly-coupled inflaton sector.
Note that while the Layer~1 floor is the same as in the
non-inflationary early-MD case, the two scenarios are
physically distinct: here $\sigma$ is not a free initial
condition but is generated by the inflaton potential,
and the additional cost of achieving the required $\sigma$
is captured by Layer~2.

\textit{High reheating ($T_{\rm reh} \gg T_{\rm form}$).}
In the standard scenario where reheating completes
well before the formation scale re-enters the horizon,
the background is radiation-dominated at
re-entry~\CITE{Green:1997}.
The collapse fraction follows the Press--Schechter form
with $\delta_c \approx 0.45$--$0.55$~\CITE{Kalaja:2019}:
\begin{equation}
  \beta_{\rm RD}(\sigma)
  \;=\;
  \tfrac{1}{2}\,\mathrm{erfc}\!\left(
    \frac{\delta_c}{\sqrt{2}\,\sigma}
  \right)
  \;\approx\;
  \frac{\sigma}{\delta_c\sqrt{2\pi}}\,
  \exp\!\left(-\frac{\delta_c^2}{2\sigma^2}\right),
  \label{eq:beta_RD}
\end{equation}
where the approximation holds for
$\delta_c/(\sqrt{2}\,\sigma) \gg 1$.
Log-differentiating Eq.~(\ref{eq:beta_RD}):
\begin{equation}
  \Delta_\sigma^{\rm (RD)}
  \;=\;
  \left|\frac{\partial\ln\beta_{\rm RD}}{\partial\ln\sigma}
  \right|
  \;=\;
  1 + \frac{\delta_c^2}{\sigma^2}.
  \label{eq:delta_sigma_layer1}
\end{equation}
For asteroid-mass PBH production one requires
$\beta_{\rm RD} \sim 10^{-9}$--$10^{-15}$, giving
$\delta_c^2/(2\sigma^2) \approx 20$--$35$,
so $\sigma \approx 0.05$--$0.07$
for $\delta_c = 0.45$--$0.55$.
Evaluating Eq.~(\ref{eq:delta_sigma_layer1}):
\begin{equation}
  \Delta_\sigma^{\rm (RD)}
  \;\approx\; 57\text{--}70
  \qquad
  (\sigma \approx 0.06,\;
   \delta_c \approx 0.45\text{--}0.55),
  \label{eq:delta_sigma_layer1_RD}
\end{equation}
approximately five times larger than the matter-dominated
value of Eq.~(\ref{eq:delta_sigma_layer1_MD_inf}),
reflecting the much higher collapse threshold in radiation
domination.

\begin{figure*}[t]
  \includegraphics[width=\textwidth]{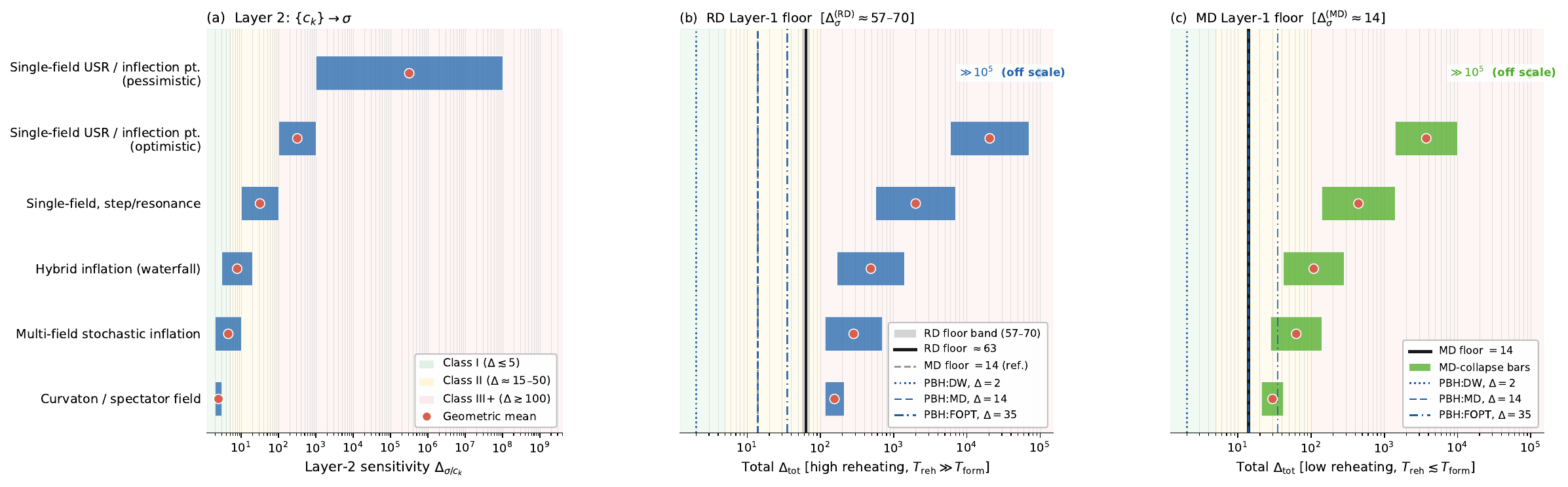}
  \caption{%
    Comparison of Layer~2 sensitivity
    ($\Delta_{\sigma/c_k}$, panel~a) and total fine-tuning
    ($\Delta_{\rm tot} = \Delta_\sigma \times
    \Delta_{\sigma/c_k}$, panels~b and~c) across six
    inflationary PBH model classes, ordered from most
    natural (bottom) to most tuned (top).
    Bars show the estimated range within each class; the
    dot marks the geometric mean.
    Background shading indicates qualitative fine-tuning
    tier bands (Class~I, II, and~III; defined in
    Sec.~\ref{subsec:classes}).
    \textit{Panel~(b):} total fine-tuning with the
    matter-dominated HYKN Layer~1 floor
    $\Delta_\sigma^{\rm (MD)} \approx 14$
    (Eq.~\ref{eq:delta_sigma_layer1_MD_inf}), applicable
    when $T_{\rm reh} \lesssim T_{\rm form}$ (low
    reheating, inflaton-oscillation MD era at re-entry).
    \textit{Panel~(c):} total fine-tuning with the
    radiation-dominated Press--Schechter Layer~1 floor
    $\Delta_\sigma^{\rm (RD)} \approx 57$--$70$
    (Eq.~\ref{eq:delta_sigma_layer1_RD}), applicable when
    $T_{\rm reh} \gg T_{\rm form}$ (standard high
    reheating, RD background at re-entry).
    The $x$-axes of panels~(b) and~(c) are identical
    (capped at $10^5$) to make the factor-of-five shift
    in the Layer~1 floor directly visible; the top row
    (single-field USR pessimistic,
    $\Delta_{\rm tot} \gtrsim 10^9$--$10^{10}$) is off
    scale and indicated by an arrow.
    Thick vertical gray lines mark the respective
    Layer~1 floor in each total-tuning panel; in
    panel~(c) the gray band spans the $57$--$70$ range.
    Blue vertical lines show the three non-inflationary
    PBH benchmarks (dotted: DW gravity-fixed,
    $\Delta = 2$; dashed: early MD, $\Delta = 14$;
    dash-dotted: FOPT, $\Delta = 35$, single-exponential
    approximation; see Sec.~\ref{subsec:caveats} for the
    more accurate estimate $\Delta \approx 265$).
    All ranges are estimates based on analytic scaling
    arguments and the results of
    Refs.~\CITE{Hertzberg:2017, Cole:2023, Kohri:2007, Bugaev:2011,
    Vennin:2020, Clesse:2015, Ballesteros:2018,
    Motohashi:2017, Kannike:2017}; precise values are
    model-dependent.
    \label{fig:inf_comparison}
  }
\end{figure*}

The two Layer~1 floors bracket the Layer~1 contribution
to the fine-tuning of inflationary models:
\begin{equation}
  \Delta_\sigma^{\rm (inf)}
  \;\in\;
  \bigl[\,\Delta_\sigma^{\rm (MD)},\;
         \Delta_\sigma^{\rm (RD)}\,\bigr]
  \;\approx\; [14,\; 57\text{--}70],
  \label{eq:delta_sigma_layer1_range}
\end{equation}
depending entirely on the reheating history.
Figure~\ref{fig:inf_comparison} displays $\Delta_{\rm tot}$
for both cases side by side, with the same $x$-axis scale
to make the factor-of-five shift in the Layer~1 floor
directly visible; note that for model classes with large
Layer~2 sensitivity the total tuning is dominated by
Layer~2 and the factor-of-five shift between panels
becomes a small fractional effect.
The enhanced power spectrum required for inflationary
PBH production carries the same unavoidable
scalar-induced gravitational-wave (SIGW) background
prediction discussed in Sec.~\ref{subsec:md}, with the
additional feature that the peak frequency and amplitude
directly encode the inflaton potential
shape~\CITE{Franciolini:2022, Ananda:2007, Baumann:2007},
making it a particularly powerful discriminant between
inflationary PBH models.

\textit{Layer~2: $\{c_k\} \to \sigma$.}
The additional tuning specific to inflationary models
enters through the mapping from the inflaton potential
coefficients $\{c_k\}$ --- e.g.\ the height of the
inflection point, its slope, and the field value at
which USR begins --- to the power spectrum amplitude
$\sigma$ on the formation scale.
Near an inflection point, the inflaton velocity $\dot\phi$
passes through a local minimum whose depth is controlled
by the local slope $c_{\rm slope}$ of the potential.
The power spectrum amplitude depends on $\dot\phi^{-2}$
at horizon crossing, so it is exponentially sensitive to
$c_{\rm slope}$:
\begin{equation}
  \sigma
  \;\propto\;
  \exp\!\left(-\frac{\rm const}{c_{\rm slope}}\right).
  \label{eq:sigma_inf}
\end{equation}
A fractional change $\delta c_{\rm slope}/c_{\rm slope}$
therefore produces a fractional change
$\delta\sigma/\sigma \propto
\delta c_{\rm slope}/c_{\rm slope}^2$,
an exponentially amplified response.
Hertzberg and Yamada~\CITE{Hertzberg:2017} quantified
this explicitly for polynomial inflation potentials:
a change in the potential coefficients by one part in
$10^2$--$10^8$ (depending on the specific model) shifts
the power spectrum peak amplitude by an order-one factor,
implying $\Delta_{\sigma/c_k} \sim 10^2$--$10^8$.
A systematic ``fine-tuning audit'' of single-field PBH
models by Cole, Gow, Byrnes, and Patil~\CITE{Cole:2023}
reached the same conclusion across three representative
classes of single-field potential (superposed-feature,
polynomial, and non-polynomial), finding that the potential
coefficients must be tuned to between one part in $\sim
10^2$ and one part in $\sim 10^8$ to hold the
power-spectrum peak amplitude fixed, in direct agreement
with the Layer~2 range adopted here.
Notably, Ref.~\CITE{Cole:2023} also separates this from the
additional sensitivity of the PBH abundance to the
power-spectrum amplitude, reporting that the latter
contributes a further factor of $\rho \approx 22$--$38$
(for radiation-dominated collapse, with a value
$\rho \approx \delta_c^2/2\sigma^2$ that maps onto the
Layer~1 floor $\Delta_\sigma^{\rm (RD)} \approx 57$--$70$ of
Eq.~(\ref{eq:delta_sigma_layer1_RD}), reduced to
$\rho \approx 5/4$ in an early-matter-dominated era).
Their two-stage decomposition of the tuning is precisely
the Layer~2~$\times$~Layer~1 structure formalized in this
paper, and provides an independent cross-check of both
factors.
The total BG measure for the potential coefficients is
therefore the product of the two layers,
\begin{equation}
  \Delta_{c_k}
  \;\sim\;
  \Delta_\sigma
  \;\times\;
  \Delta_{\sigma/c_k}
  \;\sim\;
  (14\text{--}70) \;\times\; (10^2\text{--}10^8),
  \label{eq:delta_inf_total}
\end{equation}
placing single-field inflationary collapse in
Class~III or beyond (as defined in
Sec.~\ref{subsec:classes}) for essentially all
realistic potentials~\CITE{Kalaja:2019, Stamou:2024},
with the Layer~1 factor ranging from $14$
(low-reheating MD collapse) to $57$--$70$
(high-reheating RD collapse) as discussed above.
This double-exponential structure has no analog in the
three mechanisms analyzed above, where the fundamental
parameters ($m_\phi$, $V_b$, $T_*$, $\beta/H_*$) map
onto $f_{\rm PBH}$ through at most one exponential, and
not through any hidden exponential in the
parameter-to-$\sigma$ map.

Note that the magnitude of Layer~2 varies enormously
across inflationary model classes, as summarized in
Fig.~\ref{fig:inf_comparison} and reviewed in detail
in the next section.
For single-field USR and inflection-point potentials
$\Delta_{\sigma/c_k} \sim 10^2$--$10^8$, placing the
total fine-tuning firmly in Class~III or beyond
(Sec.~\ref{subsec:classes}) regardless of the reheating
history.
For multi-field and spectator-field models
$\Delta_{\sigma/c_k} \sim 2$--$10$; the total tuning
then depends on whether collapse occurs in RD
($\Delta_{\rm tot} \approx 115$--$700$, Class~III for
standard high reheating) or MD
($\Delta_{\rm tot} \approx 21$--$140$, Class~II to lower
Class~III for low reheating with
$T_{\rm reh} \lesssim T_{\rm form}$).
The severe fine-tuning of single-field USR inflationary
PBH production is a specific pathology of models that
rely on an exponentially sensitive slow-roll feature to
amplify the power spectrum; the reheating history then
sets the overall tuning scale for all other inflationary
model classes.

The recent analysis of Iovino and Riotto~\CITE{Iovino:2024}
reaches a more optimistic conclusion by applying Wilson's
naturalness criterion to the potential coefficients
rather than to $\sigma$.
Their measure assigns a value close to unity even for
fine-tuned inflaton potentials because it is constructed
to be insensitive to the overall normalization of the
abundance: a potential that predicts $f_{\rm PBH} = 0.001$
and one that predicts $f_{\rm PBH} = 1$ are treated as
equally natural.
The BG measure is more conservative in this respect:
it registers the full sensitivity of $f_{\rm PBH}$ to
each parameter along the observable contour
$f_{\rm PBH} = 1$, and therefore captures the
double-exponential cost of maintaining that condition
via an inflection-point potential.
Both approaches are internally consistent, and answer
complementary questions.
The BG question is: ``how precisely must the potential
coefficient $c_k$ be specified to keep
$f_{\rm PBH} \approx 1$?''
The Wilson question is: ``does a technically natural UV
completion exist in which $c_k$ takes its current
value?''
For the cross-paradigm comparison of this paper, the BG
question is the relevant one, since it is the same
question being asked of all other dark matter mechanisms
in the comparison.

\subsubsection{Fine-tuning across inflationary model classes}
\label{subsubsec:inf_classes}

The analysis above shows that the double-exponential
structure of Layer~2 is a specific property of
single-field inflection-point or USR models, not a
universal feature of inflationary PBH production.
Figure~\ref{fig:inf_comparison} quantifies Layer~2
($\Delta_{\sigma/c_k}$, panel~a) and the total
fine-tuning $\Delta_{\rm tot}$ for two choices of
Layer~1 floor (panels~b and~c): the matter-dominated
HYKN floor $\Delta_\sigma^{\rm (MD)} \approx 14$ that
applies when $T_{\rm reh} \lesssim T_{\rm form}$ (low
reheating, inflaton-oscillation era at re-entry), and
the radiation-dominated Press--Schechter floor
$\Delta_\sigma^{\rm (RD)} \approx 57$--$70$ that applies
when $T_{\rm reh} \gg T_{\rm form}$ (standard high
reheating).
The $x$-axes of panels~(b) and~(c) are identical to
facilitate direct comparison; the pessimistic USR row
($\Delta_{\rm tot} \gtrsim 10^9$--$10^{10}$) is off the
shared scale and marked by an arrow.
For reference, the three non-inflationary PBH benchmarks
are shown as vertical blue lines in both total-tuning
panels at $\Delta = 2$, $14$, and $35$.

I now discuss each model class in turn.

\paragraph{Curvaton and spectator-field models
($\Delta_{\sigma/c_k} \approx 2$--$3$;
$\Delta_{\rm tot} \approx 115$--$210$ [RD] or
$21$--$42$ [MD]).}
In the curvaton mechanism, a light spectator field $\chi$
with mass $m_\chi \ll H_{\rm inf}$ acquires perturbations
$\delta\chi \simeq H_{\rm inf}/(2\pi)$ during inflation
and later converts them into curvature perturbations when
it decays~\CITE{Lyth:2002, Enqvist:2001}.
PBH production requires an enhanced curvature perturbation
on sub-CMB scales, which can be arranged if $\chi$ rolls
along a self-interacting potential that generates a
non-Gaussian or scale-dependent signal on those
scales~\CITE{Kohri:2007, Bugaev:2011}.
The key structural property is that the curvature power
spectrum scales as
$\mathcal{P}_\mathcal{R} \propto (H_{\rm inf}/m_\chi)^{2n}$
for some integer $n \geq 1$ depending on the interaction
structure.
The BG sensitivity of $\sigma$ to the curvaton mass is
therefore $\Delta_{\sigma/m_\chi} = n \approx 1$--$2$,
and to the self-coupling $\Delta_{\sigma/\lambda} \approx 1$.
The combined Layer~2 is
$\Delta_{\sigma/c_k} \approx 2$--$3$.
Multiplied by the RD Layer~1 floor
$\Delta_\sigma^{\rm (RD)} \approx 57$--$70$
(Eq.~\ref{eq:delta_sigma_layer1_RD}), the total
fine-tuning is $\Delta_{\rm tot} \approx 115$--$210$
(Class~III, Sec.~\ref{subsec:classes}) for standard
high-reheating scenarios.
For low-reheating scenarios where collapse occurs during
the inflaton-oscillation matter-dominated era, the MD
Layer~1 floor applies and the total reduces to
$\Delta_{\rm tot} \approx 21$--$42$
(Class~II, Sec.~\ref{subsec:classes}), comparable to the
early-matter-domination mechanism of
Sec.~\ref{subsec:md}.
The reheating temperature is therefore a critical
parameter that determines whether curvaton PBH models
belong to Class~II or Class~III.

\paragraph{Multi-field stochastic inflation
($\Delta_{\sigma/c_k} \approx 2$--$10$;
$\Delta_{\rm tot} \approx 115$--$700$ [RD] or
$28$--$140$ [MD]).}
When the inflaton traverses an approximately flat
direction, quantum diffusion can compete with or dominate
the classical drift, and the statistics of perturbations
must be described by a Fokker--Planck equation rather
than the standard linear perturbation
theory~\CITE{Starobinsky:1994}.
In multi-field models where a secondary field undergoes
stochastic diffusion, the power spectrum amplitude on a
given scale is controlled by the ratio
$H_{\rm inf}/M_{\rm flat}$, where $M_{\rm flat}$ is the
effective mass of the flat
direction~\CITE{Vennin:2020, Pattison:2021}.
Since $\mathcal{P}_\mathcal{R} \propto
(H_{\rm inf}/M_{\rm flat})^4$, the Layer~2 sensitivity
is $\Delta_{\sigma/M_{\rm flat}} \approx 2$ per
parameter, giving a combined
$\Delta_{\sigma/c_k} \approx 2$--$10$ when multiple
parameters of the flat-direction potential contribute.
Multiplied by the RD Layer~1 floor
$\Delta_\sigma^{\rm (RD)} \approx 57$--$70$
(Eq.~\ref{eq:delta_sigma_layer1_RD}), the total
fine-tuning range is
$\Delta_{\rm tot} \approx 115$--$700$ (Class~III,
Sec.~\ref{subsec:classes}) for high-reheating scenarios;
with the MD floor the range reduces to
$\Delta_{\rm tot} \approx 28$--$140$, straddling the
Class~II/III boundary.
The upper end of this range arises when the stochastic
enhancement requires the flat direction to remain nearly
degenerate with the inflationary trajectory for a
precise number of e-folds~\CITE{Vennin:2020}, introducing
additional sensitivity that stretches the Layer~2
estimate toward its upper bound.

\paragraph{Hybrid inflation with waterfall transition
($\Delta_{\sigma/c_k} \approx 3$--$20$;
$\Delta_{\rm tot} \approx 170$--$1400$ [RD] or
$42$--$280$ [MD]).}
In hybrid inflation, the inflaton $\phi$ rolls toward a
critical value $\phi_c$ at which a ``waterfall'' field
$\psi$ becomes tachyonic and triggers the end of
inflation~\CITE{GarciaBellido:1996}.
Near $\phi_c$, the perturbations of $\psi$ are
dramatically amplified because its effective mass
$m^2_\psi \approx g^2(\phi - \phi_c)M^{-2} - \mu^2$
approaches zero, and regions where the transition is
delayed can collapse into PBHs~\CITE{Clesse:2015}.
The curvature power spectrum amplitude scales as a power
law in the coupling $g$ and the mass parameter $\mu$:
$\mathcal{P}_\mathcal{R} \propto g^{-2n}\,\mu^{2m}$
for integers $n$, $m$ depending on the potential shape,
giving $\Delta_{\sigma/g} \approx n$ and
$\Delta_{\sigma/\mu} \approx m$.
For a mild waterfall ($|\phi - \phi_c|$ not exponentially
small), $n + m \approx 3$--$5$ and the Layer~2 tuning
is modest.
However, as $\phi \to \phi_c$ more closely, the
enhancement duration becomes exponentially sensitive to
the proximity to the critical point, stretching
$\Delta_{\sigma/c_k}$ toward $10$--$20$ at the upper end.
Multiplied by the RD Layer~1 floor, the resulting total
tuning is $\Delta_{\rm tot} \approx 170$--$1400$
(Class~III, Sec.~\ref{subsec:classes}) for
high-reheating scenarios; with the MD floor the range
is $\Delta_{\rm tot} \approx 42$--$280$, in the lower
part of Class~III.

\paragraph{Single-field with step or resonance feature
($\Delta_{\sigma/c_k} \approx 10$--$100$;
$\Delta_{\rm tot} \approx 570$--$7000$ [RD] or
$140$--$1400$ [MD]).}
A sharp feature such as a step, a bump, or a periodic
modulation in the single-field inflaton potential can
resonantly amplify perturbations as modes cross the
feature~\CITE{Ballesteros:2018, Cai:2019}.
For a step of height $h$ at field value $\phi_s$, the
peak power spectrum scales roughly as
$\mathcal{P}_\mathcal{R}^{\rm peak} \propto h^2$,
a polynomial relationship giving
$\Delta_{\sigma/h} \approx 1$.
The dominant source of Layer~2 tuning is therefore not
the feature amplitude but its position: the resonance
condition requires that precisely the modes corresponding
to the asteroid-mass scale cross the horizon when the
inflaton is at the step location.
A small change $\delta\phi_s$ in the step position shifts
the resonant scale by $\delta k/k \sim
\delta\phi_s/(\phi_s - \phi_{\rm CMB})$, changing which
physical scales are enhanced.
In terms of the inflaton potential coefficients,
$\Delta_{\sigma/\phi_s} \sim 10$--$50$ because $\phi_s$
must be tuned at the $2$--$10\%$ level to target the
asteroid-mass window specifically.
The combined Layer~2 is
$\Delta_{\sigma/c_k} \approx 10$--$100$; multiplied by
the RD Layer~1 floor this gives
$\Delta_{\rm tot} \approx 570$--$7000$ (Class~III,
Sec.~\ref{subsec:classes}) for high-reheating scenarios,
or $\Delta_{\rm tot} \approx 140$--$1400$ with the MD
floor.
A key advantage of this class of models is that the
prediction is falsifiable: the power spectrum feature
also produces a distinctive oscillatory signature in
the SIGW background~\CITE{Cai:2019} that is detectable
by LISA for formation temperatures in the asteroid-mass
range.

\paragraph{Single-field USR / inflection point, optimistic
($\Delta_{\sigma/c_k} \approx 10^2$--$10^3$;
$\Delta_{\rm tot} \approx 6000$--$7\times10^4$ [RD] or
$1400$--$10^4$ [MD]).}
For a single inflaton field with an inflection point,
the slow-roll velocity passes through a local minimum
controlled by the local slope $c_{\rm slope}$ of the
potential.
In the USR regime, quantum perturbations grow as
$\delta\phi \propto e^{3Ht}$ rather than remaining
constant, amplifying the power spectrum by a factor
$e^{6\Delta N_{\rm USR}}$ over $\Delta N_{\rm USR}$
e-folds of USR evolution~\CITE{Motohashi:2017}.
The number of USR e-folds required to achieve
$\mathcal{P}_\mathcal{R} \sim 10^{-2}$ from a CMB-scale
amplitude $\mathcal{P}_\mathcal{R}^{\rm CMB} \sim
2\times10^{-9}$ is $\Delta N_{\rm USR} \approx 2.7$,
which in turn requires the potential slope to satisfy
$c_{\rm slope} \approx {\rm const}/\Delta N_{\rm USR}$.
In the optimistic case, corresponding to potentials
where the inflection point is only mildly fine-tuned
and the coefficients are of order unity,
$\Delta_{\sigma/c_{\rm slope}} \approx
{\rm const}/c_{\rm slope} \sim 10^2$--$10^3$
(the Hertzberg--Yamada lower-end
result~\CITE{Hertzberg:2017}).
Multiplied by the RD Layer~1 floor, the total tuning
is $\Delta_{\rm tot} \approx 6000$--$7\times10^4$
(high-reheating) or
$\Delta_{\rm tot} \approx 1400$--$10^4$ (low-reheating),
in either case deep in Class~III
(Sec.~\ref{subsec:classes}); a change of order $0.1\%$
in the inflaton potential slope is sufficient to miss
the target abundance by a factor of order unity.

\paragraph{Single-field USR / inflection point, pessimistic
($\Delta_{\sigma/c_k} \approx 10^3$--$10^8$;
$\Delta_{\rm tot} \gtrsim 10^{10}$ [RD] or
$\gtrsim 10^9$ [MD]).}
For a generic polynomial inflaton potential, the
relation between the potential coefficients at the
inflection point and the resulting power spectrum
amplitude is far more sensitive than the optimistic
case suggests.
A detailed analysis of polynomial potentials of the
form $V(\phi) = V_0[1 + a_2(\phi/M)^2 +
a_3(\phi/M)^3 + \ldots]$ by Hertzberg and
Yamada~\CITE{Hertzberg:2017} and by Kannike, Raidal,
and Veerman~\CITE{Kannike:2017} shows that the required
amplitude $\mathcal{P}_\mathcal{R} \sim 10^{-2}$ at
$k_{\rm form}$ is consistent with the CMB constraint
$\mathcal{P}_\mathcal{R} \sim 2\times10^{-9}$ at
$k_{\rm CMB}$ only if the potential coefficients
satisfy a set of near-cancellation conditions.
In the generic case, achieving both constraints
simultaneously requires tuning the coefficients to
within one part in $10^3$--$10^8$, giving
$\Delta_{\sigma/c_k} \sim 10^3$--$10^8$.
Multiplied by the RD Layer~1 floor, the resulting
total fine-tuning reaches
$\Delta_{\rm tot} \gtrsim 10^{10}$ (high-reheating)
or $\Delta_{\rm tot} \gtrsim 10^9$ (low-reheating),
far exceeding any other scenario considered in this
paper in either case.
This pessimistic range represents the majority of the
parameter space for generic polynomial inflaton
potentials; the optimistic range is accessible only
in models with additional symmetry or structure that
correlates the potential coefficients in a way that
reduces the cancellation requirement.

The overall message of Fig.~\ref{fig:inf_comparison}
is stark: the fine-tuning of inflationary PBH
production spans more than seven orders of magnitude
depending on the model class, in either reheating
scenario.
For low reheating ($T_{\rm reh} \lesssim T_{\rm form}$,
MD Layer~1 floor), curvaton and spectator-field models
($\Delta_{\rm tot} \approx 21$--$42$) fall in Class~II,
comparable to the non-inflationary early-MD mechanism
and to coannihilating-WIMP dark matter.
For high reheating ($T_{\rm reh} \gg T_{\rm form}$,
RD Layer~1 floor), every inflationary class falls in
Class~III or beyond; the most natural class
(curvaton/spectator, $\Delta_{\rm tot} \approx
115$--$210$) is roughly five times more tuned than the
corresponding low-reheating estimate, reflecting the
factor-of-five ratio of the two Layer~1 floors --- a
shift that is clearly visible for model classes with
small Layer~2 sensitivity, but becomes a minor
correction when Layer~2 dominates.
In both cases the Layer~2 sensitivity
$\Delta_{\sigma/c_k}$ spans five orders of magnitude
from curvaton models to single-field USR potentials,
so the mechanism for enhancing the power spectrum
remains the most physically informative quantity
regardless of the reheating history.

\section{\label{sec:particle}Particle Dark Matter Benchmarks}

I now apply the same BG protocol to seven particle dark matter
benchmarks spanning four production paradigms: thermal-relic
WIMPs (B1--B3), freeze-in (B4), asymmetric dark matter (B5),
and axions (B6, B7).
The purpose is not to provide a comprehensive survey of the
particle dark matter landscape, but to place specific,
representative benchmarks into the same quantitative framework
used for the PBH constructions in Sec.~\ref{sec:pbh}.
The benchmarks are chosen to span the full range of BG
fine-tuning observed in the particle literature and to
correspond to physically well-motivated, phenomenologically
viable scenarios.

\subsection{Thermal-relic WIMPs}
\label{subsec:wimp}

Thermal relics freeze out when their annihilation rate falls
below the Hubble rate at temperature
$T_{\rm fo} = M_\chi/x_F$ with $x_F \approx 25$
\CITE{Jungman:1996, Kolb:1990}.
The relic abundance is then determined by the Lee--Weinberg
relation
\begin{equation}
  \Omega_\chi h^2
  \;\simeq\;
  \frac{1.07\times10^9\,{\rm GeV}^{-1}\,x_F}
       {g_*^{1/2}\,\Mpl\,\langle\sigma v\rangle_{\rm fo}},
  \label{eq:LeeWeinberg}
\end{equation}
where $g_* \approx 86$ counts the effective relativistic
degrees of freedom at freeze-out and
$\langle\sigma v\rangle_{\rm fo}$ is the thermally-averaged
annihilation cross section evaluated at $T_{\rm fo}$.
The ``WIMP miracle'' is the observation that for
$M_\chi \sim 100$~GeV and gauge-coupling-strength
$\langle\sigma v\rangle \sim \alpha_W^2/M_W^2
\sim 6.7\times10^{-9}$~GeV$^{-2}$,
Eq.~(\ref{eq:LeeWeinberg}) gives $\Omega h^2 \approx 0.12$
\CITE{Jungman:1996}.
I analyze three qualitatively distinct WIMP benchmarks that
span the range of BG fine-tuning encountered in this paradigm.

\subsubsection{Off-resonance Higgs-portal singlet (B1)}
\label{subsubsec:b1}

The simplest renormalizable portal between the Standard Model
and a dark sector is the real singlet scalar $S$ with
Lagrangian
\begin{equation}
  \mathcal{L} \supset
  \tfrac{1}{2}(\partial S)^2
  - \tfrac{1}{2}M_S^2\,S^2
  - \tfrac{\lambda_{HS}}{4}\,S^2\,H^\dagger H,
  \label{eq:singlet_lagrangian}
\end{equation}
where $\lambda_{HS}$ is the Higgs-portal coupling
\CITE{Silveira:1985, McDonald:1994, Burgess:2001}.
After electroweak symmetry breaking, $S$ acquires an effective
coupling to the physical Higgs $h$ and annihilates through
$SS \to hh$, $WW$, $ZZ$, $\bar{f}f$ channels.
In the off-resonance regime $M_S \gg M_h/2$, the
thermally-averaged cross section for these channels scales as
\begin{equation}
  \langle\sigma v\rangle
  \;\simeq\;
  \frac{\lambda_{HS}^2}{8\pi M_S^2}
  \;\equiv\;
  \lambda_{HS}^2\,\sigma_0\,(2\,{\rm TeV}/M_S)^2,
  \label{eq:sv_singlet_offres}
\end{equation}
where $\sigma_0 = 6.7\times10^{-9}$~GeV$^{-2}$ is
calibrated so that the benchmark B1
($M_S = 2$~TeV, $\lambda_{HS} = 1.0$) satisfies
$\Omega h^2 = 0.12$ exactly.
Substituting into Eq.~(\ref{eq:LeeWeinberg}),
\begin{equation}
  \Omega h^2
  \;\propto\;
  \frac{M_S^2}{\lambda_{HS}^2},
  \label{eq:Omega_singlet_offres}
\end{equation}
a pure power-law in both parameters.
Log-differentiation gives immediately
\begin{equation}
  \bgdelta_{M_S}^{(\rm B1)} \;=\; 2,
  \qquad
  \bgdelta_{\lambda_{HS}}^{(\rm B1)} \;=\; 2,
  \qquad
  \bgdelta^{(\rm B1)} \;=\; 2.
  \label{eq:delta_b1}
\end{equation}
These are constants, independent of $(M_S, \lambda_{HS})$,
for the same structural reason as the domain-wall PBH:
the abundance is a monomial in the parameters.

The spin-independent direct-detection cross section
at benchmark B1 is $\sigma_{\rm SI} \approx 1.3\times10^{-49}$~cm$^2$,
more than two orders of magnitude below the LZ~2024 limit
at $M_S = 2$~TeV.
Benchmark B1 is therefore currently unconstrained by
direct detection, and lies in the ``decoupling'' regime
of the singlet parameter space where the WIMP is heavy
enough to satisfy both relic density and direct-detection
constraints without fine-tuning.

\subsubsection{Higgs-funnel resonant singlet (B2)}
\label{subsubsec:b2}

When $M_S \approx M_h/2 \approx 62.5$~GeV, the
annihilation proceeds through an $s$-channel Higgs resonance.
The thermally-averaged cross section is enhanced relative to
the off-resonance value by a Breit--Wigner factor,
\begin{equation}
  \langle\sigma v\rangle_{\rm res}
  \;\simeq\;
  \langle\sigma v\rangle_{\rm off}\times
  \frac{R_{\rm max}\,(\Gamma_h M_h)^2}
       {(s_{\rm eff} - M_h^2)^2 + (\Gamma_h M_h)^2},
  \label{eq:sv_BW}
\end{equation}
where $s_{\rm eff} = 4M_S^2(1 + 3/2x_F)$ is the
center-of-mass energy squared evaluated at the freeze-out
temperature, $\Gamma_h = 4.07\times10^{-3}$~GeV is the
total Higgs width, and $R_{\rm max} \approx 10^4$
is the peak resonance enhancement after thermal averaging
\CITE{Griest:1991, Ibe:2008}.
Benchmark B2 ($M_S = 60.7$~GeV, $\lambda_{HS} = 0.026$)
is calibrated so that $\Omega h^2 = 0.12$ on the
near-resonance tail.

The BG measure in the resonant regime is dominated by the
sensitivity of $s_{\rm eff} - M_h^2$ to $M_S$:
a fractional shift $\epsilon$ in $M_S$ shifts
the denominator in Eq.~(\ref{eq:sv_BW}) by
$\delta(s_{\rm eff} - M_h^2) \approx 8M_S^2\,\epsilon$,
producing a fractional change in $\langle\sigma v\rangle$ of
order $8M_S^2/(\Gamma_h M_h)$.
More precisely, numerical evaluation at B2 gives
\begin{equation}
  \bgdelta_{M_S}^{(\rm B2)} \;\approx\; 6500,
  \qquad
  \bgdelta_{\lambda_{HS}}^{(\rm B2)} \;=\; 2,
  \qquad
  \bgdelta^{(\rm B2)} \;\approx\; 6500.
  \label{eq:delta_b2}
\end{equation}
The coupling sensitivity remains at 2 because
$\langle\sigma v\rangle \propto \lambda_{HS}^2$ regardless
of whether the resonance is resolved, but the mass
sensitivity is set by the inverse fractional pole width:
$\bgdelta_{M_S}^{(\rm B2)} \sim 8M_S^2/(\Gamma_h M_h)
\approx 4M_h/\Gamma_h \approx 1.2\times10^5$,
reduced to $\sim 6500$ by thermal averaging at finite
$x_F$.
This is the ``funnel tuning'' that appears generically in
models where the relic density is set by a resonant
annihilation channel, including the $A$-funnel of the MSSM
\CITE{Griest:1991} and the $Z$-funnel of Higgs-portal
dark matter.

The benchmark B2 is not yet excluded by LZ~2024
($\sigma_{\rm SI} \approx 9\times10^{-50}$~cm$^2$,
well below the sensitivity limit at 60~GeV),
but the combination of small coupling required by the relic
density and the extreme sensitivity of the abundance to
$M_S$ makes this benchmark qualitatively distinct from B1:
it is natural in $\lambda_{HS}$ but requires the singlet
mass to be fixed at the sub-percent level.

\subsubsection{Coannihilating MSSM bino (B3)}
\label{subsubsec:b3}

When the lightest supersymmetric particle (LSP) is a
bino-like neutralino $\chi$ with mass $M_\chi$, its
self-annihilation cross section is suppressed by the
hypercharge coupling and the typically large sfermion masses
appearing in $t$-channel exchange diagrams.
The relic density from bino self-annihilation alone
generically exceeds $\Omega h^2 = 0.12$ by orders of
magnitude.
Coannihilation with a nearly-degenerate next-to-lightest
supersymmetric particle (NLSP) resolves this by augmenting
the effective annihilation rate with the NLSP's own
(typically gauge-strength) cross section,
weighted by the Boltzmann-suppressed NLSP number density at
freeze-out \CITE{Griest:1991, Edsjo:1997}:
\begin{align}
  \langle\sigma v\rangle_{\rm eff}
  \;\approx\;&\;
  \langle\sigma v\rangle_{\chi\chi}
  \notag\\
  &+ 2\,g_{\rm eff}\,f(x_F)\,
    \sqrt{\langle\sigma v\rangle_{\chi\chi}\,
          \langle\sigma v\rangle_{\rm NLSP}}
  \notag\\
  &+ g_{\rm eff}^2\,f^2(x_F)\,
    \langle\sigma v\rangle_{\rm NLSP},
  \label{eq:sveff_coann}
\end{align}
where $g_{\rm eff} \approx 4$ counts the NLSP internal
degrees of freedom relative to the bino,
and $f(x_F) = (1 + \delta)^{3/2}\exp(-\delta\,x_F)$
with $\delta = (M_{\rm NLSP} - M_\chi)/M_\chi$
is the Boltzmann suppression factor.
The benchmark B3 ($M_\chi = 1$~TeV,
$M_{\rm NLSP} = 1.05$~TeV, i.e.\ $\delta = 0.05$)
is calibrated so that $\Omega h^2 \approx 0.12$
with $\langle\sigma v\rangle_{\rm NLSP}
= 0.66\,\sigma_0\,(1\,{\rm TeV}/M_\chi)^2$.

As discussed in Sec.~\ref{subsec:parameters}, the BG
derivatives must be taken with respect to the fundamental
soft masses $(M_\chi, M_{\rm NLSP})$ rather than the
derived splitting $\delta$.
The exponential factor $f(x_F) = (1+\delta)^{3/2}
e^{-\delta x_F}$ gives, at fixed $M_\chi$,
\begin{equation}
  \frac{\partial \ln \langle\sigma v\rangle_{\rm eff}}
       {\partial \ln M_{\rm NLSP}}
  \;\approx\;
  -\,\delta\,x_F\,\frac{M_{\rm NLSP}}{M_\chi}
  \;=\;
  -\,\delta\,x_F\,(1 + \delta),
  \label{eq:coann_deriv}
\end{equation}
and the BG measure on $\Omega$ (which is inversely
proportional to $\langle\sigma v\rangle_{\rm eff}$)
satisfies
\begin{equation}
  \bgdelta_{M_{\rm NLSP}}^{(\rm B3)}
  \;\approx\;
  x_F\,\frac{M_{\rm NLSP}}{M_\chi}
  \;\approx\;
  26\,\frac{M_{\rm NLSP}}{M_\chi}.
  \label{eq:delta_b3_MNLSP}
\end{equation}
At benchmark B3, $M_{\rm NLSP}/M_\chi = 1.05$, giving
$\bgdelta_{M_{\rm NLSP}} \approx 48$, while
$\bgdelta_{M_\chi} \approx 46$ from the overall
$\langle\sigma v\rangle_{\rm NLSP} \propto M_\chi^{-2}$
dependence.
The total measure is $\bgdelta^{(\rm B3)} = 48$.

\begin{figure*}[t]
  \includegraphics[width=\textwidth]{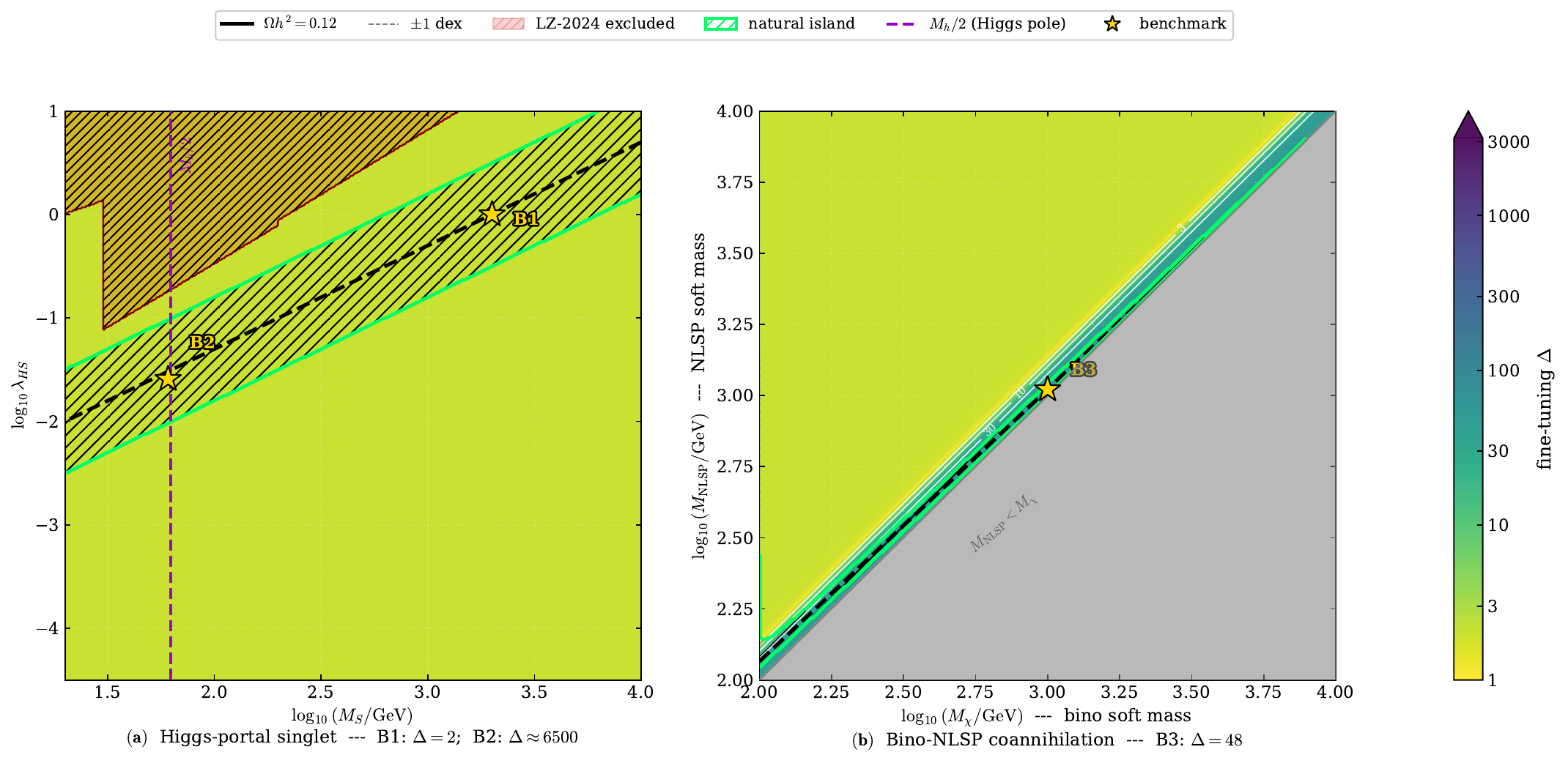}
  \caption{%
    Fine-tuning of thermal-relic WIMP dark matter.
    Color scale, contours, and natural island (green hatching)
    as in Fig.~\ref{fig:pbh}, with $\Omega h^2 = 0.12$ replacing
    $\fPBH = 1$ as the abundance target.
    The red-shaded region is excluded at 90\% CL by LZ~2024
    spin-independent direct detection \protect\CITE{LZ:2024}.
    \textbf{(a)}~Real singlet Higgs-portal WIMP in the
    $(M_S, \lambda_{HS})$ plane
    (Eq.~\protect\ref{eq:singlet_lagrangian}).
    The cyan dotted vertical line marks $M_S = M_h/2$
    (Higgs pole).
    Benchmark B1 (off-resonance, gold star):
    $M_S = 2$~TeV, $\lambda_{HS} = 1$; $\bgdelta = 2.0$.
    Benchmark B2 (Higgs funnel, gold star):
    $M_S = 60.7$~GeV, $\lambda_{HS} = 0.026$;
    $\bgdelta \approx 6500$ (color saturated to dark purple).
    The contrast between B1 (uniformly yellow stripe) and
    B2 (deep purple near the resonance) illustrates the
    Class~I vs.\ Class~III distinction.
    \textbf{(b)}~MSSM bino-NLSP coannihilation in the fundamental
    soft-mass plane $(M_\chi, M_{\rm NLSP})$, using
    Eqs.~(\protect\ref{eq:sveff_coann})--(\protect\ref{eq:delta_b3_MNLSP}).
    The gray-shaded triangle is the unphysical region
    $M_{\rm NLSP} < M_\chi$.
    The natural island (narrow teal strip) hugs the diagonal
    at constant $\bgdelta \approx 48 = x_F M_{\rm NLSP}/M_\chi$.
    Benchmark B3: $M_\chi = 1$~TeV, $M_{\rm NLSP} = 1.05$~TeV;
    $\bgdelta = 48$.
    \label{fig:wimp}
  }
\end{figure*}

This result places the coannihilating WIMP solidly in
the same tier as FOPT-PBH dark matter ($\bgdelta = 35$),
and for the same structural reason: both constructions
contain a single exponential whose argument is forced
by the requirement $\Omega = \Omega_{\rm DM}$ to equal
$\ln(T_{\rm form}/T_{\rm eq})$ or $x_F\,\delta$,
respectively (see Sec.~\ref{subsec:identity}).
The BG landscape of Fig.~\ref{fig:wimp}(b), computed
in the $(M_\chi, M_{\rm NLSP})$ plane, shows the
natural island as a narrow teal-colored strip
hugging the diagonal $M_{\rm NLSP} = M_\chi$ at a
fixed $\Delta \approx 48$ set by $x_F M_{\rm NLSP}/M_\chi$
across the full mass range $100$~GeV to $10$~TeV.

\subsection{Freeze-in (FIMP)}
\label{subsec:fimp}

\subsubsection{Production mechanism and abundance map}
\label{subsubsec:fimp_mechanism}

Feebly interacting massive particles (FIMPs)
\CITE{Hall:2010, McDonald:2002}
never reach thermal equilibrium with the Standard Model
bath; instead, they are produced gradually in the
forward direction through infrequent collisions or
decays.
In the IR-dominated freeze-in scenario relevant here,
production is dominated by decays of a heavy Standard Model
or dark-sector mediator $B$ with mass $M_B$ and
decay coupling $y$ to two DM particles,
$B \to {\rm DM} + X$.
Following Hall et al.~\CITE{Hall:2010}, the yield
produced by this mechanism from $T \gg M_B$ down to
$T = 0$ is
\begin{equation}
  Y_{\rm FI}
  \;=\;
  \frac{45\,\mathcal{X}_{\rm FI}}{\pi^4\,g_*^{3/2}}
  \cdot
  \frac{y^2\,\Mpl}{M_B}
  \;\equiv\;
  K_{\rm FI}\,\frac{y^2\,\Mpl}{M_B},
  \label{eq:YFI}
\end{equation}
where $K_{\rm FI} \approx 1.7\times10^{-3}$ for
$g_* = 100$ and $\mathcal{X}_{\rm FI}$ is a
dimensionless integral of order unity that depends mildly
on the spin of $B$.
The present-day dark matter abundance is then
\begin{equation}
  \Omega_{\rm DM} h^2
  \;=\;
  M_{\rm DM}\,Y_{\rm FI}\,
  \frac{s_0}{\rho_c/h^2}
  \;\propto\;
  M_{\rm DM}\,y^2\,\frac{\Mpl}{M_B},
  \label{eq:Omega_FI}
\end{equation}
where $s_0/(\rho_c/h^2) \approx 2.7\times10^8$~GeV$^{-1}$
converts yield to $\Omega h^2$
\CITE{Kolb:1990}.
I fix the mediator mass $M_B = 100$~GeV (electroweak scale)
as a representative choice; the scaling
$\Omega h^2 \propto M_B^{-1}$ means that changing
$M_B$ by a factor of 10 shifts the natural value of $y$
by a factor of $\sqrt{10}$, but does not change the BG
measure.

Benchmark B4 is defined by $M_{\rm DM} = 1$~GeV and
$y = 1.46\times10^{-12}$, giving $\Omega h^2 = 0.120$.

\subsubsection{Fine-tuning structure}
\label{subsubsec:fimp_ft}

Since Eq.~(\ref{eq:Omega_FI}) is a pure monomial in
$(M_{\rm DM}, y)$ with fixed $M_B$,
\begin{equation}
  \bgdelta_{M_{\rm DM}}^{(\rm B4)} \;=\; 1,
  \qquad
  \bgdelta_y^{(\rm B4)} \;=\; 2,
  \qquad
  \bgdelta^{(\rm B4)} \;=\; 2
  \label{eq:delta_b4}
\end{equation}
exactly, everywhere in the $(M_{\rm DM}, y)$ plane.
The mass sensitivity is unity because $\Omega \propto M_{\rm DM}$
linearly; the coupling sensitivity is two because
$\Omega \propto y^2$.
These are constants, independent of the benchmark point,
for the same structural reason as benchmark B1 and the
biased-domain-wall PBH: the abundance map is a power-law
monomial with no exponential, resonance, or cancellation.

The FIMP scenario is thus as natural as the off-resonance WIMP
by the BG criterion, despite the coupling being smaller by
nine orders of magnitude.
The feebleness of the coupling is not fine-tuning in the BG
sense; it is, rather, the \emph{prediction} of the freeze-in
mechanism, which requires $y \sim 10^{-12}$ to avoid
thermalizing the DM sector.
What the BG measure quantifies is not the smallness of $y$
itself, but the sensitivity of $\Omega h^2$ to
\emph{fractional variations} of $y$; a factor-of-two
change in $y$ changes $\Omega h^2$ by a factor of four,
which is $\bgdelta_y = 2$, perfectly natural.

\subsubsection{Parameter-space heatmap}
\label{subsubsec:fimp_map}

\begin{figure*}[t]
  \includegraphics[width=\textwidth]{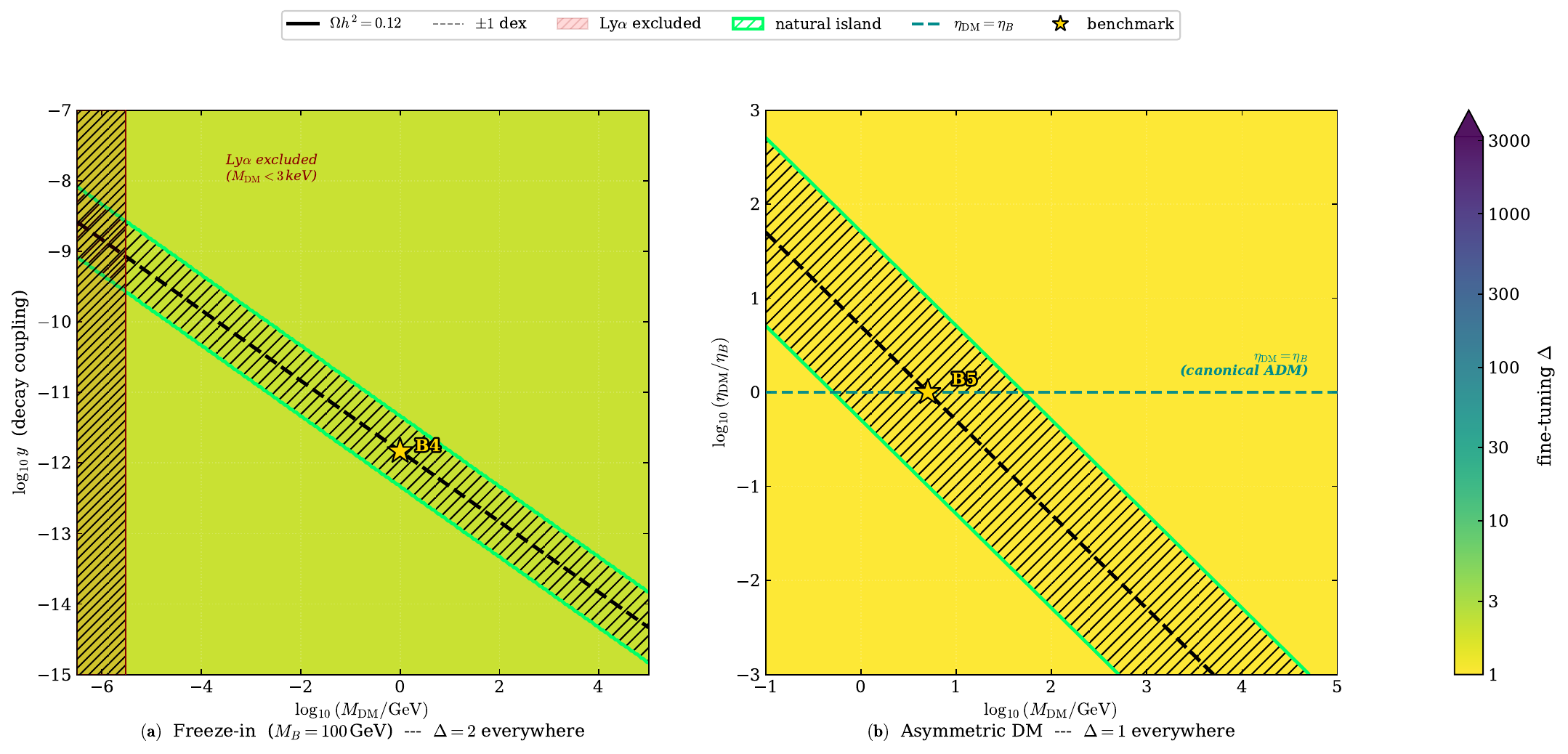}
  \caption{%
    Fine-tuning of non-thermal dark matter production.
    Color scale and conventions as in Fig.~\ref{fig:pbh}.
    Both panels are uniformly pale yellow, reflecting
    power-law abundance maps with $\bgdelta \leq 2$ everywhere.
    \textbf{(a)}~Freeze-in via heavy-mediator decay
    $B \to {\rm DM} + X$ with fixed mediator mass $M_B = 100$~GeV,
    in the $(M_{\rm DM}, y)$ plane
    (Eqs.~\protect\ref{eq:YFI}--\protect\ref{eq:Omega_FI}).
    The red-shaded region at $M_{\rm DM} \lesssim 3$~keV
    is excluded by Lyman-$\alpha$ forest observations
    \protect\CITE{Ballesteros:2020, Decant:2022}.
    The natural island is the diagonal stripe
    $y \propto M_{\rm DM}^{-1/2}$ (from $\Omega \propto M_{\rm DM} y^2$).
    $\bgdelta = 2$ throughout the entire accessible parameter space
    (Eq.~\protect\ref{eq:delta_b4}).
    Benchmark B4: $M_{\rm DM} = 1$~GeV,
    $y = 1.46\times10^{-12}$; $\bgdelta = 2$.
    \textbf{(b)}~Asymmetric dark matter in the
    $(M_{\rm DM}, \eta_{\rm DM}/\eta_B)$ plane
    (Eq.~\protect\ref{eq:ADM}).
    The cyan dashed horizontal line marks the ``canonical ADM''
    locus $\eta_{\rm DM} = \eta_B$.
    The natural island is the hyperbola
    $M_{\rm DM}\,(\eta_{\rm DM}/\eta_B) = 4.7$~GeV.
    $\bgdelta = 1$ throughout (Eq.~\protect\ref{eq:delta_b5}),
    the minimum possible BG measure for a non-trivial
    abundance map.
    Benchmark B5: $M_{\rm DM} = 5$~GeV, $\eta_{\rm DM}/\eta_B = 1$;
    $\bgdelta = 1$.
    \label{fig:fimp_adm}
  }
\end{figure*}

Figure~\ref{fig:fimp_adm}(a) shows $\log_{10}\bgdelta$
in the $(M_{\rm DM}, y)$ plane for $M_B = 100$~GeV.
The panel is uniformly pale yellow at the minimum of the
color scale ($\bgdelta = 2$) throughout the physically
accessible region, with the natural island
(green hatching along the $\Omega h^2 = 0.12$ contour)
forming a clean diagonal stripe of slope $-1/2$ in the
log--log plane, as expected from
$y \propto M_{\rm DM}^{-1/2}$ at fixed $\Omega$.
The red-shaded region at $M_{\rm DM} \lesssim 3$~keV
marks the Lyman-$\alpha$ forest exclusion for freeze-in
production from a 100-GeV mediator
\CITE{Ballesteros:2020, Decant:2022}.
Benchmark B4 (gold star) lies at the center of the
natural island in the phenomenologically well-motivated
GeV mass range.
The uniformity of the heatmap is perhaps the most
striking visual feature of the FIMP panel: there are
no gradients, no tuned corners, and no preferred points
other than those set by external constraints.

\subsection{Asymmetric dark matter}
\label{subsec:adm}

\subsubsection{Abundance map and fine-tuning}
\label{subsubsec:adm_ft}

Asymmetric dark matter (ADM) \CITE{Nussinov:1985, Kaplan:2009}
posits that the present-day dark matter abundance is set not
by thermal freeze-out but by a primordial asymmetry
between DM particles and antiparticles, generated by the
same baryogenesis mechanism that produced the baryon
asymmetry.
If the DM asymmetry per comoving entropy is
$\eta_{\rm DM} = (n_{\rm DM} - n_{\bar{\rm DM}})/s$,
and all $\bar{\rm DM}$ is annihilated away efficiently
(as required for the mechanism to work), then
\begin{equation}
  \frac{\Omega_{\rm DM}}{\Omega_b}
  \;=\;
  \frac{M_{\rm DM}}{m_p}\,R,
  \qquad
  R \;\equiv\; \frac{\eta_{\rm DM}}{\eta_B},
  \label{eq:ADM}
\end{equation}
where $m_p = 0.938$~GeV is the proton mass and
$\eta_B \approx 8.7\times10^{-11}$ is the baryon
asymmetry \CITE{Planck:2018}.
With $\Omega_b h^2 = 0.0224$ and
$\Omega_{\rm DM} h^2 = 0.120$, the observed ratio
$\Omega_{\rm DM}/\Omega_b = 5.4$ requires
\begin{equation}
  M_{\rm DM}\,R \;\simeq\; 5.0\,m_p \;\approx\; 4.7\,{\rm GeV}.
  \label{eq:ADM_constraint}
\end{equation}
The natural island in the $(M_{\rm DM}, R)$ plane is the
hyperbola $M_{\rm DM} \times R = 4.7$~GeV
(or within a factor of two thereof),
ranging from the proton mass at $R = 5$
to multi-TeV masses at $R \sim 10^{-3}$.
Benchmark B5 is defined by $M_{\rm DM} = 5.0$~GeV, $R = 1$
(equal DM and baryon asymmetries), a canonical choice
motivated by models in which baryogenesis and DM
asymmetry generation proceed via the same operator
\CITE{Kaplan:2009}.

Since Eq.~(\ref{eq:ADM}) is linear in both $M_{\rm DM}$
and $R$,
\begin{equation}
  \bgdelta_{M_{\rm DM}}^{(\rm B5)} \;=\; 1,
  \qquad
  \bgdelta_R^{(\rm B5)} \;=\; 1,
  \qquad
  \bgdelta^{(\rm B5)} \;=\; 1.
  \label{eq:delta_b5}
\end{equation}
This is the minimum possible BG measure for a non-trivial
abundance map: $\bgdelta = 1$ means the abundance is simply
proportional to the parameter, and a factor-of-two change
in either $M_{\rm DM}$ or $R$ produces a factor-of-two
change in $\Omega_{\rm DM} h^2$.
Asymmetric dark matter is the most natural dark matter
scenario in the full comparison precisely because its
abundance is set by a ratio of number densities ($R$)
that is itself a direct observable, and by the particle
mass $M_{\rm DM}$, both entering linearly.

Figure~\ref{fig:fimp_adm}(b) shows the ADM parameter space
in the $(M_{\rm DM}, R)$ plane.
The panel is uniformly yellow at $\bgdelta = 1$ throughout,
with the natural island forming the $M_{\rm DM}\times R = 4.7$~GeV
diagonal.
The cyan dashed line at $R = 1$ marks the ``canonical ADM''
locus where DM and baryon asymmetries are equal; benchmark
B5 sits at its intersection with the natural island.

\subsection{Axions}
\label{subsec:axions}

The QCD axion \CITE{Peccei:1977, Weinberg:1978, Wilczek:1978}
acquires a mass $m_a \simeq 5.7\,\mu{\rm eV}
(10^{12}\,{\rm GeV}/f_a)$ from non-perturbative QCD effects,
where $f_a$ is the Peccei--Quinn (PQ) symmetry-breaking
scale.
Its cosmological abundance depends critically on whether
PQ symmetry breaks before or after inflation, leading to
two qualitatively distinct scenarios with different numbers
of free parameters and different BG structures.

\subsubsection{Misalignment (pre-inflationary PQ breaking)}
\label{subsubsec:axion_mis}

If PQ symmetry breaks before the end of inflation,
the axion field $\theta = a/f_a$ is homogenized over
superhorizon scales and takes a single value
$\theta_i \in [-\pi, \pi]$ throughout our observable Universe
(the ``misalignment angle'').
When the Hubble rate falls below the axion mass at
temperature $T_{\rm osc} \sim \Lambda_{\rm QCD}$,
the field begins to oscillate coherently around $\theta = 0$,
and the energy density of these oscillations constitutes
the dark matter relic.
Including the anharmonic correction $F(\theta_i)$
that becomes important as $\theta_i \to \pi$
\CITE{Visinelli:2009},
\begin{equation}
  \Omega_a h^2
  \;\simeq\;
  0.18\,\theta_i^2\,
  \left(\frac{f_a}{10^{12}\,{\rm GeV}}\right)^{\!1.19}
  F(\theta_i),
  \label{eq:Omega_mis}
\end{equation}
where the exponent $1.19$ arises from the temperature
dependence of the axion mass near the QCD transition
\CITE{Borsanyi:2016} and $F(\theta_i)$ is a
smooth, monotonically increasing function satisfying
$F(\theta_i) \to 1$ as $\theta_i \to 0$ and
$F(\theta_i) \to \infty$ as $\theta_i \to \pi$
\CITE{Visinelli:2009}.
Benchmark B6 is defined by $f_a = 10^{12}$~GeV and
$\theta_i = 0.82$, which gives $\Omega h^2 = 0.131$ with
$F(\theta_i) \approx 1$ (anharmonic corrections are
$\lesssim 10\%$ at this angle); this is within 9\% of
the observed value, consistent with theoretical uncertainties
in the relic-abundance calculation.

The BG sensitivities follow from log-differentiation of
Eq.~(\ref{eq:Omega_mis}):
\begin{align}
  \bgdelta_{f_a}^{(\rm B6)}
  &\;=\; 1.19,
  \label{eq:delta_b6_fa}
  \\[4pt]
  \bgdelta_{\theta_i}^{(\rm B6)}
  &\;=\;
  2 + \theta_i\,\frac{\partial \ln F}{\partial \theta_i}
  \;\approx\; 2.2
  \quad (\theta_i = 0.82),
  \label{eq:delta_b6_theta}
\end{align}
giving $\bgdelta^{(\rm B6)} = \max(1.19, 2.2) = 2.2$.

This places the misalignment axion with $\theta_i \sim 1$
comfortably in the natural tier (Class~I of
Sec.~\ref{subsec:classes}).
However, there is an important subtlety that the BG measure
alone does not capture.
If $f_a$ is large (e.g., $f_a = 10^{16}$~GeV as in
models with a GUT-scale or string-theoretic PQ scale),
then $\Omega h^2 = 0.12$ requires a small misalignment
angle $\theta_i \simeq (0.18\,(f_a/10^{12})^{1.19})^{-1/2}
\approx 3\times10^{-3}$, well below the natural value
of order unity.
The local BG sensitivity at this point remains
$\bgdelta \approx 2$, but since $\theta_i$ is a
cosmological initial condition with a
flat prior on $[-\pi, \pi]$, the prior probability of
landing at $\theta_i \lesssim 3\times10^{-3}$ is
$\sim 10^{-3}$, corresponding to a ``prior tuning''
of $\sim 300$ that the BG measure does not register.
This distinction between local sensitivity and prior
probability is discussed in detail in
Sec.~\ref{subsec:prior}.

Figure~\ref{fig:axions}(a) displays the $(f_a, \theta_i)$
parameter space.
The BG heatmap transitions from yellow-green at moderate $f_a$
(where $\bgdelta \approx 2.2$, dominated by $\bgdelta_{\theta_i}$)
to yellow at small $\theta_i$ (where anharmonic contributions
vanish and $\bgdelta \to \max(1.19, 2) = 2$).
Near $\theta_i \to \pi$ (magenta dash-dotted line),
$\bgdelta_{\theta_i}$ diverges with the anharmonic
enhancement, visible as a color gradient toward the top
of the panel.
The natural island (green hatching) runs diagonally across
the panel following the $\Omega h^2 = 0.12$ contour,
from large $f_a$ and small $\theta_i$ to small $f_a$ and
$\theta_i \sim 1$.
The benchmark B6 (gold star, $f_a = 10^{12}$~GeV,
$\theta_i = 0.82$) and the anthropic benchmark B6$_{\rm anthr}$
(orange square, $f_a = 10^{16}$~GeV,
$\theta_i \approx 3\times10^{-3}$)
both lie on the $\Omega h^2 = 0.12$ contour,
both with $\bgdelta \approx 2$, illustrating that
the BG measure does not distinguish between them.

\subsubsection{Post-inflationary PQ (string decay)}
\label{subsubsec:axion_pi}

If PQ symmetry breaks \emph{after} inflation, different
Hubble patches independently choose their phase when the
$U(1)_{\rm PQ}$ symmetry breaks, and the axion field has
no single misalignment angle.
Instead, a network of cosmic strings forms at the PQ
transition and radiates axions as it evolves; the axion
relic density is dominated by the string-network emission
rather than by misalignment \CITE{Davis:1986, Battye:1994}.
In this scenario the yield depends only on $f_a$ through
the string tension $\mu \propto f_a^2$ and the mass
$m_a \propto f_a^{-1}$, with no free initial condition
$\theta_i$.
Large-scale numerical simulations \CITE{Buschmann:2022}
find
\begin{equation}
  \Omega_a h^2
  \;\simeq\;
  \mathcal{C}\,
  \left(\frac{f_a}{5\times10^{10}\,{\rm GeV}}\right)^{\!1.19},
  \label{eq:Omega_PI}
\end{equation}
where $\mathcal{C} \approx 0.12$ is calibrated so that
$f_a = 5\times10^{10}$~GeV gives the observed abundance,
and the exponent $1.19$ is the same as in the
misalignment case.
The simulation of Ref.~\CITE{Buschmann:2022} gives a
preferred range $f_a \simeq (2\text{--}8)\times10^{10}$~GeV
at 95\% confidence, incorporating uncertainties from the
string network logarithm and from the QCD mass modeling.

The BG measure for the post-inflationary scenario
has only a single parameter:
\begin{equation}
  \bgdelta_{f_a}^{(\rm B7)} \;=\; 1.19
  \quad \forall\; f_a.
  \label{eq:delta_b7}
\end{equation}
This is the lowest BG fine-tuning of any construction
in the comparison after pure asymmetric dark matter
($\bgdelta = 1$), and it applies globally
--- not just at a benchmark, but over the entire
parameter space.
The post-inflationary axion scenario is also the most
\emph{predictive} dark matter construction in the comparison:
it has a single parameter $f_a$ that is constrained from
above by the requirement $\Omega h^2 \leq 0.12$, from below
by astrophysical bounds on the axion--photon coupling
\CITE{Ayala:2014}, and is now being
targeted directly by cavity haloscope experiments
\CITE{Du:2018, HAYSTAC:2022} and by the CMB Stage-4
spectral distortion measurements expected to probe
$f_a \lesssim 3\times10^{11}$~GeV \CITE{CMB-S4:2022}.
Benchmark B7 is defined by $f_a = 5\times10^{10}$~GeV.

\begin{figure*}[t]
  \includegraphics[width=\textwidth]{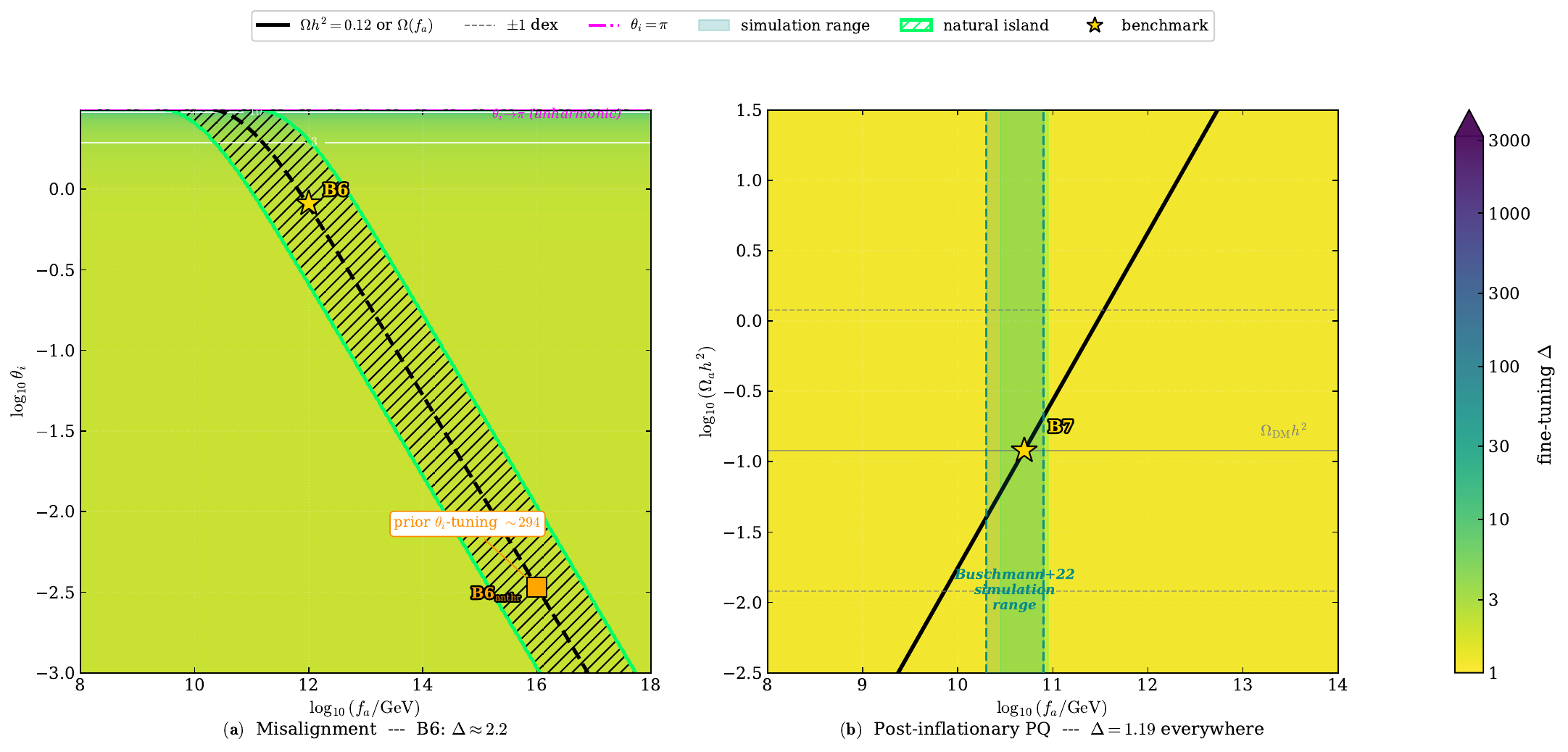}
  \caption{%
    Fine-tuning of QCD axion dark matter for two cosmological
    histories of PQ symmetry breaking.
    Color scale and conventions as in Fig.~\ref{fig:pbh}.
    \textbf{(a)}~Misalignment mechanism (pre-inflationary PQ breaking)
    in the $(f_a, \theta_i)$ plane
    (Eq.~\protect\ref{eq:Omega_mis}).
    The magenta dash-dotted horizontal line at $\theta_i = \pi$
    marks the onset of the anharmonic divergence $F(\theta_i) \to \infty$.
    The color gradient near this line reflects the diverging
    $\bgdelta_{\theta_i}$.
    The natural island (green hatching) runs diagonally from
    small $f_a$, $\theta_i \sim 1$ to large $f_a$, small $\theta_i$.
    Gold star B6: $f_a = 10^{12}$~GeV, $\theta_i = 0.82$; $\bgdelta = 2.2$.
    Orange square B6$_{\rm anthr}$: $f_a = 10^{16}$~GeV,
    $\theta_i \approx 3\times10^{-3}$; local $\bgdelta = 2.0$,
    but prior tuning $\approx 300$ (see Sec.~\protect\ref{subsec:prior}).
    \textbf{(b)}~Post-inflationary PQ (string-network scenario),
    showing $\Omega_a h^2$ vs.\ $f_a$ (black curve)
    against the uniform $\bgdelta = 1.19$ background
    (Eq.~\protect\ref{eq:Omega_PI}).
    The entire panel is uniformly pale yellow: $\bgdelta_{f_a} = 1.19$
    everywhere, the lowest fine-tuning of any construction in
    the comparison other than pure asymmetric dark matter.
    The cyan shaded band marks the preferred $f_a$ range
    $(2$--$8)\times10^{10}$~GeV from the large-scale string-network
    simulations of Ref.~\protect\CITE{Buschmann:2022}.
    The green shaded band marks the natural island.
    Gray horizontal lines show $\Omega_{\rm DM} h^2 = 0.12$
    (solid) and $\pm 1$~dex (dashed).
    Gold star B7: $f_a = 5\times10^{10}$~GeV; $\bgdelta = 1.19$.
    \label{fig:axions}
  }
\end{figure*}

Figure~\ref{fig:axions}(b) displays $\Omega_a h^2$
as a function of $f_a$ (black curve) against the
constant $\bgdelta = 1.19$ background.
The entire panel is a uniform pale yellow, the
minimum of the shared color scale, confirming that
the post-inflationary axion scenario sits at the
floor of BG fine-tuning across all scales.
The cyan band marks the preferred $f_a$ range from the
Buschmann et al.~\CITE{Buschmann:2022} simulations,
which corresponds to the portion of the natural island
(green shading) constrained by current simulations.
The benchmark B7 (gold star) lies within both the
green and cyan bands simultaneously.

\section{\label{sec:comparison}Comparison and Structural Insights}
 
\subsection{The fine-tuning hierarchy}
\label{subsec:hierarchy}
 
Table~\ref{tab:hierarchy} ranks all twelve scenarios by
$\Delta_{\rm BG}$, ordered from most natural to most
tuned, and assigns each to one of the three universality
classes defined in Sec.~\ref{subsec:classes}.
The data reveal a clear three-tier structure whose
boundaries are set by the analytic form of the abundance
map, not by the nature of the dark matter candidate.
\begin{table}[htb]
\centering
\caption{Barbieri--Giudice fine-tuning measure at benchmarks.
         Scenarios are listed in order of increasing $\Delta$.
         The ``type'' column indicates the production paradigm.
         The inflationary-collapse entry is a lower-end estimate
         (Sec.~\ref{subsec:inflation_ft}); the corresponding bar
         in Fig.~\ref{fig:hierarchy} is shown truncated.
         For the first-order phase transition mechanism the
         transition strength parameter $\alpha$ has
         $\Delta_\alpha \approx 3$--$4$ (Class~I) and does not
         drive the dominant tuning; only the nucleation rate
         $\beta/H_*$ is reported here.
         All inflationary model classes
         (Fig.~\ref{fig:inf_comparison}) fall in Class~III or
         beyond for standard high-reheating scenarios
         ($T_{\rm reh} \gg T_{\rm form}$, RD Layer~1 floor
         $\Delta_\sigma^{\rm (RD)} \approx 57$--$70$);
         curvaton and spectator-field models move to Class~II
         for low-reheating scenarios
         ($T_{\rm reh} \lesssim T_{\rm form}$, MD Layer~1
         floor $\Delta_\sigma^{\rm (MD)} \approx 14$).
         The two first-order phase transition benchmarks
         ($\alpha = 0.3$ and $\alpha = 1.0$) are listed
         separately in Table~\ref{tab:benchmarks} to
         demonstrate $\alpha$-independence, but are counted
         as a single mechanism here.
         \label{tab:hierarchy}}
\begin{ruledtabular}
\begin{tabular}{lccc}
Scenario (benchmark)              & Type  & $\Delta$           & Tier         \\
\hline
Asymmetric DM (B5)                & ADM   & 1.0                & I            \\
Axion: post-inflationary (B7)     & Axion & 1.19               & I            \\
PBH: DW, gravity-fixed bias       & PBH   & 2.0                & I            \\
WIMP: off-resonance (B1)          & WIMP  & 2.0                & I            \\
FIMP: freeze-in (B4)              & FIMP  & 2.0                & I            \\
Axion: misalignment (B6)          & Axion & 2.2                & I            \\
PBH: DW, free $V_b$               & PBH   & 4.5                & I            \\
\hline
PBH: early matter dom.\           & PBH   & 14                 & II           \\
PBH: phase transition             & PBH   & $35^{\,a}$         & II/III$^{\,a}$ \\
WIMP: coannihilation (B3)         & WIMP  & 48                 & II           \\
\hline
WIMP: Higgs funnel (B2)           & WIMP  & 6500               & III          \\
PBH: inflation, single-field      & PBH   & $\gtrsim 10^3$     & III$^{+,b}$  \\
\end{tabular}
\end{ruledtabular}
\begin{minipage}{\columnwidth}
\footnotesize
\vspace{4pt}
$^a$ $\Delta \approx 35$ uses the single-exponential
approximation Eq.~(\ref{eq:beta_fopt}); the more accurate
super-exponential collapse probability of
Ref.~\cite{Gouttenoire:2024} gives $\Delta \approx 265$
(Class~III) at the benchmark $\beta/H_* \approx 8$, and
tracing the abundance to the underlying scalar potential
gives $\Delta \sim 10^3$--$10^4$
(Sec.~\ref{subsec:caveats}).\\[2pt]
$^b$ The quoted value is a lower-end estimate for
single-field ultra-slow-roll and inflection-point models.
Curvaton and spectator-field inflationary models are
substantially more natural: $\Delta \approx 21$--$42$
(Class~II) for low-reheating scenarios
($T_{\rm reh} \lesssim T_{\rm form}$, MD Layer~1 floor),
rising to $\Delta \approx 115$--$210$ (Class~III) for
standard high-reheating scenarios
($T_{\rm reh} \gg T_{\rm form}$, RD Layer~1 floor);
see Sec.~\ref{subsec:inflation_ft} and
Fig.~\ref{fig:inf_comparison}.
\end{minipage}
\end{table}
 
\subsection{Three universality classes}
\label{subsec:classes}
 
The twelve scenarios fall into three qualitatively distinct
tiers distinguished by the analytic form of their abundance
maps; the precise definitions and membership are given below,
and Fig.~\ref{fig:hierarchy} summarizes the hierarchy visually.

\begin{figure*}[t]
  \includegraphics[width=\textwidth]{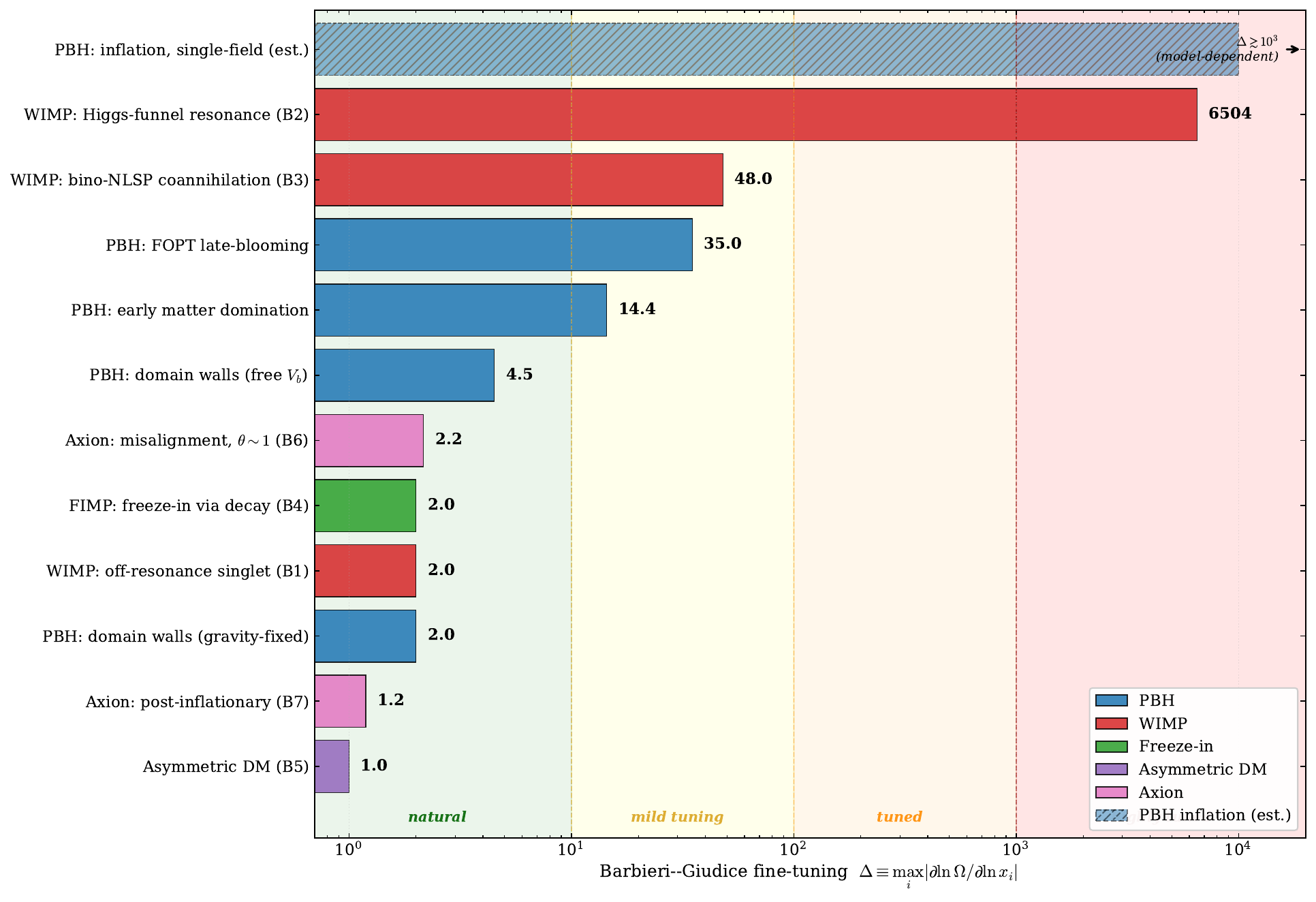}
  \caption{%
    Summary of the Barbieri--Giudice fine-tuning measure $\bgdelta$
    at the benchmark points listed in Table~\ref{tab:hierarchy},
    ordered from most natural (bottom) to most tuned (top).
    Bar colors indicate the production paradigm:
    \textit{blue} = PBH,
    \textit{red} = thermal-relic WIMP,
    \textit{green} = freeze-in (FIMP),
    \textit{purple} = asymmetric dark matter,
    \textit{pink} = axion.
    Background shading indicates qualitative tuning tiers:
    \textit{green} ($\bgdelta < 10$, natural),
    \textit{yellow} ($10 \leq \bgdelta < 100$, mild tuning),
    \textit{orange} ($100 \leq \bgdelta < 1000$, tuned),
    \textit{red} ($\bgdelta \geq 1000$, highly tuned).
    The topmost bar (PBH: inflation, single-field) is shown
    hatched and truncated with an arrow: the value is a
    lower-end estimate $\bgdelta \gtrsim 10^3$ derived from
    Eq.~(\ref{eq:delta_inf_total}), with the precise value
    model-dependent and potentially orders of magnitude larger
    (see Secs.~\ref{subsec:inflation_ft} and~\ref{subsec:classes}).
    The FOPT-PBH bar shows $\bgdelta = 35$ from the single-exponential
    approximation Eq.~(\ref{eq:beta_fopt}); the more accurate
    super-exponential collapse probability~\cite{Gouttenoire:2024}
    gives $\bgdelta \approx 265$ at the benchmark (Class~III),
    and the microphysical estimate gives $\bgdelta \sim 10^3$--$10^4$
    (Sec.~\ref{subsec:caveats}, Eq.~\ref{eq:delta_fopt_GV_bench}).
    All inflationary model classes fall in Class~III or beyond
    once the correct radiation-dominated Layer~1 floor is applied
    (Sec.~\ref{subsec:inflation_ft} and Sec.~\ref{subsec:caveats}).
    \label{fig:hierarchy}
  }
\end{figure*}
  
Before discussing each tier, I establish that the
classification is robust to the choice of fine-tuning
measure.
Table~\ref{tab:benchmarks} reports all three measures, $\bgdelta_{\rm BG}$, $\bgdelta_{\rm SR}$, and
$\epsilon$, for every benchmark.
The key observations are:
(i)~$\bgdelta_{\rm SR}$ and $\bgdelta_{\rm BG}$ agree
to within 40\% for all scenarios except B3, where both
soft masses are nearly equally sensitive
($\bgdelta_{M_\chi} = 46$, $\bgdelta_{M_{\rm NLSP}} = 48$),
giving $\bgdelta_{\rm SR} = \sqrt{46^2+48^2} \approx 67$,
a factor of $\sqrt{2}$ above $\bgdelta_{\rm BG} = 48$.
For all single-parameter-dominated scenarios (B2, the
two FOPT benchmarks, the MD case), SR and BG are
identical to three significant figures.
(ii)~The island half-width $\epsilon$ is monotonically
related to $\bgdelta_{\rm BG}$ through $\epsilon
\approx 0.69/\bgdelta_{\rm BG}$ (Eq.~\ref{eq:eps}),
with values ranging from $\epsilon = 75\%$ for ADM
to $\epsilon < 0.1\%$ for the Higgs-funnel WIMP.
(iii)~No scenario changes tier under either alternative
measure.
The three-tier classification is therefore a structural
property of the abundance maps, independent of whether
one uses the max (BG), quadrature (SR), or geometric
($\epsilon$) convention.
 
\begin{figure*}[t]
  \includegraphics[width=\textwidth]{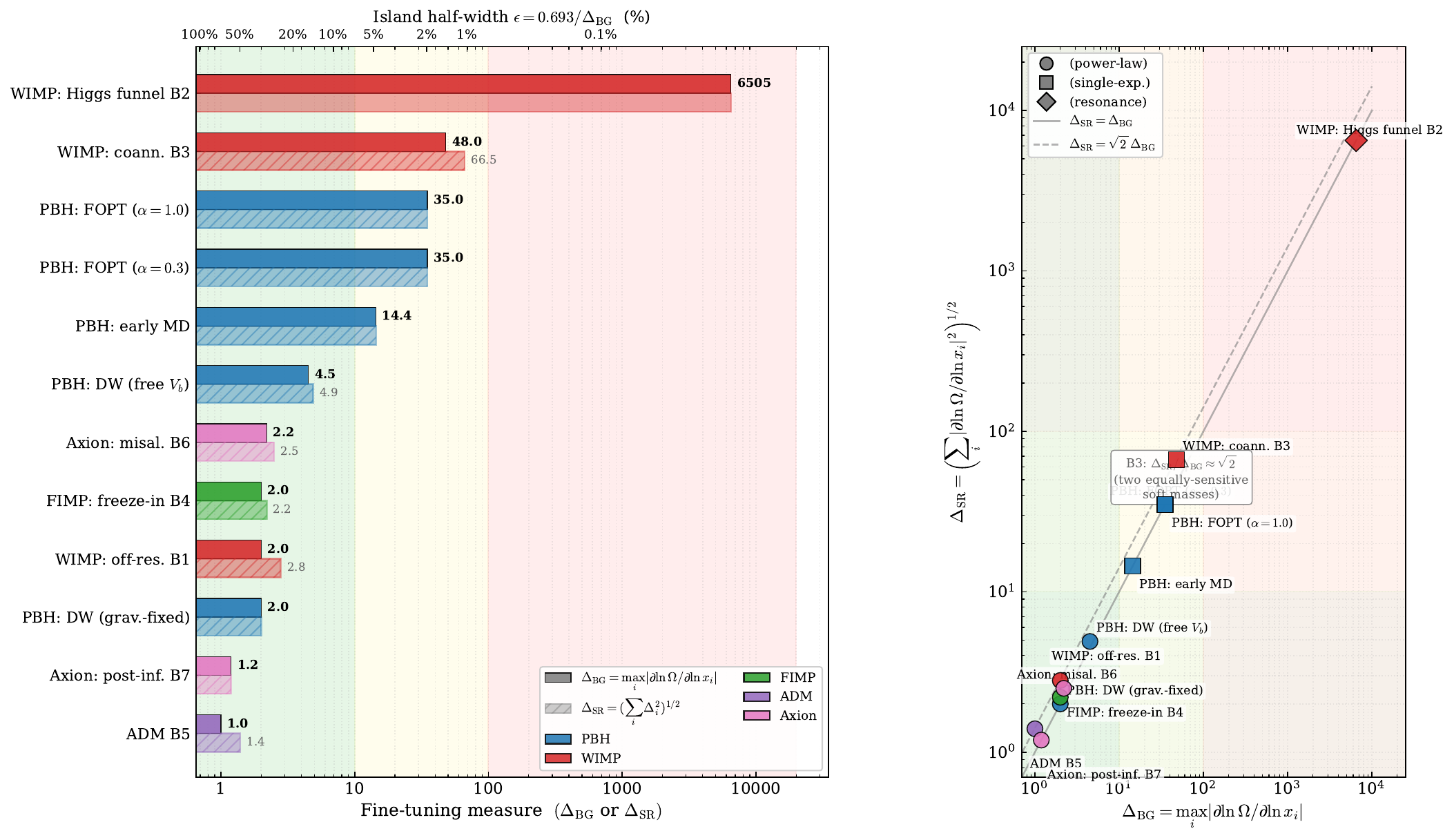}
  \caption{%
    Comparison of the Barbieri--Giudice ($\bgdelta_{\rm BG}$),
    Strumia--Rattazzi ($\bgdelta_{\rm SR}$), and island
    half-width ($\epsilon$) fine-tuning measures across
    all benchmarks of Table~\ref{tab:benchmarks}.
    \textit{Left:} Paired horizontal bars for
    $\bgdelta_{\rm BG}$ (solid) and $\bgdelta_{\rm SR}$
    (hatched, lighter); the top axis reads off
    $\epsilon = 0.693/\bgdelta_{\rm BG}$ directly as a percentage.
    Bar colors follow the paradigm convention of Fig.~\ref{fig:hierarchy}.
    Background shading bands mark the three tiers.
    \textit{Right:} $\bgdelta_{\rm SR}$ vs.\ $\bgdelta_{\rm BG}$
    on log--log axes.
    Marker shapes distinguish Class~I (circles),
    Class~II (squares), and Class~III (diamond).
    The solid gray diagonal is $\bgdelta_{\rm SR} = \bgdelta_{\rm BG}$;
    the dashed line shows $\bgdelta_{\rm SR} =
    \sqrt{2}\,\bgdelta_{\rm BG}$, the expectation for
    two equally-sensitive parameters.
    All scenarios lie on the solid diagonal except B3
    (WIMP coannihilation, red square), which sits on the
    dashed line because both soft masses contribute equally.
    No scenario changes tier under either measure.
    \label{fig:ft_comparison}
  }
\end{figure*}

Figure~\ref{fig:ft_comparison} displays this comparison
visually.
The left panel shows paired horizontal bars for
$\bgdelta_{\rm BG}$ (solid) and $\bgdelta_{\rm SR}$
(hatched, same color) for all twelve benchmarks on a
logarithmic scale, with the island half-width $\epsilon$
read off the top axis.
The sole exception is benchmark~B3 (WIMP coannihilation),
whose SR bar (66.5, hatched) visibly extends beyond
its BG bar (48.0, solid): the two soft masses $M_\chi$
and $M_{\rm NLSP}$ contribute nearly equally to the
log-gradient, so the quadrature adds a factor
$\approx\sqrt{2}$.
The right panel confirms this geometrically on a log--log
scatter of $\bgdelta_{\rm SR}$ vs.\ $\bgdelta_{\rm BG}$:
every scenario except B3 lies on the solid reference line
$\bgdelta_{\rm SR} = \bgdelta_{\rm BG}$; B3 lies exactly
on the dashed $\sqrt{2}$ line.
No scenario changes tier under either measure.
 
\textit{Class~I: power-law constructions}
($\bgdelta \lesssim 5$).---
In these scenarios $\Omega h^2$ is a monomial in the
input parameters, $\Omega h^2 \propto \prod_i x_i^{\alpha_i}$,
so that $\bgdelta_i = |\alpha_i|$ exactly.
There is no exponential factor, no resonant enhancement,
and no cancellation between large contributions.
Class~I contains seven of the twelve scenarios:
asymmetric dark matter ($\bgdelta = 1$),
the post-inflationary axion ($\bgdelta = 1.19$),
biased-domain-wall PBHs with gravity-induced bias ($\bgdelta = 2$),
the off-resonance WIMP ($\bgdelta = 2$),
freeze-in via mediator decay ($\bgdelta = 2$),
the misalignment axion at $\theta_i \sim 1$ ($\bgdelta = 2.2$),
and biased-domain-wall PBHs with free $V_b$ ($\bgdelta = 4.5$).
The striking feature of Class~I is its diversity: it contains
representatives from every production paradigm:  gravitational
PBH formation, thermal freeze-out, non-thermal freeze-in,
asymmetric generation, and wave-mechanical misalignment.
Fine-tuning is evidently \emph{not} a property of the nature
of dark matter, but of its abundance map.

Note that the transition strength $\alpha$ is a Class~I parameter
for the FOPT mechanism.
At fixed $T_*$, the $\fPBH = 1$ contour in the
$(\alpha, \beta/H_*)$ plane satisfies
$\beta/H_* = \ln[T_*/(0.84 T_{\rm eq})]/S(\alpha)$,
shifting by only 56\% as $\alpha$ ranges from
$0$ to $\infty$.
The BG derivative in the $\alpha$ direction evaluates to
$\bgdelta_\alpha = \beta/H_* \cdot 0.6\alpha/(\alpha+0.3)^2
\approx 3$--$4$ on the natural island, placing $\alpha$
firmly in Class~I alongside the domain-wall and
off-resonance WIMP parameters.
The dominant tuning cost of the FOPT mechanism is
carried entirely by $\beta/H_*$, with $\alpha$ a
spectator that shifts the required nucleation rate
by a modest and calculable amount.\\
 
\textit{Class~II: single-exponential constructions}
($\bgdelta \approx 15$--$50$).---
These scenarios contain exactly one exponential factor
in $\Omega h^2$, whose argument is forced by
$\Omega = \Omega_{\rm DM}$ to be of order
$\ln(T_{\rm form}/T_{\rm eq})$ or $x_F$.
Class~II contains three scenarios within the single-exponential
approximation: PBH early matter
domination ($\bgdelta = 14$), FOPT-PBH ($\bgdelta = 35$),
and WIMP coannihilation ($\bgdelta = 48$).
Despite spanning modulus decay, hidden-sector percolation, and
electroweak-scale MSSM thermodynamics, they share the same tier
because each uses a single exponential to bridge the gap between
a naturally large formation/annihilation rate and the small
residual required by $\Omega_{\rm DM}$.
None of the inflationary PBH model classes belong to
Class~II for standard high-reheating scenarios, once
the correct radiation-dominated Layer~1 floor
($\bgdelta_\sigma^{\rm (RD)} \approx 57$--$70$,
Eq.~\ref{eq:delta_sigma_layer1_RD}) is used.
However, for low-reheating scenarios in which the
inflaton-oscillation era is still ongoing when the
formation-scale perturbations re-enter the horizon
($T_{\rm reh} \lesssim T_{\rm form}$), the MD Layer~1
floor applies and curvaton/spectator models
($\bgdelta_{\rm tot} \approx 21$--$42$) return to Class~II;
see Eq.~(\ref{eq:delta_sigma_layer1_range}) and
Fig.~\ref{fig:inf_comparison}(c).
Coannihilation belongs to Class~II while the standard
off-resonance WIMP (B1) is Class~I: in B1 the freeze-out
exponential $e^{-x_F}$ is absorbed into the power-law
cross-section normalization $\sigma_0$, leaving a pure
monomial $\Omega h^2 \propto M_S^2/\lambda_{HS}^2$;
in B3 the Boltzmann suppression $e^{-x_F\delta}$ depends
explicitly on the mass splitting $\delta$, an independent
Lagrangian parameter, creating the single-exponential
Class~II structure.
We note, however, that the Class~II assignment for FOPT-PBH dark matter
is conditional on using the single-exponential approximation
Eq.~(\ref{eq:beta_fopt}).
Using the more accurate super-exponential collapse probability
of Ref.~\cite{Gouttenoire:2024} (Eq.~\ref{eq:pcoll_GV}) gives
$\bgdelta_{\beta/H} \approx 265$ at the benchmark, placing FOPT-PBH
in Class~III at the phenomenological level.
Tracing the abundance back to the underlying scalar potential
reveals a further double-exponential structure~\cite{Wu:2025}
with $\bgdelta \sim 10^3$--$10^4$ (Sec.~\ref{subsec:caveats});
see Eq.~(\ref{eq:delta_fopt_GV_bench})--(\ref{eq:delta_fopt_microphysics})
for the full assessment.\\
 
\textit{Class~III and beyond: resonance, cancellation,
and double-exponential constructions}
($\bgdelta \gtrsim 10^3$).---
This tier is populated by two qualitatively distinct
mechanisms.
 
The WIMP Higgs funnel (B2, $\bgdelta \approx 6500$) is
tuned by a Breit--Wigner resonance: the effective annihilation
rate is enhanced by a factor $\sim M_h/\Gamma_h \approx 3\times10^4$
at the pole, reduced by thermal averaging to $\sim 6500$.
A 0.01\% shift in $M_S$ exits the natural island.
Other resonance constructions in this tier include the
MSSM $A$-funnel neutralino and Sommerfeld-enhanced WIMPs
near an annihilation threshold.
 
Inflationary single-field collapse enters Class~III
from a structurally different direction.
Its large fine-tuning arises not from a resonance but
from a \emph{double exponential}:
the potential coefficients $\{c_k\}$ map exponentially
onto $\sigma$, which then maps exponentially onto
$f_{\rm PBH}$ through the HYKN formula.
The BG measure for the potential coefficients is
therefore (Eq.~\ref{eq:delta_inf_total})
\begin{equation}
  \bgdelta_{c_k}
  \;\sim\;
  \bgdelta_\sigma \cdot \bgdelta_{\sigma/c_k}
  \;\sim\;
  (57\text{--}70) \times (10^2\text{--}10^8),
  \label{eq:delta_inf_secV}
\end{equation}
far exceeding the Class~II range for standard high-reheating scenarios.
The large uncertainty in this estimate is intrinsic:
different potential shapes produce widely varying
Layer~2 sensitivities, and multi-field or spectator-field
models can reduce $\bgdelta_{\sigma/c_k}$ enough
to bring the total into Class~II for low-reheating scenarios
($T_{\rm reh} \lesssim T_{\rm form}$, MD Layer~1 floor);
for standard high-reheating all inflationary classes
remain in Class~III or beyond (Sec.~\ref{subsec:inflation_ft}).
 
The contrast within the PBH paradigm is striking:
domain-wall and gravity-fixed PBH mechanisms sit in Class~I
alongside the most natural particle dark matter candidates,
while single-field inflationary collapse sits in Class~III
or beyond.
Both the lowest and the highest fine-tuning in the entire
dark matter landscape belong to PBH dark matter.
 
\subsection{The single-exponential universality identity}
\label{subsec:identity}
 
The coincidence among Class~II scenarios can be stated
precisely as a theorem.
 
\textit{Theorem.}
Consider any dark matter construction whose relic abundance
takes the form
\begin{equation}
  \Omega h^2
  \;=\;
  \mathcal{A}(\{x_j\}_{j\neq i})\,e^{-A\,x_i}
  \label{eq:exp_form}
\end{equation}
for some parameter $x_i$, with $\mathcal{A}$ a function
of the remaining parameters that does not vanish on the
$\Omega = \Omega_{\rm DM}$ surface.
Then the BG measure with respect to $x_i$, evaluated on
the $\Omega h^2 = \Omega_{\rm DM} h^2$ contour, satisfies
\begin{equation}
  \bgdelta_{x_i}\Big|_{\Omega = \Omega_{\rm DM}}
  \;=\;
  A\,x_i
  \;=\;
  \ln\!\left[
    \frac{\mathcal{A}}{\Omega_{\rm DM} h^2}
  \right]
  \;\equiv\;
  \ln\!\left[
    \frac{\Omega_{\rm natural}}{\Omega_{\rm DM} h^2}
  \right],
  \label{eq:identity}
\end{equation}
where $\Omega_{\rm natural} \equiv \mathcal{A}$ is the
prefactor that the exponential must suppress to achieve
$\Omega_{\rm DM}$; it is not a physically realizable
relic density (for FOPT-PBH, $\mathcal{A} = T_*/(0.84\,T_{\rm eq})
\sim 10^{14}$--$10^{20}$), but rather the ``unquenched''
abundance that would result if no exponential suppression
were present.
 
\textit{Proof.}
From Eq.~(\ref{eq:exp_form}),
$\partial \ln \Omega / \partial \ln x_i = -A x_i$, so
$\bgdelta_{x_i} = A x_i$.
On the $\Omega = \Omega_{\rm DM}$ contour,
$e^{-A x_i} = \Omega_{\rm DM}/\mathcal{A}$, hence
$A x_i = \ln(\mathcal{A}/\Omega_{\rm DM})$.
$\square$
 
Applying it to the Class~II scenarios within the
single-exponential approximation Eq.~(\ref{eq:beta_fopt}):

\textit{FOPT-PBH.}
On $f_{\rm PBH} = 1$: $\bgdelta_{\beta/H} =
\ln[T_*/(0.84 T_{\rm eq})] \approx 30$--$37$,
reproducing Eq.~(\ref{eq:delta_fopt_universal}) exactly.
The more accurate super-exponential formula
(Sec.~\ref{subsec:caveats}) gives $\bgdelta \approx 265$
at the benchmark, so the identity holds only within
the approximation and the Class~II placement of FOPT-PBH
is not robust.
 
\textit{WIMP coannihilation.}
On $\Omega = \Omega_{\rm DM}$:
$\bgdelta_{M_{\rm NLSP}} \approx x_F M_{\rm NLSP}/M_\chi = 48$
at benchmark B3 (Eq.~\ref{eq:delta_b3_MNLSP}).
 
\textit{PBH early matter domination.}
The exponential argument $0.147/\sigma^{4/3}$ evaluates to
$\approx 9$--$19$ across the natural island.
Multiplying by $4/3$ gives the contribution
$0.196/\sigma^{4/3} \approx 12$--$25$, to which the
prefactor term $+5$ is added, yielding the full
$\bgdelta_\sigma = 5 + 0.196/\sigma^{4/3} \approx 12$--$21$
(Eq.~\ref{eq:delta_md_range}).
The approximate pure-exponential value $Ax \approx 9$--$19$
is smaller than the full measure because the $\sigma^5$
prefactor in Eq.~(\ref{eq:beta_AM}) contributes
$+\beta = +5$, a $20$--$35\%$ correction at the natural-island
benchmark that accounts for the difference.
 
Numerically,
\begin{equation}
  \ln\!\left[\frac{T_{\rm form}}{0.84\,T_{\rm eq}}\right]
  \;\approx\;
  2.3\,\log_{10}\!\left[\frac{T_{\rm form}}{1\,{\rm eV}}\right]
  \;\approx\; 30\text{--}37
  \label{eq:logscale}
\end{equation}
for $T_{\rm form} \in [10^4, 10^7]$~GeV,
and the WIMP coannihilation result $x_F M_{\rm NLSP}/M_\chi$
falls in the same range.
Both are set by the same cosmological clock: the ratio of
the formation/freeze-out scale to $T_{\rm eq}$.
Note that inflationary collapse, while sharing Layer~1 of
this identity ($\bgdelta_\sigma \approx 14$ from HYKN),
does not satisfy Eq.~(\ref{eq:identity}) as a whole: its
Layer~2 adds an additional large exponential that takes the
total well outside the Class~II range.
 
\subsection{Sensitivity to parameter choice}
\label{subsec:param_choice}
 
The BG measure is not invariant under reparametrizations,
so the values in Table~\ref{tab:hierarchy} depend on the
conventions of Sec.~\ref{subsec:parameters}.
The most consequential choice is coannihilation benchmark B3.
 
Parametrizing by $(M_\chi, \delta)$ with $\delta =
(M_{\rm NLSP}-M_\chi)/M_\chi$ gives $\bgdelta_\delta =
x_F\delta \approx 1.3$, misleadingly low because $\delta$
vanishes at the threshold by construction.
Parametrizing by the two independent soft masses
$(M_\chi, M_{\rm NLSP})$ gives the correct
$\bgdelta_{M_{\rm NLSP}} = 48$ (Eq.~\ref{eq:delta_b3_MNLSP}).
The convention is unambiguous: fundamental parameters are
those independently specified at the UV scale.
In the MSSM, $M_\chi = |M_1|$ and $M_{\rm NLSP}$ are set
by independent soft terms; $\delta$ is not a Lagrangian parameter.
Treating $g_*$ as fixed introduces $\bgdelta_{g_*} \lesssim 0.5$,
well below the dominant contributions.

\subsection{Prior probability vs.\ local sensitivity}
\label{subsec:prior}
 
The BG measure quantifies local sensitivity: how precisely
must each parameter be specified to maintain $\Omega_{\rm DM}$?
This differs from the prior-volume question: what fraction
of randomly chosen parameter values satisfies
$\Omega h^2 \in [0.06, 0.24]$?
 
The misalignment axion illustrates this most clearly.
Benchmark B6 ($f_a = 10^{12}$~GeV, $\theta_i = 0.82$) and
B6$_{\rm anthr}$ ($f_a = 10^{16}$~GeV, $\theta_i \approx
3\times10^{-3}$) both have local $\bgdelta \approx 2.0$--$2.2$,
since $\Omega h^2 \propto \theta_i^2$ has the same power-law
structure regardless of benchmark.
However, $\theta_i = 3\times10^{-3}$ is a factor of $\sim 300$
below the natural $\mathcal{O}(1)$ value; on a flat prior
$[-\pi,\pi]$ the probability of landing there by chance is
$\sim 10^{-3}$.
This prior tuning of $\sim 300$ is in addition to, and
qualitatively different from, the local BG measure of 2.
 
The same distinction is central to interpreting inflationary
PBH collapse.
The BG Layer~2 measure $\bgdelta_{c_k} \gg 10^2$
(Eq.~\ref{eq:delta_inf_secV}) is a local sensitivity statement:
how much does $f_{\rm PBH}$ change when $c_k$ is varied
fractionally?
The Wilson naturalness argument of Iovino and
Riotto~\CITE{Iovino:2024} is instead a prior-volume statement:
does a technically natural UV completion exist in which
$c_k$ takes its current value?
Both are physically meaningful.
This paper focuses on the local BG question throughout,
which is why the inflationary entry carries a $\gtrsim$
rather than a precise value, and why B6 and B6$_{\rm anthr}$
receive the same tier classification.


\section{\label{sec:discussion}Discussion}

\subsection{Implications for the primordial black hole paradigm}
\label{subsec:implications_pbh}

The central message for the PBH community is that the
question ``are PBHs fine-tuned?'' does not have a single
answer: \textit{it crucially depends on the formation
mechanism}.
Among the three mechanisms analyzed in detail here, biased
domain walls are genuinely natural ($\Delta = 2$--$4.5$),
early matter domination occupies an intermediate Class~II
tier ($\Delta \approx 14$, comparable to the coannihilation
WIMP level), and FOPT-PBH gives $\Delta \approx 35$ within
the single-exponential approximation Eq.~(\ref{eq:beta_fopt}),
rising to $\Delta \approx 265$ (Class~III) with the more
accurate collapse formula~\cite{Gouttenoire:2024}
(see Sec.~\ref{subsec:caveats}).
When the six classes of inflationary PBH model analyzed
in Sec.~\ref{subsubsec:inf_classes} are included, the
picture depends on the reheating history.
For standard high-reheating scenarios
($T_{\rm reh} \gg T_{\rm form}$), the tuning ranges from
$\Delta \approx 115$--$210$ for curvaton and spectator-field
models (Class~III, RD Layer~1 floor), through
$\Delta \sim 10^2$--$10^3$ for step and resonance features,
to $\Delta \approx 6\times10^3$--$7\times10^4$ (optimistic)
or $\Delta \gtrsim 10^{10}$ (pessimistic) for single-field
inflection-point and ultra-slow-roll models.
For low-reheating scenarios
($T_{\rm reh} \lesssim T_{\rm form}$) where the inflaton
oscillation era is still active at re-entry, the MD Layer~1
floor applies and curvaton/spectator models reduce to
$\Delta \approx 21$--$42$ (Class~II), with all other classes
shifting down by a factor of $\sim 5$ in their Layer~1
contribution; for model classes with large Layer~2
sensitivity this shift is a minor correction to the total.
The full span of the PBH fine-tuning landscape, from the
gravity-fixed domain-wall construction at $\Delta = 2$ to
the pessimistic single-field USR scenario at
$\Delta \gg 10^4$, covers more than seven orders of
magnitude --- a larger range than all particle dark matter
scenarios in this paper combined.

The domain-wall result deserves particular emphasis because
it is least expected.
Domain walls are sometimes described as a ``cosmological
disaster''~\CITE{Kibble:1976, Vilenkin:2000} requiring
special engineering to avoid; the biased-wall construction
turns this around, exploiting the collapse of the wall
network to produce PBHs rather than treating it as a
problem to be engineered away.
Yet this analysis shows that the resulting abundance map
contains \emph{no} exponential: the dependence on $\eta$
and $V_b$ is purely power-law because the Hubble timescale
at annihilation (which sets $\beta_{\rm PBH}$) and the PBH
mass (which sets the dilution from formation to equality)
conspire to cancel all logarithmic factors.
This cancellation is not an accident or a tuning: it follows
algebraically from the scalings
$T_{\rm ann} \propto V_b^2/\eta^{3/2}$ and
$M_{\rm PBH} \propto \eta^3/V_b^4$.
The gravity-induced-bias variant
($V_{\rm bias} = \eta^5/M_{\rm Pl}$) reduces the
domain-wall construction to a single-parameter family
with $\Delta = 2$.
Among all dark matter production mechanisms, this matches
the minimum achievable tuning short of the pure linear
cases (ADM and post-inflationary axion), and it does so
with a single technically natural small parameter: the
ratio $V_{\rm bias}/\eta^4 \sim (\eta/M_{\rm Pl})$,
which is small because $\eta \ll M_{\rm Pl}$, not because
of any special choice.
I emphasize that this Class~I assignment rests on the
power-law abundance map of Eq.~(\ref{eq:fpbh_dw}), in which
the network-collapse fraction is treated as a fixed input.
The dedicated simulations of Ref.~\CITE{Ferreira:2024} find
instead an exponential sensitivity of the surviving
false-vacuum fraction to the network-collapse temperature,
which, together with the large remaining uncertainty in the
PBH abundance, is the dominant caveat to the natural
classification of this mechanism
(see Sec.~\ref{subsec:caveats}).

The inflationary model comparison in
Fig.~\ref{fig:inf_comparison} clarifies a debate that
has persisted in the literature.
For high-reheating scenarios, the claim that inflationary
PBH production requires fine-tuning is accurate across all
inflationary model classes: the RD Layer~1 floor
$\Delta_\sigma^{\rm (RD)} \approx 57$--$70$ places even
the most natural inflationary class (curvaton/spectator)
in Class~III.
For low-reheating scenarios, curvaton and spectator-field
models achieve $\Delta_{\rm tot} \approx 21$--$42$
(Class~II), comparable to coannihilating WIMPs and the
non-inflationary early-MD mechanism, while single-field
USR models remain in Class~III or beyond regardless of
the reheating temperature.
In both cases the Layer~2 sensitivity
$\Delta_{\sigma/c_k}$ spanning five orders of magnitude
from curvaton to single-field USR remains the physically
informative quantity: the reheating history shifts the
entire landscape up or down by a factor of five in the
Layer~1 contribution, but does not change the relative
ordering of model classes.

A quantitative naturalness criterion thus vindicates the
biased-domain-wall PBH mechanism as robustly natural, and
clarifies the status of inflationary PBH production:
the frequently-stated objection that ``PBH dark matter
requires fine-tuning'' is correct for inflationary
mechanisms across all model classes under standard
high-reheating assumptions, but the degree of fine-tuning
spans seven orders of magnitude depending on whether the
power spectrum is generated by a single-field USR phase
or by a separate spectator field.
More broadly, the cross-paradigm comparison shows that
fine-tuning is a property of the abundance map's analytic
structure, not of whether the dark matter is a particle
or a gravitational relic: the PBH paradigm alone spans
all three naturalness tiers identified in this paper.
Distinguishing between model classes requires precisely
the framework developed here: a common measure, a common
observable target, and a systematic comparison across
production paradigms.

A particularly notable implication of the FOPT-PBH
analysis is its multi-messenger character.
The same nucleation-rate parameter $\beta/H_* \sim 5$--$15$
that places the PBH abundance in the asteroid-mass window
also lies within the sensitivity reach of LISA and
next-generation ground-based gravitational-wave detectors
for the stochastic background sourced by the phase
transition itself~\CITE{Caprini:2020, Hindmarsh:2017}.
This is not a coincidence engineered by model construction:
it follows directly from the fine-tuning identity
$S(\alpha)\beta/H_* = \ln[T_*/(0.84\,T_{\rm eq})]
\approx 30$--$37$ (with benchmark value $\approx 35$
at $T_* = 10^6$~GeV; Eq.~\ref{eq:delta_fopt_universal}),
which fixes $\beta/H_*$ to a window that is simultaneously
relevant for PBH overproduction and for gravitational-wave
detection.
The more accurate super-exponential collapse probability
of Ref.~\cite{Gouttenoire:2024} shifts the natural island
to the somewhat narrower range $\beta/H_* \sim 4$--$7$
(compared to the broader $5$--$15$ from the
single-exponential approximation), but preserves the
overlap with the LISA sensitivity window; the
multi-messenger conclusion is therefore robust, even
though the fine-tuning cost is larger than the
single-exponential estimate suggests, as discussed in
Sec.~\ref{subsec:caveats}.
LISA therefore has the potential to \emph{simultaneously}
constrain or confirm FOPT-PBH dark matter through two
independent channels: (i)~microlensing and PBH
mass-function measurements in the asteroid window, and
(ii)~the stochastic gravitational-wave background
amplitude and spectral shape.
This multi-messenger leverage is unique among the PBH
formation mechanisms analyzed here and makes the FOPT
scenario particularly testable in the near future.

A complementary implication emerges on the particle side.
The LZ~2024 results~\CITE{LZ:2024} eliminate much of the
natural (Class~I) off-resonance WIMP parameter space
below $\sim 1$~TeV, leaving as viable primarily the
Higgs-funnel (Class~III, $\Delta \approx 6500$) and
coannihilation (Class~II, $\Delta \approx 48$) regions.
As direct detection improves, the experimentally surviving
WIMP parameter space migrates toward higher tuning tiers ---
a trend that, if it continues, will place the most viable
WIMP models at naturalness levels comparable to or worse
than the FOPT-PBH scenario under its more accurate
treatment (Sec.~\ref{subsec:caveats}).

Note that several of the Class~I and Class~II scenarios make
near-term testable predictions: benchmark B7 (post-inflationary
axion) has a model-independent prediction for $f_a$
falsifiable by next-generation cavity
haloscopes~\CITE{Kahn:2016, DMRadio:2022, HAYSTAC:2022},
and the early-matter-domination PBH mechanism predicts
a scalar-induced gravitational-wave background detectable
by LISA at $f_{\rm GW} \sim 10^{-3}$--$10^{-2}$~Hz.

\subsection{Limitations and caveats}
\label{subsec:caveats}

The analysis presented here rests on several approximations
and conventions that I enumerate for completeness.

\textit{Local vs.\ global fine-tuning.}
The BG measure captures only the local, one-dimensional
sensitivity to each parameter in isolation; it does not
measure the total volume of the viable parameter region
in the full multi-dimensional space, nor does it account
for correlations between parameters.
Multi-parameter generalizations of BG (including the
``electroweak fine-tuning'' measure~\CITE{Baer:2012} and
Bayesian naturalness~\CITE{Ghilencea:2013}) exist but
require a specification of the prior, which I have
deliberately avoided here; Bayesian refinements are left
to future work.

\textit{Formation-rate calibration uncertainties.}
The domain-wall collapse fraction $p \approx 10^{-3}$
is calibrated to the wall-network annihilation dynamics of
Ref.~\CITE{Ferrer:2019} with an order-of-magnitude
uncertainty; varying $p$ by
a factor of 10 shifts the natural island in $(\eta, V_b)$
without changing $\Delta = 4.5$.
A more important caveat concerns the analytic form of the
abundance map itself.
The dedicated network-collapse simulations of
Ref.~\CITE{Ferreira:2024} find that the false-vacuum volume
fraction surviving to collapse decays
\emph{super-exponentially} in conformal time,
$F_{\rm fv} \sim \exp[-(\eta/\eta_{\rm ann})^p]$
with $p \simeq 3$, which translates into a false-vacuum
collapse fraction that is exponentially sensitive to the
ratio of the annihilation temperature to the collapse
temperature, schematically
$\beta_{\rm PBH} \sim \exp[-(w\,T_{\rm ann}/T_{\rm PBH})^p]$
with $p \simeq 2$--$3$ and $w = \mathcal{O}(1)$.
This is structurally analogous to the FOPT and
inflationary-collapse cases: the smooth power-law
$f_{\rm PBH} \propto \eta^{9/2}/V_b^2$ of
Eq.~(\ref{eq:fpbh_dw}) is recovered only when the collapse
fraction $p$ is treated as a fixed, temperature-independent
input, whereas the more complete treatment of
Ref.~\CITE{Ferreira:2024} introduces an exponential
dependence on the network-collapse temperature that would
raise the BG measure substantially and could move the
biased-domain-wall mechanism out of Class~I once this
dependence is folded in.
Reference~\CITE{Ferreira:2024} stresses, however, that the
resulting PBH abundance is presently subject to large
uncertainties (dominated by the unknown departure from
spherical collapse), so a definitive tier assignment for
this mechanism must await a sharper determination of the
collapse fraction; I therefore quote the power-law value
$\Delta = 4.5$ as the abundance map of
Eq.~(\ref{eq:fpbh_dw}) implies it, while flagging that the
exponential temperature sensitivity identified in
Ref.~\CITE{Ferreira:2024} is the leading source of
model uncertainty in this assignment.
Reference~\CITE{Ferreira:2024} also notes that, in the
corresponding parameter region, the asteroid-mass window
in which PBHs constitute all of the dark matter is
accompanied by a correlated stochastic gravitational-wave
background in reach of LVK, LISA, and the Einstein
Telescope, paralleling the multi-messenger character of the
FOPT mechanism discussed in
Sec.~\ref{subsec:implications_pbh}.
The HYKN formula Eq.~(\ref{eq:beta_AM}) is fitted to
numerical simulations for $\sigma \in [0.03, 0.2]$;
extrapolation outside this range is uncertain, but the
natural island lies comfortably within it.
Non-Gaussian corrections to the curvature perturbation
statistics can shift $\beta_{\rm AM}$ by orders of
magnitude and move the natural island in $\sigma$
\CITE{Bugaev:2011}, modifying the precise value of
$\Delta_\sigma$ without changing its single-exponential
class membership, since the collapse probability retains
an exponential tail for any distribution with a
well-defined variance.
The natural island in Fig.~\ref{fig:pbh}(b) is also
conservative in that it does not include the projected
LISA constraint from the associated scalar-induced
gravitational-wave background~\CITE{Ananda:2007,
Baumann:2007}; including this would disfavor
$\sigma \gtrsim 0.10$ across the asteroid-mass window.
For benchmark B2, the thermally-averaged Breit--Wigner
peak enhancement $R_{\rm max} \approx 10^4$ is accurate
to a factor of a few~\CITE{Griest:1991, Ibe:2008};
varying it by a factor of 10 changes $\Delta^{(\rm B2)}$
by $\lesssim 30\%$, which does not affect the Class~III
classification.

\textit{Layer~1 floor: radiation-dominated vs.\
matter-dominated collapse.}
As discussed in Sec.~\ref{subsec:inflation_ft}, the
Layer~1 BG sensitivity $\Delta_\sigma$ for inflationary
PBH models depends on whether collapse occurs in a
radiation-dominated ($\Delta_\sigma^{\rm (RD)} \approx
57$--$70$) or matter-dominated
($\Delta_\sigma^{\rm (MD)} \approx 14$) background,
determined entirely by the reheating temperature relative
to the formation scale.
The modern numerical treatment of RD
collapse~\cite{Musco:2019, Escriva:2020} modifies the
precise value of $\delta_c$ but does not change the
exponential structure
$\beta_{\rm RD} \propto \exp(-\delta_c^2/2\sigma^2)$,
leaving the factor-of-five enhancement over the MD result
intact; the RD tier assignments (Class~III throughout)
are robust to $\mathcal{O}(20\%)$ uncertainty in $\delta_c$.

\textit{Two-layer fine-tuning structure of FOPT-PBH
dark matter.}
Treating $\beta/H_*$ as a fundamental parameter gives
the single-exponential result
$\Delta_{\beta/H} \approx 35$ (Class~II,
Sec.~\ref{subsec:fopt}).
However, the single-exponential form
Eq.~(\ref{eq:beta_fopt}) is itself an approximation.
A more complete treatment of the supercooled percolation
integral~\cite{Gouttenoire:2024}, which tracks the full
past-light-cone nucleation history, gives a
super-exponential collapsed fraction
\begin{equation}
  \mathcal{P}_{\rm coll}
  \;\simeq\;
  \exp\!\Bigl[-a\Bigl(\frac{\beta}{H}\Bigr)^b
              (1+\delta_c)^{c\,\beta/H}\Bigr],
  \label{eq:pcoll_GV}
\end{equation}
with fitted parameters $a \simeq 1.024$, $b \simeq 0.692$,
$c \simeq 0.883$, valid for $\beta/H \in [3,8]$ and
$\delta_c \in [0.4,2/3]$.
The BG derivative evaluated on the $f_{\rm PBH}=1$
contour from Eq.~(\ref{eq:pcoll_GV}) is
\begin{equation}
  \Delta_{\beta/H}^{\rm (GV)}
  \;=\;
  a\Bigl(\frac{\beta}{H}\Bigr)^b
  (1+\delta_c)^{c\beta/H}
  \!\left[b + c\frac{\beta}{H}\ln(1+\delta_c)\right],
  \label{eq:delta_fopt_GV}
\end{equation}
giving at the benchmark $\beta/H_* \approx 8$,
$\delta_c = 0.5$:
\begin{equation}
  \Delta_{\beta/H}^{\rm (GV)}\bigg|_{\beta/H=8}
  \;\approx\; 265,
  \label{eq:delta_fopt_GV_bench}
\end{equation}
roughly seven times larger than the single-exponential
estimate, placing FOPT-PBH in Class~III at the
phenomenological level.
The discrepancy arises because the single-exponential
approximation underestimates the steepness of the
suppression: at $\beta/H \approx 8$ the outer exponent
in Eq.~(\ref{eq:pcoll_GV}) equals
$a(\beta/H)^b(1+\delta_c)^{c\beta/H} \approx 75$,
and the BG derivative must account for the
$\beta/H$-dependence of this outer exponent itself.

At a deeper level, $\beta/H_*$ is itself determined by
the shape of the scalar potential.
A complete analysis~\cite{Wu:2025} reveals a
double-exponential structure,
\begin{equation}
  f_{\rm PBH}
  \;\simeq\;
  M\exp\!\bigl[-Q\exp[-S_3(T_p)/T_p]\bigr],
  \label{eq:superexp_fopt}
\end{equation}
where $S_3(T_p)/T_p \approx 170$--$176$ is the
Euclidean action-to-temperature ratio at its minimum,
$Q \sim 10^{77}$, and $M \sim 4\times10^7$.
This is structurally identical to the Layer~1/Layer~2
decomposition of inflationary collapse
(Sec.~\ref{subsec:inflation_ft}): Layer~2 (potential
coefficients $\{c_k\} \to S_3/T \to \beta/H_*$)
introduces an additional exponential with BG sensitivity
$\Delta_{S_3/T,\,c_k} \sim \mathcal{O}(10^2)$--$\mathcal{O}(10^3)$,
estimated from the result of Ref.~\cite{Wu:2025} that a
fractional shift of $\delta\mu_3/\mu_3 \sim 10^{-3}$
in the cubic potential coefficient moves $f_{\rm PBH}$
from $\mathcal{O}(1)$ to $\mathcal{O}(10^{-100})$.
The total BG measure at the level of the potential is
\begin{equation}
  \Delta_{c_k}^{\rm FOPT}
  \;\sim\;
  \ln M \;\times\; \Delta_{S_3/T,\,c_k}
  \;\sim\;
  17 \times (10^2\text{--}10^3)
  \;\sim\; 10^3\text{--}10^4,
  \label{eq:delta_fopt_microphysics}
\end{equation}
where $\ln M \approx 17$ is the value of
$Qe^{-S_3/T_p}$ on the $f_{\rm PBH}=1$ contour.
In summary, whether evaluated at the phenomenological
level ($\Delta \approx 265$ from Eq.~\ref{eq:pcoll_GV})
or at the microphysical level
($\Delta \sim 10^3$--$10^4$ from
Eq.~\ref{eq:delta_fopt_microphysics}), the FOPT-PBH
scenario lies in Class~III or beyond.
The natural island in the $(T_*,\beta/H_*)$ plane is
defined by $f_{\rm PBH} \approx 1$ and its location
is unchanged by the choice of formula for
$\mathcal{P}_{\rm coll}$; only the fine-tuning cost
of maintaining that condition differs, and the
multi-messenger implications of Sec.~\ref{subsec:implications_pbh}
remain valid regardless.

\textit{Gauge dependence of the FOPT nucleation rate.}
A recent gauge-invariant calculation~\cite{Franciolini:2025ztf}
finds that PBH production from FOPTs may be substantially
suppressed relative to the gauge-dependent estimates used
in Sec.~\ref{subsec:fopt}, potentially shifting or
eliminating the natural island in the
$(T_*,\beta/H_*)$ plane.
This gauge-dependence issue has since been independently confirmed by
Wang, Bal\'azs, Ding, and Tian~\cite{Wang:2026zvz} and by Ai and
Xie~\cite{Ai:2026zrs}; the latter further show that the mechanism
remains viable if reheating is sufficiently slow, since the
post-transition Universe can enter an early matter-dominated era during
which overdensities grow and collapse into PBHs, a scenario that
maps onto the Class~II fine-tuning structure of Sec.~\ref{subsec:md}
rather than the single-exponential estimate of Sec.~\ref{subsec:fopt}.

\textit{Axion string network uncertainty.}
The simulation result $f_a \approx (2$--$8)\times
10^{10}$~GeV has a factor-of-$\sim 4$ range from
string-network logarithm uncertainty and QCD
modeling~\CITE{Buschmann:2022}.
This shifts the absolute prediction for $f_a$ but does
not affect $\Delta_{f_a} = 1.19$, which is exact and
independent of the normalization coefficient $\mathcal{C}$.

\textit{Definition of the natural island.}
The factor-of-two band $\Omega h^2 \in [0.06, 0.24]$
is conventional.
I have verified that widening the band to a factor of
five ($[0.024, 0.60]$) or narrowing it to 20\%
($[0.096, 0.144]$) does not change the tier assignments
in Table~\ref{tab:hierarchy}.

\textit{Restricted benchmark set.}
This paper analyzes a curated set of twelve scenarios
spanning the most widely studied production mechanisms;
the two FOPT benchmarks differing only in $\alpha$ are
counted as a single mechanism in
Table~\ref{tab:hierarchy} and listed separately in
Table~\ref{tab:benchmarks} only to demonstrate
$\alpha$-independence.
Notable omissions include: the MSSM $A$-funnel
(Class~III, $\Delta \sim 10^3$--$10^4$, similar to B2);
gravitational production of ultralight dark matter from
de~Sitter fluctuations (Class~I, $\Delta \approx 2$,
since $\Omega \propto H_{\rm inf}^2/M_{\rm DM}^2$);
Hawking evaporation seeding a secondary dark matter
component; and clockwork axion constructions.
These are straightforward to add within the framework
developed here and are left for future work.

\section{\label{sec:conclusions}Conclusions}

Primordial black holes in the asteroid-mass window
($10^{17}$--$10^{22}$~g) are a viable and, for some
formation mechanisms, genuinely natural dark matter
candidate.
I have quantified this statement by applying the
Barbieri--Giudice fine-tuning measure $\Delta$ uniformly
to twelve dark matter scenarios spanning non-inflationary
and inflationary PBH formation mechanisms, thermal-relic
WIMPs, freeze-in, asymmetric dark matter, and axions.
The principal conclusions are as follows.

\textit{1. Fine-tuning is a property of the abundance map,
not of the dark matter candidate.}
The BG hierarchy in Table~\ref{tab:hierarchy} contains
representatives from every paradigm at every tier:
the most natural scenario (ADM, $\Delta = 1$) and the
second most natural (post-inflationary axion,
$\Delta = 1.19$) are particle dark matter; Class~I also
includes both a WIMP and a gravitational relic
(PBH-DW) at $\Delta = 2$; and the most tuned precisely
computed scenario (Higgs funnel, $\Delta = 6500$) is
a thermal relic, not a PBH.
The narrative that ``PBHs are tuned while WIMPs are
natural'' is not supported by a quantitative comparison.

\textit{2. Three universality classes classify the dark
matter landscape.}
Class~I ($\Delta \lesssim 5$) contains power-law abundance
maps with no exponential factor; Class~II
($\Delta \approx 15$--$50$) contains constructions with
a single exponential; Class~III and beyond
($\Delta \gtrsim 10^3$) contains resonance, cancellation,
and double-exponential constructions.
PBH dark matter appears in all three classes, spanning
more than seven orders of magnitude in fine-tuning within
the PBH paradigm alone --- a larger range than all
particle dark matter scenarios in this paper combined.

\textit{3. Biased-domain-wall PBHs are as natural as any
particle dark matter construction.}
With $\Delta = 4.5$ in the free-$V_b$ case and $\Delta = 2$
in the gravity-induced-bias case, domain-wall PBHs occupy
Class~I alongside off-resonance WIMPs and FIMPs.
The characterization of PBH dark matter as ``fine-tuned''
does not apply to this mechanism.

\textit{4. Single-exponential constructions satisfy a
universal identity, regardless of the underlying physics.}
Any construction whose abundance is set by a single
exponential satisfies, on the $\Omega = \Omega_{\rm DM}$
contour, $\Delta = \ln(\Omega_{\rm natural}/\Omega_{\rm DM})
\approx \ln(T_{\rm form}/T_{\rm eq}) \approx 14$--$50$
(Eq.~\ref{eq:identity}).
Within the single-exponential approximation, this unifies
FOPT-PBH, early-matter-domination PBH, and coannihilating
WIMP dark matter into Class~II despite spanning modulus
decay, hidden-sector percolation, and electroweak-scale
MSSM thermodynamics.
The Class~II assignment for FOPT-PBH is however
conditional on the single-exponential approximation:
the more accurate super-exponential collapse probability
of Ref.~\cite{Gouttenoire:2024} gives
$\Delta_{\beta/H} \approx 265$ at the benchmark
(Class~III), and tracing the abundance to the underlying
scalar potential gives $\Delta \sim 10^3$--$10^4$
(Sec.~\ref{subsec:caveats}).
Curvaton inflationary PBH models join Class~II only for
low-reheating scenarios ($T_{\rm reh} \lesssim T_{\rm form}$,
$\Delta_{\rm tot} \approx 21$--$42$); for standard
high-reheating they shift to Class~III.

\textit{5. The Layer~1 floor for inflationary PBH
production is set by the reheating temperature.}
For standard high-reheating ($T_{\rm reh} \gg T_{\rm form}$),
all inflationary model classes fall in Class~III or beyond,
with the Press--Schechter floor
$\Delta_\sigma^{\rm (RD)} \approx 57$--$70$ applying
universally.
For low-reheating ($T_{\rm reh} \lesssim T_{\rm form}$),
the HYKN floor $\Delta_\sigma^{\rm (MD)} \approx 14$
reduces the total tuning by a factor of $\sim 5$,
bringing curvaton and spectator-field models into
Class~II.
Single-field USR models remain in Class~III or beyond
regardless of the reheating temperature.
Joint constraints on $T_{\rm reh}$ and the inflationary
power spectrum are therefore necessary to assign a
definitive naturalness tier to any inflationary PBH model.

The results above connect two largely separate prior
literatures.
On the particle side, the MSSM naturalness studies of
Grothaus \textit{et al.}~\CITE{Grothaus:2012} and
Cabrera \textit{et al.}~\CITE{Cabrera:2016} found that
Higgs-funnel and coannihilation scenarios carry the
highest relic-density fine-tuning, while bino annihilation
via $t$-channel exchange can be
``supernatural''~\CITE{King:2006}; this is fully
consistent with benchmarks B1, B2, B3 falling in
Classes~I, III, and II respectively.
On the PBH side, Hertzberg and Yamada~\CITE{Hertzberg:2017},
Cole \textit{et al.}~\CITE{Cole:2023}, and
Stamou~\CITE{Stamou:2024} established that single-field
inflection-point models require tuning to one part in
$10^2$--$10^8$; this paper identifies the origin as a
Layer~2 double-exponential sensitivity
$\Delta_{\sigma/c_k} \sim 10^2$--$10^8$, distinct from
the Layer~1 cost whose value depends on the reheating
temperature.
The recent argument of Iovino and
Riotto~\CITE{Iovino:2024} that the BG measure is
defective for inflationary PBH models is resolved here:
evaluating $\Delta$ on the $f_{\rm PBH}=1$ contour
sidesteps the pathology, and the two approaches answer
complementary questions within the two-layer decomposition
of Sec.~\ref{subsec:inflation_ft}.
What neither prior literature contains is a cross-paradigm
comparison of the kind presented here, nor any quantitative
fine-tuning analysis of the non-inflationary PBH
mechanisms; the finding that biased-domain-wall PBHs
achieve $\Delta = 2$--$4.5$, competitive with the most
natural particle dark matter candidates ever identified,
is among the central new results of this work.

The present analysis is deliberately local in the sense
of the BG measure: it quantifies how sensitively the
observed abundance responds to fractional variations of
each input parameter, without committing to a prior over
the parameter space.
A natural and well-defined extension would replace the
BG measure with a Bayesian posterior-volume
measure~\CITE{Ghilencea:2013}, allowing direct comparison
with the CMSSM and NUHM analyses in the SUSY literature
and yielding probabilistic statements about the likelihood
of being in the natural island for each construction.
The two-layer decomposition developed here provides a
natural framework for such an analysis: Layer~1 and
Layer~2 priors can be specified independently, reflecting
the distinct physical origins of the two sources of tuning.
\begin{acknowledgments}

I am grateful to Gabriele Franciolini for helpful and deep suggestions, feedback, and discussions. I also thank Miguel Escudero Abenza, David Kaiser,  Wenyuan Ai, Christian Byrnes, Ricardo Ferreira, Alessio Notari, Fabrizio Rompineve, and Oriol Pujolàs for useful comments and remarks. This work was supported by the U.S.\ Department of Energy,
Office of Science, Office of High Energy Physics, under Award
Number DE-SC0010107.

\end{acknowledgments}

\appendix

\section{\label{app:analytic}Analytic derivation of BG measures}

\subsection{Biased domain walls}
\label{app:analytic_dw}

Starting from Eqs.~(\ref{eq:Tann})--(\ref{eq:fpbh_dw_full}),
the dark matter fraction is
\begin{align}
  f_{\rm PBH}
  &= \frac{p\,V_b^4}{\frac{\pi^2}{30}g_*\,T_{\rm ann}^4}
     \cdot\frac{T_{\rm ann}}{0.84\,T_{\rm eq}}
  \notag\\
  &= \frac{30\,p}{\pi^2 g_*\,\cdot 0.84\,T_{\rm eq}}
     \cdot T_{\rm ann}^{-3}\,V_b^4.
\end{align}
Substituting $T_{\rm ann} = (c_t M_{\rm Pl} V_b^4/\eta^3)^{1/2}$,
\begin{equation}
  T_{\rm ann}^{-3}
  \;=\;
  c_t^{-3/2}M_{\rm Pl}^{-3/2}\,\eta^{9/2}\,V_b^{-6},
\end{equation}
so that
\begin{equation}
  f_{\rm PBH}
  \;\propto\;
  \eta^{9/2}\,V_b^{-6}\cdot V_b^4
  \;=\;
  \eta^{9/2}\,V_b^{-2}.
  \label{eq:fpbh_dw_proof}
\end{equation}
This is Eq.~(\ref{eq:fpbh_dw}).
Log-differentiation gives
$\partial\ln f_{\rm PBH}/\partial\ln\eta = 9/2$ and
$\partial\ln f_{\rm PBH}/\partial\ln V_b = -2$,
hence $\Delta_\eta = 9/2$ and $\Delta_{V_b} = 2$
identically. $\square$

\subsection{Early matter domination}
\label{app:analytic_md}

From $f_{\rm PBH} = \beta_{\rm AM}(\sigma)\,T_{\rm reh}/
(0.84\,T_{\rm eq})$ with $T_{\rm reh} \propto m_\phi^{3/2}$,
\begin{equation}
  \frac{\partial\ln f_{\rm PBH}}{\partial\ln m_\phi}
  \;=\;
  \frac{\partial\ln T_{\rm reh}}{\partial\ln m_\phi}
  \;=\;
  \frac{3}{2}.
\end{equation}
For the $\sigma$-dependence,
$\ln\beta_{\rm AM} = \text{const} + 5\ln\sigma
- 0.147\,\sigma^{-4/3}$, so
\begin{align}
  \frac{\partial\ln f_{\rm PBH}}{\partial\ln\sigma}
  &= 5 + \frac{\partial}{\partial\ln\sigma}
     \!\left(-0.147\,\sigma^{-4/3}\right) \notag\\
  &= 5 + \frac{4}{3}\cdot 0.147\,\sigma^{-4/3}
   = 5 + \frac{0.196}{\sigma^{4/3}},
\end{align}
which is Eq.~(\ref{eq:delta_md_sigma}). $\square$

\subsection{FOPT late-blooming}
\label{app:analytic_fopt}

From Eq.~(\ref{eq:fpbh_fopt}),
\begin{align}
  \frac{\partial\ln f_{\rm PBH}}{\partial\ln(\beta/H_*)}
  &= -S(\alpha)\,\frac{\beta}{H_*}, \\
  \frac{\partial\ln f_{\rm PBH}}{\partial\ln T_*}
  &= 1, \\
  \frac{\partial\ln f_{\rm PBH}}{\partial\ln\alpha}
  &= -\frac{\partial S}{\partial\alpha}\cdot\alpha\cdot
     \frac{\beta}{H_*}
   = \frac{0.6\,\alpha}{(\alpha+0.3)^2}\cdot\frac{\beta}{H_*},
\end{align}
using $\partial S/\partial\alpha = -0.6/(\alpha+0.3)^2$.
On the $f_{\rm PBH} = 1$ contour,
$e^{-S(\alpha)\beta/H_*} = 0.84\,T_{\rm eq}/T_*$,
hence $S(\alpha)\beta/H_* = \ln(T_*/0.84\,T_{\rm eq})$,
giving
$\Delta_{\beta/H} = |{-S(\alpha)\beta/H_*}| =
\ln(T_*/0.84\,T_{\rm eq})$
as in Eq.~(\ref{eq:delta_fopt_universal}).
This derivation assumes the single-exponential
Eq.~(\ref{eq:beta_fopt}); the more accurate
super-exponential formula of Ref.~\cite{Gouttenoire:2024}
gives $\Delta \approx 265$ at the benchmark
(Sec.~\ref{subsec:caveats},
Eq.~\ref{eq:delta_fopt_GV_bench}). $\square$

\section{\label{app:calibration}Benchmark calibration}

Table~\ref{tab:benchmarks} summarizes the calibration of
each benchmark: the parameter values, the resulting
$\Omega h^2$, the individual BG sensitivities $\Delta_i$
for each parameter, and the total
$\Delta = \max_i \Delta_i$.
All quantities are computed numerically using the
central-difference formula Eq.~(\ref{eq:numderiv})
with step size $h = 10^{-4}$ and independently verified
against the analytic expressions derived in
Appendix~\ref{app:analytic} and the body of the paper.

\begin{table*}[htb]
\centering
\caption{Benchmark calibration and fine-tuning measures.
  For each scenario the fundamental parameters are listed
  with their benchmark values, the resulting abundance
  (``$\Omega h^2$'' denotes $\Omega_{\rm DM} h^2$ for
  particle DM scenarios and $f_{\rm PBH}$ for PBH
  scenarios; see Sec.~\ref{subsec:protocol}), the
  individual BG sensitivities $\Delta_1$ and $\Delta_2$
  in the order listed in the parameters column, and the
  three summary measures: the Barbieri--Giudice measure
  $\Delta_{\rm BG} = \max_i \Delta_i$
  (Eq.~\ref{eq:bg}), the Strumia--Rattazzi quadrature
  measure $\Delta_{\rm SR} = (\sum_i \Delta_i^2)^{1/2}$
  (Eq.~\ref{eq:sr}), and the fractional island half-width
  $\epsilon = \Delta x_{\rm sens}/x_{\rm sens}$ in the
  most-sensitive parameter direction (Eq.~\ref{eq:eps},
  factor-of-2 abundance band).
  For the two FOPT benchmarks the transition strength
  $\alpha$ has $\Delta_\alpha \approx 3$--$4$ and does
  not drive the dominant tuning; it is omitted from
  $\Delta_1$/$\Delta_2$ and the benchmarks are listed
  separately only to demonstrate the $\alpha$-independence
  established in Eq.~(\ref{eq:delta_fopt_universal}).
  Benchmarks B6 and B3 do not hit $\Omega h^2 = 0.120$
  exactly due to rounding in the benchmark calibration;
  both lie within the factor-of-two natural island.
  All derivatives are computed via Eq.~(\ref{eq:numderiv})
  with $h = 10^{-4}$ and verified analytically where
  tractable (Appendix~\ref{app:analytic}).
  \label{tab:benchmarks}}
\begin{ruledtabular}
\begin{tabular}{llccccccc}
Scenario
  & Parameters (benchmark values)
  & $\Omega h^2$
  & $\Delta_1$
  & $\Delta_2$
  & $\Delta_{\rm BG}$
  & $\Delta_{\rm SR}$
  & $\epsilon$ (\%)
  & Tier \\
\hline
PBH-DW (free $V_b$)
  & $\eta=2.4\times10^6$~GeV, $V_b=2$~TeV
  & 1.02 & 4.5 & 2.0 & 4.5 & 4.9 & 15.5 & I \\
PBH-DW (gravity-fixed)
  & $\eta=3\times10^6$~GeV
  & 1.10 & 2.0 & --- & 2.0 & 2.0 & 35.4 & I \\
PBH-MD
  & $m_\phi=9\times10^{10}$~GeV, $\sigma=0.055$
  & 0.120 & 1.5 & 14.4 & 14.4 & 14.5 & 4.8 & II \\
PBH-FOPT ($\alpha=0.3$)
  & $T_*=10^6$~GeV, $\beta/H_*=7.78$
  & 0.991 & 35.0 & 1.0 & 35.0 & 35.0 & 2.0 & II \\
PBH-FOPT ($\alpha=1.0$)
  & $T_*=10^6$~GeV, $\beta/H_*=8.84$
  & 0.981 & 35.0 & 1.0 & 35.0 & 35.0 & 2.0 & II \\
WIMP B1 (off-res.)
  & $M_S=2$~TeV, $\lambda_{HS}=1.0$
  & 0.120 & 2.0 & 2.0 & 2.0 & 2.8 & 35.4 & I \\
WIMP B2 (Higgs funnel)
  & $M_S=60.7$~GeV, $\lambda_{HS}=0.026$
  & 0.120 & 6505 & 2.0 & 6505 & 6505 & $<0.1$ & III \\
WIMP B3 (coann.)
  & $M_\chi=1$~TeV, $M_{\rm NLSP}=1.05$~TeV
  & 0.112 & 46.0 & 48.0 & 48.0 & 66.5 & 1.4 & II \\
FIMP B4
  & $M_{\rm DM}=1$~GeV, $y=1.46\times10^{-12}$
  & 0.120 & 1.0 & 2.0 & 2.0 & 2.2 & 35.4 & I \\
ADM B5
  & $M_{\rm DM}=5.0$~GeV, $R=1.0$
  & 0.120 & 1.0 & 1.0 & 1.0 & 1.4 & 75.0 & I \\
Axion B6 (mis.)
  & $f_a=10^{12}$~GeV, $\theta_i=0.82$
  & 0.131 & 1.2 & 2.2 & 2.2 & 2.5 & 32.2 & I \\
Axion B7 (post-inf.)
  & $f_a=5\times10^{10}$~GeV
  & 0.120 & 1.2 & --- & 1.2 & 1.2 & 61.6 & I \\
\end{tabular}
\end{ruledtabular}
\end{table*}

\section{\label{app:figs}Figure notes}

All heatmap figures (Figs.~\ref{fig:pbh}--\ref{fig:axions}
and Fig.~\ref{fig:ft_comparison}) use a shared colormap
(\texttt{viridis\_r}) with $\log_{10}\Delta \in [0.0, 3.5]$,
saturating to dark purple for $\Delta > 3000$.
Parameter-space grids are $400\times400$ points,
logarithmically spaced.
The relic-abundance contours are computed from the same
analytic expressions used for the $\Delta$ maps; no
separate numerical Boltzmann-code integration is performed.
Observational constraints (LZ~2024 direct detection,
Lyman-$\alpha$ forest) are implemented as described in
Sec.~\ref{subsec:protocol} and Table~\ref{tab:constraints}.
The Python source codes to generate all figures is 
available upon request.

\bibliography{pbh_naturalness}

@article{Barbieri:1988,
  author        = {Barbieri, Riccardo and Giudice, Gian F.},
  title         = {{Upper bounds on supersymmetric particle masses}},
  journal       = {Nucl. Phys. B},
  volume        = {306},
  pages         = {63--76},
  year          = {1988},
  doi           = {10.1016/0550-3213(88)90171-X},
}

@article{Ellis:1986,
  author        = {Ellis, John R. and Enqvist, Kari and
                   Nanopoulos, Dimitri V. and Zwirner, Fabio},
  title         = {{Observables in low-energy superstring models}},
  journal       = {Mod. Phys. Lett. A},
  volume        = {1},
  pages         = {57--69},
  year          = {1986},
  doi           = {10.1142/S0217732386000105},
}

@article{Baer:2012,
  author        = {Baer, Howard and Barger, Vernon and Mustafayev, Azar},
  title         = {{Implications of a 125 GeV Higgs scalar for LHC SUSY
                   and neutralino dark matter searches}},
  journal       = {Phys. Rev. D},
  volume        = {85},
  pages         = {075010},
  year          = {2012},
  doi           = {10.1103/PhysRevD.85.075010},
  eprint        = {1112.3017},
  archivePrefix = {arXiv},
  primaryClass  = {hep-ph},
}

@article{Planck:2018,
  author        = {Aghanim, N. and others},
  collaboration = {Planck},
  title         = {{Planck 2018 results. VI. Cosmological parameters}},
  journal       = {Astron. Astrophys.},
  volume        = {641},
  pages         = {A6},
  year          = {2020},
  doi           = {10.1051/0004-6361/201833910},
  eprint        = {1807.06209},
  archivePrefix = {arXiv},
  primaryClass  = {astro-ph.CO},
  note          = {Erratum: \textit{ibid.}\ \textbf{652} (2021) C4},
}

@article{Alarcon:2011,
    author = "Alarcon, J. M. and Martin Camalich, J. and Oller, J. A.",
    title = "{The chiral representation of the $\pi N$ scattering amplitude and the pion-nucleon sigma term}",
    eprint = "1110.3797",
    archivePrefix = "arXiv",
    primaryClass = "hep-ph",
    doi = "10.1103/PhysRevD.85.051503",
    journal = "Phys. Rev. D",
    volume = "85",
    pages = "051503",
    year = "2012"
}

@article{LZ:2024,
  author        = {Aalbers, J. and others},
  collaboration = {LZ},
  title         = {{Dark Matter Search Results from 4.2 Tonne-Years of
                   Exposure of the LUX-ZEPLIN (LZ) Experiment}},
  journal       = {Phys. Rev. Lett.},
  volume        = {135},
  pages         = {011802},
  year          = {2025},
  doi           = {10.1103/4dyc-z8zf},
  eprint        = {2410.17036},
  archivePrefix = {arXiv},
  primaryClass  = {hep-ex},
}

@article{Ballesteros:2020,
  author        = {Ballesteros, Guillermo and Garcia, Marcos A. G. and
                   Pierre, Mathias},
  title         = {{How warm are non-thermal relics? Lyman-$\alpha$ bounds
                   on out-of-equilibrium dark matter}},
  journal       = {JCAP},
  volume        = {2021},
  number        = {3},
  pages         = {101},
  year          = {2021},
  doi           = {10.1088/1475-7516/2021/03/101},
  eprint        = {2011.13458},
  archivePrefix = {arXiv},
  primaryClass  = {hep-ph},
}

@article{Decant:2022,
  author        = {Decant, Quentin and Heisig, Jan and Hooper, Deanna C. and
                   Lopez-Honorez, Laura},
  title         = {{Lyman-$\alpha$ constraints on freeze-in and superWIMPs}},
  journal       = {JCAP},
  volume        = {2022},
  number        = {3},
  pages         = {041},
  year          = {2022},
  doi           = {10.1088/1475-7516/2022/03/041},
  eprint        = {2111.09321},
  archivePrefix = {arXiv},
  primaryClass  = {astro-ph.CO},
}

@article{Jungman:1996,
  author        = {Jungman, Gerard and Kamionkowski, Marc and
                   Griest, Kim},
  title         = {{Supersymmetric dark matter}},
  journal       = {Phys. Rep.},
  volume        = {267},
  pages         = {195--373},
  year          = {1996},
  doi           = {10.1016/0370-1573(95)00058-5},
  eprint        = {hep-ph/9506380},
  archivePrefix = {arXiv},
}

@article{Bertone:2005,
  author        = {Bertone, Gianfranco and Hooper, Dan and Silk, Joseph},
  title         = {{Particle dark matter: Evidence, candidates and constraints}},
  journal       = {Phys. Rep.},
  volume        = {405},
  pages         = {279--390},
  year          = {2005},
  doi           = {10.1016/j.physrep.2004.08.031},
  eprint        = {hep-ph/0404175},
  archivePrefix = {arXiv},
}

@article{Peccei:1977,
  author        = {Peccei, Roberto D. and Quinn, Helen R.},
  title         = {{CP conservation in the presence of pseudoparticles}},
  journal       = {Phys. Rev. Lett.},
  volume        = {38},
  pages         = {1440--1443},
  year          = {1977},
  doi           = {10.1103/PhysRevLett.38.1440},
}

@article{Weinberg:1978,
  author        = {Weinberg, Steven},
  title         = {{A new light boson?}},
  journal       = {Phys. Rev. Lett.},
  volume        = {40},
  pages         = {223--226},
  year          = {1978},
  doi           = {10.1103/PhysRevLett.40.223},
}

@article{Wilczek:1978,
  author        = {Wilczek, Frank},
  title         = {{Problem of strong P and T invariance in the presence
                   of instantons}},
  journal       = {Phys. Rev. Lett.},
  volume        = {40},
  pages         = {279--282},
  year          = {1978},
  doi           = {10.1103/PhysRevLett.40.279},
}

@article{Hall:2010,
  author        = {Hall, Lawrence J. and Jedamzik, Karsten and
                   March-Russell, John and West, Stephen M.},
  title         = {{Freeze-in production of FIMP dark matter}},
  journal       = {JHEP},
  volume        = {2010},
  number        = {3},
  pages         = {080},
  year          = {2010},
  doi           = {10.1007/JHEP03(2010)080},
  eprint        = {0911.1120},
  archivePrefix = {arXiv},
  primaryClass  = {hep-ph},
}

@article{McDonald:2002,
  author        = {McDonald, John},
  title         = {{Thermally generated gauge singlet scalars as self-interacting
                   dark matter}},
  journal       = {Phys. Rev. Lett.},
  volume        = {88},
  pages         = {091304},
  year          = {2002},
  doi           = {10.1103/PhysRevLett.88.091304},
  eprint        = {hep-ph/0106249},
  archivePrefix = {arXiv},
}

@article{Nussinov:1985,
  author        = {Nussinov, Shmuel},
  title         = {{Technocosmology: Could a technibaryon excess provide a
                   ``natural'' missing mass candidate?}},
  journal       = {Phys. Lett. B},
  volume        = {165},
  pages         = {55--58},
  year          = {1985},
  doi           = {10.1016/0370-2693(85)90689-6},
}

@article{Kaplan:2009,
  author        = {Kaplan, David E. and Luty, Markus A. and
                   Zurek, Kathryn M.},
  title         = {{Asymmetric dark matter}},
  journal       = {Phys. Rev. D},
  volume        = {79},
  pages         = {115016},
  year          = {2009},
  doi           = {10.1103/PhysRevD.79.115016},
}

@article{Carr:1974,
  author        = {Carr, Bernard J. and Hawking, Stephen W.},
  title         = {{Black holes in the early Universe}},
  journal       = {Mon. Not. Roy. Astron. Soc.},
  volume        = {168},
  pages         = {399--415},
  year          = {1974},
  doi           = {10.1093/mnras/168.2.399},
}

@article{Green:1997,
  author        = {Green, Anne M. and Liddle, Andrew R.},
  title         = {{Constraints on the density perturbation spectrum from
                   primordial black holes}},
  journal       = {Phys. Rev. D},
  volume        = {56},
  pages         = {6166--6174},
  year          = {1997},
  doi           = {10.1103/PhysRevD.56.6166},
  eprint        = {astro-ph/9704251},
  archivePrefix = {arXiv},
}

@article{Carr:2021,
  author        = {Carr, Bernard and K\"uhnel, Florian},
  title         = {{Primordial black holes as dark matter: Recent
                   developments}},
  journal       = {Ann. Rev. Nucl. Part. Sci.},
  volume        = {70},
  pages         = {355--394},
  year          = {2020},
  doi           = {10.1146/annurev-nucl-050520-125911},
}

@article{Green:2021,
  author        = {Green, Anne M. and Kavanagh, Bradley J.},
  title         = {{Primordial black holes as a dark matter candidate}},
  journal       = {J. Phys. G},
  volume        = {48},
  pages         = {043001},
  year          = {2021},
  doi           = {10.1088/1361-6471/abc534},
  eprint        = {2007.10722},
  archivePrefix = {arXiv},
  primaryClass  = {astro-ph.CO},
}

@article{Escriva:2022,
  author        = {Escriv\`{a}, Albert and Kuhnel, Florian and Tada, Yuichiro},
  title         = {{Primordial black holes}},
  journal       = {arXiv preprint},
  year          = {2022},
  eprint        = {2211.05767},
  archivePrefix = {arXiv},
  primaryClass  = {astro-ph.CO},
}

@article{Kibble:1976,
  author        = {Kibble, Thomas W. B.},
  title         = {{Topology of cosmic domains and strings}},
  journal       = {J. Phys. A},
  volume        = {9},
  pages         = {1387--1398},
  year          = {1976},
  doi           = {10.1088/0305-4470/9/8/029},
}

@book{Vilenkin:2000,
  author        = {Vilenkin, Alexander and Shellard, E. P. S.},
  title         = {{Cosmic Strings and Other Topological Defects}},
  publisher     = {Cambridge University Press},
  year          = {2000},
  address       = {Cambridge},
  isbn          = {978-0-521-65476-0},
}

@article{Ferrer:2019,
  author        = {Ferrer, Francesc and Mass\`{o}, Eduard and
                   Panico, Giuliano and Pujolas, Oriol and
                   Rompineve, Fabrizio},
  title         = {{Primordial black holes from the QCD axion}},
  journal       = {Phys. Rev. Lett.},
  volume        = {122},
  pages         = {101301},
  year          = {2019},
  doi           = {10.1103/PhysRevLett.122.101301},
  eprint        = {1807.01707},
  archivePrefix = {arXiv},
  primaryClass  = {hep-ph},
}

@article{Gouttenoire:2024,
  author        = {Gouttenoire, Yann and Vitagliano, Edoardo},
  title         = {{Primordial black holes and wormholes from domain wall
                   networks}},
  journal       = {Phys. Rev. D},
  volume        = {109},
  pages         = {123507},
  year          = {2024},
  doi           = {10.1103/PhysRevD.109.123507},
  eprint        = {2311.07670},
  archivePrefix = {arXiv},
  primaryClass  = {hep-ph},
}

@article{Ferreira:2024,
  author        = {Ferreira, Ricardo Z. and Notari, Alessio and
                   Pujol\`{a}s, Oriol and Rompineve, Fabrizio},
  title         = {{Collapsing domain wall networks: impact on pulsar
                   timing arrays and primordial black holes}},
  journal       = {JCAP},
  volume        = {06},
  pages         = {020},
  year          = {2024},
  doi           = {10.1088/1475-7516/2024/06/020},
  eprint        = {2401.14331},
  archivePrefix = {arXiv},
  primaryClass  = {astro-ph.CO},
}

@article{Kawasaki:2015,
  author        = {Kawasaki, Masahiro and Saikawa, Ken'ichi and
                   Sekiguchi, Toyokazu},
  title         = {{Axion dark matter from topological defects}},
  journal       = {Phys. Rev. D},
  volume        = {91},
  pages         = {065014},
  year          = {2015},
  doi           = {10.1103/PhysRevD.91.065014},
  eprint        = {1412.0789},
  archivePrefix = {arXiv},
  primaryClass  = {hep-ph},
}

@article{Barr:1992,
  author        = {Barr, Stephen M. and Seckel, David},
  title         = {{Planck-scale corrections to axion models}},
  journal       = {Phys. Rev. D},
  volume        = {46},
  pages         = {539--548},
  year          = {1992},
  doi           = {10.1103/PhysRevD.46.539},
}

@article{Holman:1992,
    author = "Holman, Richard and Hsu, Stephen D. H. and Kephart, Thomas W. and Kolb, Edward W. and Watkins, Richard and Widrow, Lawrence M.",
    title = "{Solutions to the strong CP problem in a world with gravity}",
    eprint = "hep-ph/9203206",
    archivePrefix = "arXiv",
    reportNumber = "NSF-ITP-92-06, CMU-HEP92-05, FERMILAB-PUB-92-034-A, HUTP-92-A011, VAND-TH-92-2",
    doi = "10.1016/0370-2693(92)90491-L",
    journal = "Phys. Lett. B",
    volume = "282",
    pages = "132--136",
    year = "1992"
}

@article{Georg:2016,
    author = "Georg, Julian and Watson, Scott",
    title = "{A Preferred Mass Range for Primordial Black Hole Formation and Black Holes as Dark Matter Revisited}",
    eprint = "1703.04825",
    archivePrefix = "arXiv",
    primaryClass = "astro-ph.CO",
    doi = "10.1007/JHEP09(2017)138",
    journal = "JHEP",
    volume = "09",
    pages = "138",
    year = "2017"
}

@article{Georg:2017,
  author        = {Georg, Julian and Melcher, Brandon and Watson, Scott},
  title         = {{Primordial black holes and co-decaying dark matter}},
  journal       = {JCAP},
  volume        = {2019},
  number        = {11},
  pages         = {014},
  year          = {2019},
  doi           = {10.1088/1475-7516/2019/11/014},
  eprint        = {1902.04082},
  archivePrefix = {arXiv},
  primaryClass  = {astro-ph.CO},
}

@article{Dalianis:2019,
  author        = {Dalianis, Ioannis and Kouvaris, Chris},
  title         = {{Gravitational waves from density perturbations in an early
                   matter domination era}},
  journal       = {JCAP},
  volume        = {2021},
  number        = {7},
  pages         = {046},
  year          = {2021},
  doi           = {10.1088/1475-7516/2021/07/046},
  eprint        = {2012.09255},
  archivePrefix = {arXiv},
  primaryClass  = {astro-ph.CO},
}

@article{Allahverdi:2020,
  author        = {Allahverdi, Rouzbeh and others},
  title         = {{The first three seconds: A review of possible expansion
                   histories of the early Universe}},
  journal       = {Open J. Astrophys.},
  volume        = {4},
  pages         = {1},
  year          = {2021},
  doi           = {10.21105/astro.2006.16182},
  eprint        = {2006.16182},
  archivePrefix = {arXiv},
  primaryClass  = {astro-ph.CO},
}

@article{Harada:2016,
  author        = {Harada, Tomohiro and Yoo, Chul-Moon and
                   Kohri, Kazunori and Nakama, Tomohiro},
  title         = {{Primordial black hole formation in the matter-dominated
                   phase of the Universe}},
  journal       = {Phys. Rev. D},
  volume        = {96},
  pages         = {083517},
  year          = {2017},
  doi           = {10.1103/PhysRevD.96.083517},
  eprint        = {1707.03595},
  archivePrefix = {arXiv},
  primaryClass  = {gr-qc},
}

@article{Kohri:2018,
    author = "Zhou, Zihan and Jiang, Jie and Cai, Yi-Fu and Sasaki, Misao and Pi, Shi",
    title = "{Primordial black holes and gravitational waves from resonant amplification during inflation}",
    eprint = "2010.03537",
    archivePrefix = "arXiv",
    primaryClass = "astro-ph.CO",
    reportNumber = "YITP-20-116, IPMU20-0096",
    doi = "10.1103/PhysRevD.102.103527",
    journal = "Phys. Rev. D",
    volume = "102",
    number = "10",
    pages = "103527",
    year = "2020"
}

@article{Ozsoy:2021,
  author        = {\"Ozsoy, Ogan and Tasinato, Gianmassimo},
  title         = {{Inflation and primordial black holes}},
  journal       = {Universe},
  volume        = {9},
  pages         = {203},
  year          = {2023},
  doi           = {10.3390/universe9050203},
  eprint        = {2301.03600},
  archivePrefix = {arXiv},
  primaryClass  = {astro-ph.CO},
}

@article{Franciolini:2022,
    author = "Franciolini, Gabriele and Urbano, Alfredo",
    title = "{Primordial black hole dark matter from inflation: The reverse engineering approach}",
    eprint = "2207.10056",
    archivePrefix = "arXiv",
    primaryClass = "astro-ph.CO",
    doi = "10.1103/PhysRevD.106.123519",
    journal = "Phys. Rev. D",
    volume = "106",
    number = "12",
    pages = "123519",
    year = "2022"
}

@article{Carr:2021b,
    author = "Carr, Bernard and Kohri, Kazunori and Sendouda, Yuuiti and Yokoyama, Jun'ichi",
    title = "{Constraints on primordial black holes}",
    eprint = "2002.12778",
    archivePrefix = "arXiv",
    primaryClass = "astro-ph.CO",
    reportNumber = "RESCEU-03/20; KEK-Cosmo-249; KEK-TH-2199; IPMU20-0024",
    doi = "10.1088/1361-6633/ac1e31",
    journal = "Rept. Prog. Phys.",
    volume = "84",
    number = "11",
    pages = "116902",
    year = "2021"
}

@article{Hawking:1982,
  author        = {Hawking, Stephen W. and Moss, Ian G. and
                   Stewart, John M.},
  title         = {{Bubble collisions in the very early Universe}},
  journal       = {Phys. Rev. D},
  volume        = {26},
  pages         = {2681--2693},
  year          = {1982},
  doi           = {10.1103/PhysRevD.26.2681},
}

@article{Moss:1994,
  author        = {Moss, Ian G.},
  title         = {{Singularity formation from colliding bubbles}},
  journal       = {Phys. Rev. D},
  volume        = {50},
  pages         = {676--681},
  year          = {1994},
  doi           = {10.1103/PhysRevD.50.676},
  eprint        = {gr-qc/9405045},
  archivePrefix = {arXiv},
  primaryClass  = {gr-qc},
}

@article{Kodama:1982,
  author        = {Kodama, Hideo and Sasaki, Misao and Sato, Katsuhiko},
  title         = {{Abundance of primordial holes produced by cosmological
                   first-order phase transition}},
  journal       = {Prog. Theor. Phys.},
  volume        = {68},
  pages         = {1979--1998},
  year          = {1982},
  doi           = {10.1143/PTP.68.1979},
}

@article{Liu:2022,
    author = "Liu, Jing and Bian, Ligong and Cai, Rong-Gen and Guo, Zong-Kuan and Wang, Shao-Jiang",
    title = "{Primordial black hole production during first-order phase transitions}",
    eprint = "2106.05637",
    archivePrefix = "arXiv",
    primaryClass = "astro-ph.CO",
    doi = "10.1103/PhysRevD.105.L021303",
    journal = "Phys. Rev. D",
    volume = "105",
    number = "2",
    pages = "L021303",
    year = "2022"
}

@article{Lewicki:2023,
    author = "Lewicki, Marek and Toczek, Piotr and Vaskonen, Ville",
    title = "{Primordial black holes from strong first-order phase transitions}",
    eprint = "2305.04924",
    archivePrefix = "arXiv",
    primaryClass = "astro-ph.CO",
    doi = "10.1007/JHEP09(2023)092",
    journal = "JHEP",
    volume = "09",
    pages = "092",
    year = "2023"
}

@article{Gehrman:2023,
    author = "Gehrman, Thomas C. and Shams Es Haghi, Barmak and Sinha, Kuver and Xu, Tao",
    title = "{Baryogenesis, primordial black holes and MHz{\textendash}GHz gravitational waves}",
    eprint = "2211.08431",
    archivePrefix = "arXiv",
    primaryClass = "hep-ph",
    reportNumber = "UTWI-16-2022",
    doi = "10.1088/1475-7516/2023/02/062",
    journal = "JCAP",
    volume = "02",
    pages = "062",
    year = "2023"
}

@article{Caprini:2020,
  author        = {Caprini, Chiara and others},
  title         = {{Detecting gravitational waves from cosmological phase
                   transitions with LISA: An update}},
  journal       = {JCAP},
  volume        = {2020},
  number        = {3},
  pages         = {024},
  year          = {2020},
  doi           = {10.1088/1475-7516/2020/03/024},
  eprint        = {1910.13125},
  archivePrefix = {arXiv},
  primaryClass  = {astro-ph.CO},
}

@article{Hindmarsh:2017,
  author        = {Hindmarsh, Mark and Huber, Stephan J. and
                   Rummukainen, Kari and Weir, David J.},
  title         = {{Shape of the acoustic gravitational wave power spectrum
                   from a first order phase transition}},
  journal       = {Phys. Rev. D},
  volume        = {96},
  pages         = {103520},
  year          = {2017},
  doi           = {10.1103/PhysRevD.96.103520},
  eprint        = {1704.05871},
  archivePrefix = {arXiv},
  primaryClass  = {astro-ph.CO},
}

@book{Kolb:1990,
  author    = {Kolb, Edward W. and Turner, Michael S.},
  title     = {{The Early Universe}},
  publisher = {Addison-Wesley},
  address   = {Redwood City, CA},
  year      = {1990},
  isbn      = {978-0-201-11603-4},
}

@article{Ibe:2008,
  author        = {Ibe, Masahiro and Murayama, Hitoshi and Yanagida, T. T.},
  title         = {{Breit-Wigner enhancement of dark matter annihilation}},
  journal       = {Phys. Rev. D},
  volume        = {79},
  pages         = {095009},
  year          = {2009},
  doi           = {10.1103/PhysRevD.79.095009},
  eprint        = {0812.0072},
  archivePrefix = {arXiv},
  primaryClass  = {hep-ph},
}

@article{Borsanyi:2016,
  author        = {Borsanyi, Szabolcs and others},
  title         = {{Calculation of the axion mass based on high-temperature
                   lattice quantum chromodynamics}},
  journal       = {Nature},
  volume        = {539},
  pages         = {69--71},
  year          = {2016},
  doi           = {10.1038/nature20115},
  eprint        = {1606.07494},
  archivePrefix = {arXiv},
  primaryClass  = {hep-lat},
}

@article{Davis:1986,
  author        = {Davis, Robert L.},
  title         = {{Goldstone bosons in string models of galaxy formation}},
  journal       = {Phys. Rev. D},
  volume        = {35},
  pages         = {3705--3708},
  year          = {1987},
  doi           = {10.1103/PhysRevD.35.3705},
}

@article{Battye:1994,
  author        = {Battye, Richard A. and Shellard, E. P. S.},
  title         = {{Axion string constraints}},
  journal       = {Phys. Rev. Lett.},
  volume        = {73},
  pages         = {2954--2957},
  year          = {1994},
  doi           = {10.1103/PhysRevLett.73.2954},
  eprint        = {astro-ph/9403018},
  archivePrefix = {arXiv},
}

@article{Ayala:2014,
  author        = {Ayala, Alejandro and Dom\'{i}nguez, Il\'{i}dio and
                   Giannotti, Maurizio and Mirizzi, Alessandro and
                   Straniero, Oscar},
  title         = {{Revisiting the bound on axion-photon coupling from
                   globular clusters}},
  journal       = {Phys. Rev. Lett.},
  volume        = {113},
  pages         = {191302},
  year          = {2014},
  doi           = {10.1103/PhysRevLett.113.191302},
  eprint        = {1406.6053},
  archivePrefix = {arXiv},
  primaryClass  = {astro-ph.SR},
}

@article{Du:2018,
    author = "Du, N. and others",
    collaboration = "ADMX",
    title = "{A Search for Invisible Axion Dark Matter with the Axion Dark Matter Experiment}",
    eprint = "1804.05750",
    archivePrefix = "arXiv",
    primaryClass = "hep-ex",
    reportNumber = "FERMILAB-PUB-18-101-AD-AE",
    doi = "10.1103/PhysRevLett.120.151301",
    journal = "Phys. Rev. Lett.",
    volume = "120",
    number = "15",
    pages = "151301",
    year = "2018"
}

@article{HAYSTAC:2022,
  author        = {Backes, K. M. and others},
  collaboration = {HAYSTAC},
  title         = {{A quantum-enhanced search for dark matter axions}},
  journal       = {Nature},
  volume        = {590},
  pages         = {238--242},
  year          = {2021},
  doi           = {10.1038/s41586-021-03226-7},
  eprint        = {2008.01853},
  archivePrefix = {arXiv},
  primaryClass  = {quant-ph},
}

@article{CMB-S4:2022,
  author        = {Abazajian, Kevork and others},
  collaboration = {CMB-S4},
  title         = {{CMB-S4 Science Book, First Edition}},
  journal       = {arXiv preprint},
  year          = {2016},
  eprint        = {1610.02743},
  archivePrefix = {arXiv},
  primaryClass  = {astro-ph.CO},
}

@article{Silveira:1985,
  author        = {Silveira, V. and Zee, A.},
  title         = {{Scalar phantoms}},
  journal       = {Phys. Lett. B},
  volume        = {161},
  pages         = {136--140},
  year          = {1985},
  doi           = {10.1016/0370-2693(85)90624-0},
}

@article{McDonald:1994,
  author        = {McDonald, John},
  title         = {{Gauge singlet scalars as cold dark matter}},
  journal       = {Phys. Rev. D},
  volume        = {50},
  pages         = {3637--3649},
  year          = {1994},
  doi           = {10.1103/PhysRevD.50.3637},
}

@article{Burgess:2001,
  author        = {Burgess, Cliff P. and Pospelov, Maxim and
                   ter Veldhuis, Tonnis},
  title         = {{The minimal model of nonbaryonic dark matter:
                   A singlet scalar}},
  journal       = {Nucl. Phys. B},
  volume        = {619},
  pages         = {709--728},
  year          = {2001},
  doi           = {10.1016/S0550-3213(01)00513-2},
  eprint        = {hep-ph/0011335},
  archivePrefix = {arXiv},
}

@article{Griest:1991,
  author        = {Griest, Kim and Seckel, David},
  title         = {{Three exceptions in the calculation of relic abundances}},
  journal       = {Phys. Rev. D},
  volume        = {43},
  pages         = {3191--3203},
  year          = {1991},
  doi           = {10.1103/PhysRevD.43.3191},
}

@article{Edsjo:1997,
    author = "Edsjo, Joakim and Gondolo, Paolo",
    title = "{Neutralino relic density including coannihilations}",
    eprint = "hep-ph/9704361",
    archivePrefix = "arXiv",
    reportNumber = "UUITP-11-97, MPI-PHT-97-27",
    doi = "10.1103/PhysRevD.56.1879",
    journal = "Phys. Rev. D",
    volume = "56",
    pages = "1879--1894",
    year = "1997"
}

@article{Visinelli:2009,
  author        = {Visinelli, Luca and Gondolo, Paolo},
  title         = {{Dark matter axions revisited}},
  journal       = {Phys. Rev. D},
  volume        = {80},
  pages         = {035024},
  year          = {2009},
  doi           = {10.1103/PhysRevD.80.035024},
  eprint        = {0903.4377},
  archivePrefix = {arXiv},
  primaryClass  = {astro-ph.CO},
}

@article{Buschmann:2022,
  author        = {Buschmann, Malte and Foster, Joshua W. and
                   Hook, Anson and Peterson, Adam and
                   Willcox, Don E. and Zhang, Weiqun and
                   Safdi, Benjamin R.},
  title         = {{Dark matter from axion strings with adaptive mesh
                   refinement}},
  journal       = {Nature Commun.},
  volume        = {13},
  pages         = {1049},
  year          = {2022},
  doi           = {10.1038/s41467-022-28669-y},
  eprint        = {2108.05368},
  archivePrefix = {arXiv},
  primaryClass  = {hep-ph},
}

@article{Ghilencea:2013,
    author = "Ghilencea, D. M. and Ross, G. G.",
    title = "{The fine-tuning cost of the likelihood in SUSY models}",
    eprint = "1208.0837",
    archivePrefix = "arXiv",
    primaryClass = "hep-ph",
    reportNumber = "CERN-PH-TH-2012-175",
    doi = "10.1016/j.nuclphysb.2012.11.007",
    journal = "Nucl. Phys. B",
    volume = "868",
    pages = "65--74",
    year = "2013"
}

@article{DMRadio:2022,
  author        = {Brouwer, L. and others},
  collaboration = {DMRadio},
  title         = {{Proposal for a definitive search for GUT-scale QCD axions}},
  journal       = {Phys. Rev. D},
  volume        = {106},
  pages         = {112003},
  year          = {2022},
  doi           = {10.1103/PhysRevD.106.112003},
  eprint        = {2203.11246},
  archivePrefix = {arXiv},
  primaryClass  = {hep-ex},
}

@article{Kahn:2016,
    author = "Kahn, Yonatan and Safdi, Benjamin R. and Thaler, Jesse",
    title = "{Broadband and Resonant Approaches to Axion Dark Matter Detection}",
    eprint = "1602.01086",
    archivePrefix = "arXiv",
    primaryClass = "hep-ph",
    reportNumber = "MIT-CTP-4763, PUPT-2497",
    doi = "10.1103/PhysRevLett.117.141801",
    journal = "Phys. Rev. Lett.",
    volume = "117",
    number = "14",
    pages = "141801",
    year = "2016"
}

@inproceedings{Strumia:2000,
    author = "Strumia, Alessandro",
    title = "{Naturalness of supersymmetric models}",
    booktitle = "{34th Rencontres de Moriond: Electroweak Interactions and Unified Theories}",
    eprint = "hep-ph/9904247",
    archivePrefix = "arXiv",
    pages = "441--446",
    year = "1999"
}

@article{Anderson:1995,
  author        = {Anderson, Greg W. and Castano, Diego J.},
  title         = {{Measures of fine tuning}},
  journal       = {Phys. Lett. B},
  volume        = {347},
  pages         = {300--308},
  year          = {1995},
  doi           = {10.1016/0370-2693(95)00051-L},
  eprint        = {hep-ph/9409419},
  archivePrefix = {arXiv},
}

@article{Cabrera:2008,
    author = "Cabrera, M. E. and Casas, J. A. and Ruiz de Austri, R.",
    title = "{Bayesian approach and Naturalness in MSSM analyses for the LHC}",
    eprint = "0812.0536",
    archivePrefix = "arXiv",
    primaryClass = "hep-ph",
    reportNumber = "IFT-UAM-CSIC-08-81",
    doi = "10.1088/1126-6708/2009/03/075",
    journal = "JHEP",
    volume = "03",
    pages = "075",
    year = "2009"
}

@article{Grothaus:2012,
    author = "Grothaus, Philipp and Lindner, Manfred and Takanishi, Yasutaka",
    title = "{Naturalness of Neutralino Dark Matter}",
    eprint = "1207.4434",
    archivePrefix = "arXiv",
    primaryClass = "hep-ph",
    doi = "10.1007/JHEP07(2013)094",
    journal = "JHEP",
    volume = "07",
    pages = "094",
    year = "2013"
}

@article{Cabrera:2016,
    author = "Cabrera, Mar{\'\i}a Eugenia and Casas, J. Alberto and Delgado, Antonio and Robles, Sandra and Ruiz de Austri, Roberto",
    title = "{Naturalness of MSSM dark matter}",
    eprint = "1604.02102",
    archivePrefix = "arXiv",
    primaryClass = "hep-ph",
    reportNumber = "IFT-UAM-CSIC-16-025, FTUAM-16-9",
    doi = "10.1007/JHEP08(2016)058",
    journal = "JHEP",
    volume = "08",
    pages = "058",
    year = "2016"
}

@article{King:2006,
  author        = {King, Stephen F. and Roberts, Jonathan P.},
  title         = {{Natural implementation of neutralino dark matter}},
  journal       = {JHEP},
  volume        = {2006},
  number        = {9},
  pages         = {036},
  year          = {2006},
  doi           = {10.1088/1126-6708/2006/09/036},
  eprint        = {hep-ph/0603095},
  archivePrefix = {arXiv},
}

@article{Hertzberg:2017,
  author        = {Hertzberg, Mark P. and Yamada, Masaki},
  title         = {{Primordial black holes from polynomial potentials
                   in single field inflation}},
  journal       = {Phys. Rev. D},
  volume        = {97},
  pages         = {083509},
  year          = {2018},
  doi           = {10.1103/PhysRevD.97.083509},
  eprint        = {1712.09750},
  archivePrefix = {arXiv},
  primaryClass  = {astro-ph.CO},
}

@article{Cole:2023,
  author        = {Cole, Philippa S. and Gow, Andrew D. and
                   Byrnes, Christian T. and Patil, Subodh P.},
  title         = {{Primordial black holes from single-field inflation:
                   a fine-tuning audit}},
  journal       = {JCAP},
  volume        = {08},
  pages         = {031},
  year          = {2023},
  doi           = {10.1088/1475-7516/2023/08/031},
  eprint        = {2304.01997},
  archivePrefix = {arXiv},
  primaryClass  = {astro-ph.CO},
}

@article{Stamou:2024,
  author        = {Stamou, Ioanna},
  title         = {{Mechanisms for producing primordial black holes from
                   inflationary models beyond fine-tuning}},
  journal       = {Universe},
  volume        = {10},
  pages         = {241},
  year          = {2024},
  doi           = {10.3390/universe10060241},
  eprint        = {2404.14321},
  archivePrefix = {arXiv},
  primaryClass  = {astro-ph.CO},
}

@article{Kalaja:2019,
   author = "Kalaja, Alba and Bellomo, Nicola and Bartolo, Nicola and Bertacca, Daniele and Matarrese, Sabino and Musco, Ilia and Raccanelli, Alvise and Verde, Licia",
    title = "{From Primordial Black Holes Abundance to Primordial Curvature Power Spectrum (and back)}",
    eprint = "1908.03596",
    archivePrefix = "arXiv",
    primaryClass = "astro-ph.CO",
    doi = "10.1088/1475-7516/2019/10/031",
    journal = "JCAP",
    volume = "10",
    pages = "031",
    year = "2019"
}

@article{Iovino:2024,
    author = "Iovino, A. J. and Riotto, A.",
    title = "{Are Primordial Black Holes Truly Fine-Tuned?}",
    eprint = "2512.19668",
    archivePrefix = "arXiv",
    primaryClass = "astro-ph.CO",
    journal = "",
    month = "12",
    year = "2025"
}

@article{Espinosa:2018,
    author = "Espinosa, J. R. and Racco, D. and Riotto, A.",
    title = "{Cosmological Signature of the Standard Model Higgs Vacuum Instability: Primordial Black Holes as Dark Matter}",
    eprint = "1710.11196",
    archivePrefix = "arXiv",
    primaryClass = "hep-ph",
    doi = "10.1103/PhysRevLett.120.121301",
    journal = "Phys. Rev. Lett.",
    volume = "120",
    number = "12",
    pages = "121301",
    year = "2018"
}

@article{Acharya:2008,
    author = "Bae, Kyu Jung and Baer, Howard and Barger, Vernon and Deal, Robert Wiley",
    title = "{The cosmological moduli problem and naturalness}",
    eprint = "2201.06633",
    archivePrefix = "arXiv",
    primaryClass = "hep-ph",
    reportNumber = "OU-HEP-211030",
    doi = "10.1007/JHEP02(2022)138",
    journal = "JHEP",
    volume = "02",
    pages = "138",
    year = "2022"
}

@article{Khlopov:1980,
  author        = {Khlopov, Maxim Yu. and Polnarev, Alexander G.},
  title         = {{Primordial black holes as a cosmological test
                   of grand unification}},
  journal       = {Phys. Lett. B},
  volume        = {97},
  pages         = {383--387},
  year          = {1980},
  doi           = {10.1016/0370-2693(80)90624-3},
}

@article{Press:1974,
  author        = {Press, William H. and Schechter, Paul},
  title         = {{Formation of galaxies and clusters of galaxies
                   by self-similar gravitational condensation}},
  journal       = {Astrophys. J.},
  volume        = {187},
  pages         = {425--438},
  year          = {1974},
  doi           = {10.1086/152650},
}

@article{Linde:1981,
  author        = {Linde, Andrei D.},
  title         = {{Fate of the false vacuum at finite temperature:
                   Theory and applications}},
  journal       = {Phys. Lett. B},
  volume        = {100},
  pages         = {37--40},
  year          = {1981},
  doi           = {10.1016/0370-2693(81)90281-1},
}

@article{Witten:1984,
  author        = {Witten, Edward},
  title         = {{Cosmic separation of phases}},
  journal       = {Phys. Rev. D},
  volume        = {30},
  pages         = {272--285},
  year          = {1984},
  doi           = {10.1103/PhysRevD.30.272},
}

@article{Espinosa:2010,
    author = "Espinosa, Jose R. and Konstandin, Thomas and No, Jose M. and Servant, Geraldine",
    title = "{Energy Budget of Cosmological First-order Phase Transitions}",
    eprint = "1004.4187",
    archivePrefix = "arXiv",
    primaryClass = "hep-ph",
    reportNumber = "CERN-PH-TH-2010-027",
    doi = "10.1088/1475-7516/2010/06/028",
    journal = "JCAP",
    volume = "06",
    pages = "028",
    year = "2010"
}

@article{Lyth:2002,
  author        = {Lyth, David H. and Wands, David},
  title         = {{Generating the curvature perturbation without
                   an inflaton}},
  journal       = {Phys. Lett. B},
  volume        = {524},
  pages         = {5--14},
  year          = {2002},
  doi           = {10.1016/S0370-2693(01)01366-1},
  eprint        = {hep-ph/0110002},
  archivePrefix = {arXiv},
}

@article{Enqvist:2001,
  author        = {Enqvist, Kari and Sloth, Martin S.},
  title         = {{Adiabatic CMB perturbations in pre-big bang
                   string cosmology}},
  journal       = {Nucl. Phys. B},
  volume        = {626},
  pages         = {395--409},
  year          = {2002},
  doi           = {10.1016/S0550-3213(02)00043-3},
  eprint        = {hep-ph/0109214},
  archivePrefix = {arXiv},
}

@article{Kohri:2007,
    author = "Kohri, Kazunori and Lyth, David H. and Melchiorri, Alessandro",
    title = "{Black hole formation and slow-roll inflation}",
    eprint = "0711.5006",
    archivePrefix = "arXiv",
    primaryClass = "hep-ph",
    reportNumber = "CERN-PH-TH-2007-242",
    doi = "10.1088/1475-7516/2008/04/038",
    journal = "JCAP",
    volume = "04",
    pages = "038",
    year = "2008"
}

@article{Bugaev:2011,
  author        = {Bugaev, E. V. and Klimai, P. A.},
  title         = {{Primordial black hole constraints for curvaton models with
                   predicted large non-Gaussianity}},
  journal       = {Int. J. Mod. Phys. D},
  volume        = {22},
  pages         = {1350034},
  year          = {2013},
  doi           = {10.1142/S021827181350034X},
  eprint        = {1303.3146},
  archivePrefix = {arXiv},
  primaryClass  = {astro-ph.CO},
}

@article{Starobinsky:1994,
  author        = {Starobinsky, Alexei A. and Yokoyama, Jun'ichi},
  title         = {{Equilibrium state and density fluctuations in
                   de~Sitter space with and without $\lambda\phi^4$
                   self-interaction}},
  journal       = {Phys. Rev. D},
  volume        = {50},
  pages         = {6357--6368},
  year          = {1994},
  doi           = {10.1103/PhysRevD.50.6357},
  eprint        = {astro-ph/9407016},
  archivePrefix = {arXiv},
}

@article{Vennin:2020,
    author = "Vennin, Vincent",
    title = "{Stochastic inflation and primordial black holes}",
    eprint = "2009.08715",
    archivePrefix = "arXiv",
    primaryClass = "astro-ph.CO",
    reportNumber = "tel-03027572",
    school = "AstroParticule et Cosmologie, France, U. Paris-Saclay",
    journal = "",
    month = "6",
    year = "2020"
}

@article{Pattison:2021,
    author = "Pattison, Chris and Vennin, Vincent and Assadullahi, Hooshyar and Wands, David",
    title = "{Quantum diffusion during inflation and primordial black holes}",
    eprint = "1707.00537",
    archivePrefix = "arXiv",
    primaryClass = "hep-th",
    doi = "10.1088/1475-7516/2017/10/046",
    journal = "JCAP",
    volume = "10",
    pages = "046",
    year = "2017"
}

@article{GarciaBellido:1996,
  author        = {Garc{\'\i}a-Bellido, Juan and Linde, Andrei D.
                   and Wands, David},
  title         = {{Density perturbations and black hole formation
                   in hybrid inflation}},
  journal       = {Phys. Rev. D},
  volume        = {54},
  pages         = {6040--6058},
  year          = {1996},
  doi           = {10.1103/PhysRevD.54.6040},
  eprint        = {astro-ph/9605094},
  archivePrefix = {arXiv},
}

@article{Clesse:2015,
    author = "Clesse, S{\'e}bastien and Garc{\'\i}a-Bellido, Juan",
    title = "{Massive Primordial Black Holes from Hybrid Inflation as Dark Matter and the seeds of Galaxies}",
    eprint = "1501.07565",
    archivePrefix = "arXiv",
    primaryClass = "astro-ph.CO",
    doi = "10.1103/PhysRevD.92.023524",
    journal = "Phys. Rev. D",
    volume = "92",
    number = "2",
    pages = "023524",
    year = "2015"
}

@article{Ballesteros:2018,
  author        = {Ballesteros, Guillermo and Taoso, Marco},
  title         = {{Primordial black hole dark matter from single
                   field inflation}},
  journal       = {Phys. Rev. D},
  volume        = {97},
  pages         = {023501},
  year          = {2018},
  doi           = {10.1103/PhysRevD.97.023501},
  eprint        = {1709.05565},
  archivePrefix = {arXiv},
  primaryClass  = {astro-ph.CO},
}

@article{Cai:2019,
  author        = {Cai, Yi-Fu and Chen, Xingang and
                   Namjoo, Mohammad Hossein and
                   Sasaki, Misao and
                   Wang, Dong-Gang and Wang, Ziwei},
  title         = {{Revisiting non-Gaussianity from non-attractor
                   inflation models}},
  journal       = {JCAP},
  volume        = {2018},
  number        = {5},
  pages         = {012},
  year          = {2018},
  doi           = {10.1088/1475-7516/2018/05/012},
  eprint        = {1712.09998},
  archivePrefix = {arXiv},
  primaryClass  = {astro-ph.CO},
}

@article{Motohashi:2017,
  author        = {Motohashi, Hayato and Hu, Wayne},
  title         = {{Primordial black holes and slow-roll violation}},
  journal       = {Phys. Rev. D},
  volume        = {96},
  pages         = {063503},
  year          = {2017},
  doi           = {10.1103/PhysRevD.96.063503},
  eprint        = {1706.06784},
  archivePrefix = {arXiv},
  primaryClass  = {astro-ph.CO},
}

@article{Kannike:2017,
    author = {Kannike, Kristjan and Marzola, Luca and Raidal, Martti and Veerm{\"a}e, Hardi},
    title = "{Single Field Double Inflation and Primordial Black Holes}",
    eprint = "1705.06225",
    archivePrefix = "arXiv",
    primaryClass = "astro-ph.CO",
    doi = "10.1088/1475-7516/2017/09/020",
    journal = "JCAP",
    volume = "09",
    pages = "020",
    year = "2017"
}

@article{Martin:1997,
  author        = {Martin, Stephen P.},
  title         = {{A supersymmetry primer}},
  journal       = {Adv. Ser. Direct. High Energy Phys.},
  volume        = {18},
  pages         = {1--98},
  year          = {1998},
  doi           = {10.1142/9789812839657_0001},
  eprint        = {hep-ph/9709356},
  archivePrefix = {arXiv},
}

@article{Ananda:2007,
  author        = {Ananda, Kishore N. and Clarkson, Chris and
                   Wands, David},
  title         = {{The cosmological gravitational wave background
                   from primordial density perturbations}},
  journal       = {Phys. Rev. D},
  volume        = {75},
  pages         = {123518},
  year          = {2007},
  doi           = {10.1103/PhysRevD.75.123518},
  eprint        = {gr-qc/0612013},
  archivePrefix = {arXiv},
}

@article{Baumann:2007,
  author        = {Baumann, Daniel and Steinhardt, Paul J. and
                   Takahashi, Keitaro and Ichiki, Kiyotomo},
  title         = {{Gravitational wave spectrum induced by primordial
                   scalar perturbations}},
  journal       = {Phys. Rev. D},
  volume        = {76},
  pages         = {084019},
  year          = {2007},
  doi           = {10.1103/PhysRevD.76.084019},
  eprint        = {hep-th/0703290},
  archivePrefix = {arXiv},
}

@article{Wu:2025,
    author = "Wu, Yanda and Profumo, Stefano",
    title = "{Superexponential primordial black hole production via delayed vacuum decay}",
    eprint = "2412.10666",
    archivePrefix = "arXiv",
    primaryClass = "hep-ph",
    doi = "10.1103/PhysRevD.111.103524",
    journal = "Phys. Rev. D",
    volume = "111",
    number = "10",
    pages = "103524",
    year = "2025"
}

@article{Franciolini:2025ztf,
    author = "Franciolini, Gabriele and Gouttenoire, Yann and Jinno, Ryusuke",
    title = "{Curvature Perturbations from First-Order Phase Transitions: Implications to Black Holes and Gravitational Waves}",
    eprint = "2503.01962",
    archivePrefix = "arXiv",
    primaryClass = "hep-ph",
    reportNumber = "CERN-TH-2025-044, KOBE-COSMO-25-05",
    doi = "10.1103/tfcx-kzqx",
    journal = "Phys. Rev. Lett.",
    volume = "136",
    number = "17",
    pages = "171404",
    year = "2026"
}

@article{Musco:2019,
    author = "Musco, Ilia",
    title = "{Threshold for primordial black holes: Dependence on the shape of the cosmological perturbations}",
    eprint = "1809.02127",
    archivePrefix = "arXiv",
    primaryClass = "gr-qc",
    doi = "10.1103/PhysRevD.100.123524",
    journal = "Phys. Rev. D",
    volume = "100",
    number = "12",
    pages = "123524",
    year = "2019"
}

@article{Escriva:2020,
    author = "Escriv{\`a}, Albert and Germani, Cristiano and Sheth, Ravi K.",
    title = "{Universal threshold for primordial black hole formation}",
    eprint = "1907.13311",
    archivePrefix = "arXiv",
    primaryClass = "gr-qc",
    reportNumber = "ICC-19-013",
    doi = "10.1103/PhysRevD.101.044022",
    journal = "Phys. Rev. D",
    volume = "101",
    number = "4",
    pages = "044022",
    year = "2020"
}

@article{Wang:2026zvz,
    author = "Wang, Xiao and Bal{\'a}zs, Csaba and Ding, Ran and Tian, Chi",
    title = "{How large are curvature perturbations from slow first-order phase transitions? A gauge-invariant analysis}",
    eprint = "2601.14412",
    archivePrefix = "arXiv",
    primaryClass = "hep-ph",
    doi = "10.1103/y9ks-n9sv",
    journal = "Phys. Rev. D",
    volume = "113",
    number = "12",
    pages = "123026",
    year = "2026"
}

@article{Ai:2026zrs,
    author = "Ai, Wen-Yuan and Xie, Ke-Pan",
    title = "{Reviving primordial black hole formation in slow first-order phase transitions}",
    eprint = "2605.11332",
    journal = "",
    archivePrefix = "arXiv",
    primaryClass = "hep-ph",
    month = "5",
    year = "2026"
}

@article{Geller:2022,
    author = "Geller, Sarah R. and Qin, Wenzer and McDonough, Evan and Kaiser, David I.",
    title = "{Primordial black holes from multifield inflation with nonminimal couplings}",
    eprint = "2205.04471",
    archivePrefix = "arXiv",
    primaryClass = "hep-th",
    reportNumber = "MIT-CTP/5426",
    doi = "10.1103/PhysRevD.106.063535",
    journal = "Phys. Rev. D",
    volume = "106",
    number = "6",
    pages = "063535",
    year = "2022"
}

@article{Qin:2023,
    author = "Qin, Wenzer and Geller, Sarah R. and Balaji, Shyam and McDonough, Evan and Kaiser, David I.",
    title = "{Planck constraints and gravitational wave forecasts for primordial black hole dark matter seeded by multifield inflation}",
    eprint = "2303.02168",
    archivePrefix = "arXiv",
    primaryClass = "astro-ph.CO",
    reportNumber = "MIT-CTP/5525",
    doi = "10.1103/PhysRevD.108.043508",
    journal = "Phys. Rev. D",
    volume = "108",
    number = "4",
    pages = "043508",
    year = "2023"
}

@article{Lorenzoni:2025,
    author = "Lorenzoni, Dario L. and Geller, Sarah R. and Ireland, Zachary and Kaiser, David I. and McDonough, Evan and Wittmeier, Kyle A.",
    title = "{Light Scalar Fields Foster Production of Primordial Black Holes}",
    eprint = "2504.13251",
    archivePrefix = "arXiv",
    journal = "",
    primaryClass = "astro-ph.CO",
    reportNumber = "MIT-CTP/5864",
    month = "4",
    year = "2025"
}

@article{McDonough:2025,
    author = "Lorenzoni, Dario L. and Geller, Sarah R. and Kaiser, David I. and McDonough, Evan",
    title = "{Primordial Black Holes from Inflation with a Spectator Field}",
    eprint = "2512.04199",
    archivePrefix = "arXiv",
    journal = "",
    primaryClass = "astro-ph.CO",
    reportNumber = "MIT-CTP/5973",
    month = "12",
    year = "2025"
}

\end{document}